\documentclass[aip,jcp,reprint,nobibnotes,superscriptaddress,floatfix, 
showpacs,citeautoscript,showkeys]{revtex4-1}

\usepackage{graphicx}
\usepackage{scalerel}
\usepackage[version=4]{mhchem}
\usepackage{latexsym}
\usepackage{textcomp}
\usepackage{amssymb}
\usepackage{amsbsy}
\usepackage{amsmath}
\usepackage{epstopdf}
\usepackage{bm}
\usepackage{longtable}
\usepackage{subfigure}
\usepackage{epsfig}
\usepackage{dcolumn}
\usepackage[Euler]{upgreek}
\usepackage{multirow}
\usepackage{footnote}
\usepackage{stmaryrd}
\usepackage{tabulary}
\usepackage{color}
\usepackage{xcolor}
\usepackage{tikz}
\usetikzlibrary{shapes}
\usepackage{mathtools}
\usepackage{empheq}
\usepackage{cancel}

\usepackage{breqn}

\makeatletter
\let\cat@comma@active\@empty
\makeatother

\newcommand{\markerfour}{\raisebox{0pt}{\tikz{\node[draw=blue,scale=0.6,diamond,fill=none](){};}}}

\newcommand{\markerfive}{\raisebox{0pt}{\tikz{\node[draw,scale=0.6,diamond,fill=blue!20!blue](){};}}}

\begin{document}
\title{Rheological consequences of wet and dry friction in a dumbbell model with hydrodynamic interactions and internal viscosity }
\date{\today}
\author{R. Kailasham}
\affiliation{IITB-Monash Research Academy, Indian Institute of Technology Bombay, Mumbai, Maharashtra -  400076, India}
\affiliation{Department of Chemistry, Indian Institute of Technology Bombay, Mumbai, Maharashtra -  400076, India}
\affiliation{Department of Chemical Engineering, Monash University,
Melbourne, VIC 3800, Australia}
\author{Rajarshi Chakrabarti}
\email{rajarshi@chem.iitb.ac.in}
\affiliation{Department of Chemistry, Indian Institute of Technology Bombay, Mumbai, Maharashtra -  400076, India}
\author{J. Ravi Prakash}
\email[Electronic mail: ]{ravi.jagadeeshan@monash.edu}
\affiliation{Department of Chemical Engineering, Monash University,
Melbourne, VIC 3800, Australia}


\begin{abstract}

The effect of fluctuating internal viscosity and hydrodynamic interactions on a range of rheological properties of dilute polymer solutions is examined using a finitely extensible dumbbell model for a polymer. Brownian dynamics simulations are used to compute both transient and steady state viscometric functions in shear flow. The results enable a careful differentiation of the influence, on rheological properties, of solvent-mediated friction from that of a dissipative mechanism that is independent of solvent viscosity.  In particular, hydrodynamic interactions have a significant influence on the magnitude of the stress jump at the inception of shear flow, and on the transient viscometric functions, but a negligible effect on the steady state viscometric functions at high shear rates. Zero-shear rate viscometric functions of free-draining dumbbells remain essentially independent of the internal viscosity parameter, as predicted by the Gaussian approximation, but the inclusion of hydrodynamic interactions induces a dependence on both the hydrodynamic interaction and the internal viscosity parameter. Large values of the internal viscosity parameter lead to linear viscoelastic predictions that mimic the behavior of rigid dumbbell solutions. On the other hand, steady-shear viscometric functions at high shear rates differ in general from those for rigid dumbbells, depending crucially on the finite extensibility of the dumbbell spring.

\end{abstract}

 
\maketitle

\section{\label{sec:intro}Introduction}

The influence of internal friction on the rheological response of polymer solutions has been studied for several decades, dating back to the inception of the development of coarse-grained kinetic theory models for polymer dynamics~\cite{kuhn1945bedeutung,cerf51,Peterlin1967,Booij1970}. The inclusion of
a resistive force proportional to the rate of change of the connector vector between beads in addition to the spring force has been shown to lead to a finite limiting value for the infinite frequency limit of the dynamic viscosity~\cite{Peterlin1967}, instantaneous stress jumps at the inception of steady shear flow~\cite{Manke1988}, and to a shear-rate dependent viscosity~\cite{Schieber1993,Hua1995307}. More recently, several experimental~\cite{Qiu20043398,Khatri20071825,Soranno07032017,Soranno2018,Ameseder2018}, theoretical~\cite{Pugh1975,Khatri20076770,Samanta2015,Samanta2016,Samanta2016165,Bian2017,Bian2018} and simulation studies~\cite{DeSancho2014,Echeverria2014,Schulz20154565,makarov2018} have shown that the presence of internal friction modulates conformational changes in a number of different biological contexts. This includes slowing down the process of protein folding~\cite{Qiu20043398}, affecting the dynamics of intermolecular interactions in intrinsically disordered proteins~\cite{Soranno07032017,Soranno2018}, and influencing stretching transitions in single biomolecule force spectroscopy~\cite{Khatri20071825,Khatri20076770}. In a parallel development, recent advances in modelling the non-equilibrium behaviour of polymer solutions have revealed the crucial role played by fluctuating hydrodynamic interactions (HI) in determining the dynamics of polymer chains~\cite{Prabhakar20021191,Sunthar2005,Larson2005,Schroeder2018}. Coarse-grained polymer models that include both fluctuating internal friction and hydrodynamic interactions, however, are rare~\cite{Hua19961473}, with the majority including hydrodynamic interactions in a pre-averaged manner~\cite{manke1992stress,Dasbach19924118}.
In this work, we use Brownian dynamics (BD) simulations to solve a coarse-grained model which incorporates finite chain extensibility, fluctuating internal friction and hydrodynamic interactions, in order to study the relative roles played by internal friction and hydrodynamic interactions in determining the dynamics of polymer molecules. 

The earliest models for polymer chains, proposed by Rouse~\cite{Rouse1953} and Zimm~\cite{Zimm1956}, modelled macromolecules as massless beads (which act as centres of friction) connected together by entropic springs. These models, which do not incorporate any internal mode of dissipation, or internal friction, are successful in qualitatively describing several rheological properties of polymer solutions, such as small amplitude oscillatory material functions over a range of frequencies~\cite{Bird1987b,Larson1988} and a non-zero first normal stress difference coefficient in shear flow~\cite{Bird1987b}. Refined models that account for the finite extensibility of the springs are also able to predict the shear-rate dependence of viscometric functions~\cite{Wedgewood1988}.

The Rouse and Zimm models predict that the dynamic viscosity of polymer solutions, $\eta'$ approaches the solvent viscosity, $\eta_s$, at high frequencies. However, experiments studying the viscoelastic properties of polystyrene solutions~\cite{Lamb1964,Massa1971} reveal that in the high-frequency regime, $\eta'$ plateaus at a value higher than $\eta_s$.
Polymer solutions have also been observed to exhibit a ``jump'' in stress at the inception of flow~\cite{Mackay1992, liang1993stress}, and when flow is switched off. Such a jump was found to be higher than the contribution from the Newtonian solvent. These observations have been predicted by rigid-rod models~\cite{Bird1987b}, but not by bead-spring or flexible polymer models extant at that time. On the other hand, it was found that a phenomenological incorporation of internal viscosity (IV) into the standard models~\cite{Manke1988,manke1992stress,Hua1995307} was able to successfully predict these experimental observations. Interestingly,  Gerhardt and Manke~\cite{Gerhardt1994} showed subsequently that the stress jump and the high-frequency plateau in dynamic viscosity are identically equal to each other for linear viscoelastic fluids.

The effects of internal friction have also been observed in the dynamics of biomolecules. Experiments on cold shock proteins have found that the reconfiguration time of proteins tends to a non-zero value even in the limit of zero solvent viscosity~\cite{Qiu20043398,Soranno07032017,Soranno2018}. In the context of single molecule stretching experiments on polysaccharides~\cite{Khatri20071825}, internal friction manifests as the resistance to change in the isomeric state of dextran monomers from the \textit{chair} to the \textit{boat} state, as seen from thermal noise force spectroscopy. A continuum description of the polymer chain, denoted as the Rouse model with Internal Friction (RIF) by McLeish and coworkers~\cite{Khatri20076770}, has been found to satisfactorily explain~\cite{Pugh1975,Samanta2013,Samanta2014,Samanta2016165} these experimental observations.

Despite the success of models with internal viscosity in explaining certain experimental observations, there has been a long debate about the physical origins of internal friction.  One popular contemporary view is that the resistance to dihedral angle rotation between adjacent bonds of a polymer chain is a plausible physical source of internal friction~\cite{Manke1985,DeSancho2014}. While there exists a static persistence length in polymer chains, due to the {\it{trans}} state being more energetically favourable than the {\it{gauche}}, there also exists a dynamic persistence time, associated with the mean hopping time for transition between dihedral angle states~\cite{degennes,Rubinstein2003}. de Gennes~\cite{degennes} argues that at timescales lower than this persistence time, the polymer appears to be frozen, and resists any change in its conformation, in the manner of an ``internal'' friction. There exist molecular dynamics simulation studies which appear to suggest that such transitions are the source of internal friction~\cite{DeSancho2014,Echeverria2014}.

A recent study by ~\citet{Soranno2018} compares experimental data on the reconfiguration time of proteins against the predictions of several models of internal friction, and concludes that it is not only difficult to discriminate between the predictions made by the models but also non-trivial to assign a single mechanism as the source of internal friction. There are studies which argue that internal friction seems to stem from a collection of effects which includes, but is not limited to, dihedral angle rotations~\cite{DeSancho2014,Echeverria2014}, intramolecular interactions~\cite{Alexander-Katz2009}, such as hydrodgen bonds~\cite{Schulz20154565} and disulfide linkages~\cite{Ameseder2018}, and a coupling of the translational and rotational degrees of freedom to the dihedral angle~\cite{Daldrop2018}. 

While it is not clear if the reconfiguration time can be neatly partitioned to represent a solvent-viscosity dependent contribution and another contribution that does not depend on solvent viscosity, we use the terms ``wet'' and ``dry'' friction in this work to denote the two modes of dissipation, namely, solvent drag and internal friction. The simplest models for solvent drag, such as the Rouse model, do not account for perturbations in the solvent velocity field due to hydrodynamic interactions, whereas the solvent drag considered here includes their presence and accounts for their fluctuations.
 
As suggested by Manke and Williams~\cite{Manke1985}, if polymer chains are modeled at the monomer level, by considering a full-description of the bond-lengths, bond-angles, and the barriers separating the dihedral states, there would be no need for the concept of internal viscosity. They argue that the necessity to include IV arises only because the coarse-grained description of a polymer chain lumps together several monomer segments into a ``bead". In such a picture, a barrier to torsional angle rotation can no longer be defined meaningfully. This is akin to solvent friction that only arises when the solvent degrees of freedom are coarse-grained. To describe the forces acting between the beads at this level of modeling, simple mechanical models have been used. The inclusion of a dashpot, in parallel with the spring connecting adjacent beads, provides a rate-dependent restoring force to any change in the length of the connector vector joining the two beads.

The correct form of the expression to be used for the force in the connector vector joining the two beads in models with internal viscosity was initially disputed, with most researchers using the linearized rotational velocity (LRV) approximation proposed by Cerf~\cite{cerf51} and Peterlin~\cite{Peterlin1967}. Subsequent analytical work by Williams and coworkers~\cite{Manke1988,Dasbach19924118} aimed at capturing the stress jump in polymer solutions helped conclusively identify the correct form of the force expression, which was identical to that suggested earlier by Kuhn\cite{kuhn1945bedeutung} in 1945. Furthermore, they showed that the LRV approximation for the treatment of internal viscosity was incorrect.

Schieber and coworkers have published a series of papers examining the rheological properties of dumbbell models with internal viscosity using the Gaussian approximation~\cite{Schieber1993}, and subsequently, with exact BD simulations~\cite{Hua1995307}. However, these computations were restricted to the study of internal viscosity alone. The importance of including fluctuating hydrodynamic interactions in order to obtain accurate predictions of non-linear dynamic properties of polymer solutions is well established, thanks to the development of BD simulations~\cite{Ottinger1996} that enable exact solutions of models with HI~\cite{Prabhakar20021191,Sunthar2005}. Additionally, the inclusion of HI would help to accurately identify the effect of  ``wet"  and ``dry" friction on the rheological properties of dilute polymer solutions.

In this work, we consider a dumbbell model of the polymer with a finitely extensible spring, and fluctuating internal friction and hydrodynamic interactions, and examine its properties at equilibrium, and in the presence of flow. 

Analytical studies by Manke and Williams~\cite{manke1992stress} predict that the stress jump of a model with IV and pre-averaged HI would be higher than that for a model with IV alone. A prior BD simulation study of a dumbbell model with fluctuating internal viscosity and  hydrodynamic interactions, which examined the stress and velocity fields during startup of shear-flow using the CONFESSIT approach~\cite{Hua19961473}, concludes that hydrodynamic interactions have a negligible effect on the stress field. Here, we re-examine the accuracy of this prediction, with particular attention to the magnitude of the stress jump in the presence of fluctuating hydrodynamic interactions.

The Gaussian approximation for internal viscosity~\cite{Schieber1993} concludes that it has no effect on zero-shear rate viscometric functions. The validity of this prediction is scrutinized using exact BD simulations, particularly in the presence of both fluctuating HI and IV. Building upon prior work ~\cite{Gerhardt1994,Hua1996}, we also present what we consider to be a hitherto unexplored relationship between zero-shear rate properties, the relaxation modulus and the stress jump.

Polymer solutions and melts are commonly observed to exhibit an ``overshoot'' in their rheological properties\cite{Bird1987a} when subjected to shear flow. The effect of fluctuating internal viscosity and hydrodynamic interactions on the magnitude, and the time of occurrence of the overshoot is analyzed in this work, and compared to prior observations.

It is well known that dumbbell models which limit the extensibility of the spring predict the shear-thinning of viscometric functions~\cite{Warner1972,Christiansen1977}. The effect of internal viscosity on such shear-thinning has already been studied~\cite{Hua1995307}. Here, the combined effect of fluctuating internal viscosity and hydrodynamic interactions on this phenomenon is quantitatively analyzed by comparing shear-thinning exponents for the various cases.

Manke and Williams have analytically examined the relationship between rigid dumbbells and dumbbell models with an infinite value of the internal viscosity parameter~\cite{Manke1986,Manke1989,Manke1991,Manke1993}, and predict that an ensemble of Hookean dumbbells with an infinite value of the internal viscosity parameter would resemble the viscometric functions of an ensemble of rigid dumbbells, at least qualitatively. ~\citet{Hua1996} have compared the linear viscoelastic properties of Hookean dumbbells with IV to that of rigid dumbbells, and find that the relaxation modulus of dumbbells with a high value of the IV parameter agrees remarkably well with that of rigid dumbbells with a Gaussian distribution of lengths. We use exact BD simulations to calculate the linear viscoelastic and viscometric properties of dumbbells with a high value of the IV parameter, and compare the results against prior observations. We find that the nature of the spring force law qualitatively influences the predicted viscometric functions.

The paper is structured as follows. In section II we describe the dumbbell model for a polymer, its solution methodology and the simulation details.  Section III, which summarises our results and the relevant discussions, is subdivided into four sections; the first section deals with code validation, the second discusses the transient response of dumbbells subjected to shear-flow, the third presents the steady-state results of viscometric functions, and the fourth compares the results for models with a high value of the internal viscosity parameter against that of rigid dumbbells. The key findings of this work are summarised  in the final section. Appendix A details the derivation of the appropriate Fokker-Planck and stochastic differential equations for the dumbbell model considered in our work, appendix B contains the derivation of the stress tensor expression used for calculating the viscometric functions, and appendix C describes the derivation of the relaxation modulus for an ensemble of rigid dumbbells with a FENE distribution of lengths. 

\section{\label{sec:SM} Governing Equations and Simulation Details}
A dumbbell model for the polymer with two massless beads, each of radius $a$, connected by a finitely-extensible non-linear elastic (FENE) spring in parallel with a dashpot, is used as a coarse-grained model for a polymer. The dumbbells are suspended in an incompressible, Newtonian solvent of viscosity $\eta_s$. The velocity-field in the fluid is written as
\begin{equation}
\bm{v}(\bm{r},t)=\bm{v}_0+\boldsymbol{\kappa}(t)\cdot\bm{r}
\end{equation}
where $\bm{r}$ is the position co-ordinate with respect to a fixed frame of reference, $\bm{v}_0$ is a constant vector, and 
$\boldsymbol{\kappa}\equiv(\nabla\bm{v})^{T}$ is the transpose of the fluid velocity gradient, which can be a function of time but is independent of the position, $\bm{r}$.

The co-ordinates of the beads of the dumbbell are given by $\bm{r}_1$ and $\bm{r}_2$. The connector vector joining the two beads is denoted by $\bm{Q}\equiv\bm{r}_2-\bm{r}_1$. The configurational distribution function, $\psi(\bm{Q},t)$, is the probability that the configuration of the dumbbell lies between $\bm{Q}$ and $\bm{Q}+d\bm{Q}$ at time $t$. The force in the connector vector, $\bm{F}^{(c)}$, has two components: a FENE spring contribution, $\bm{F}^{(s)}$, and a contribution due to internal viscosity, $\bm{F}^{(IV)}$, 
\begin{equation}
\bm{F}^{(c)} = \bm{F}^{(s)} + \bm{F}^{(IV)}
\end{equation}
where 
\begin{equation}
\bm{F}^{(s)} = \frac{\partial \phi}{\partial \bm{Q}} =  \frac{H\bm{Q}}{1-(Q/Q_0)^2}
\end{equation}
and
\begin{equation}
\bm{F}^{(IV)} = K\frac{\bm{QQ}}{Q^2}\cdot\llbracket{\dot {\bm{Q}}}\rrbracket
\end{equation}
Here, $H$ is the spring constant, $Q_0$, the maximum allowed length of the spring, and $\phi$, the conservative intramolecular FENE potential between the two beads. $K$ is the internal viscosity co-efficient, which has the same dimensions as that of the bead friction co-efficient, $\zeta(\coloneqq 6\pi \eta_s a)$, and $\llbracket{\dot {\bm{Q}}}\rrbracket$, is the momentum-averaged rate of change of the connector vector $\bm{Q}$.

The equation of continuity for the probability density $\psi(\bm{Q},t)$ can be written as~\cite{Bird1987b}
\begin{equation}\label{eq:diff_eqn}
\frac{\partial \psi}{\partial t} = -\frac{\partial}{\partial \bm{Q}}\cdot\left\{\psi\llbracket{\dot {\bm{Q}}}\rrbracket\right\}
\end{equation}
An expression for $\llbracket{\dot {\bm{Q}}}\rrbracket$ can be derived by considering a force balance on the beads, assuming that their masses are negligible~\cite{Bird1987b}. As shown in Appendix~\ref{sec:App_A}, this leads to
\begin{widetext}
\begin{equation}
\label{eq:dim1}
\llbracket{\dot {\bm{Q}}}\rrbracket   = \left[ \boldsymbol{\delta}   - \frac{\epsilon\beta}{\epsilon\beta  + 1}\frac{\bm{QQ}}{Q^2}\right]    \cdot \Biggl(\left[\boldsymbol{\kappa}\cdot\bm{Q}\right] 
  - \frac{k_BT}{\zeta}\bm{M}\cdot\frac{\partial}{\partial \bm{Q}}\ln \psi  
    - \frac{1}{\zeta}\bm{M}\cdot\frac{\partial \phi}{\partial \bm{Q}} \Biggr)
\end{equation}
\end{widetext}
Here $\epsilon \coloneqq 2K/\zeta$ is the IV parameter, $k_B$, the Boltzmann's constant and $T$, the absolute temperature of the solution. The dimensionless tensor $\bm{M}$ captures the effects of hydrodynamic interactions.  It is related to the HI tensor,  $\boldsymbol{\Omega}(\bm{Q})$ by
\begin{equation} \label{eq:Mtensor}
\bm{M}=2(\boldsymbol{\delta}-\zeta \boldsymbol{\Omega})
\end{equation}
where
\begin{equation}\label{eq:hitensor}
\boldsymbol{\Omega}(\bm{Q})=\frac{3a}{4\zeta Q}\left(A\boldsymbol{\delta} + B\frac{\bm{QQ}}{Q^2}\right)
\end{equation}
In Brownian dynamics simulations, the HI tensor must remain positive-definite for all values of $Q\equiv|\bm{Q}|$. One choice is the Rotne-Prager-Yamakawa (RPY) expression for the HI tensor, in which the variables $A$ and $B$ are defined as follows
\begin{align}\label{eq:more_Q}
A & = 1 + \frac{2}{3}\left(\frac{a}{Q}\right)^2, B = 1 - 2\left(\frac{a}{Q}\right)^2 \textrm{for} \hspace{3pt} Q\geq2a \\[10pt]
\label{eq:less_Q}
A & = \frac{4}{3}\left(\frac{Q}{a}\right) - \frac{3}{8}\left(\frac{Q}{a}\right)^2, B = \frac{1}{8}\left(\frac{Q}{a}\right)^2 \textrm{for} \hspace{3pt} Q<2a
\end{align}
For the special case of dumbbells, it was found that a regularization~\cite{Zylka1989} of the Oseen-Burgers expression for the HI tensor has a smooth dependence on $Q$, is positive-definite for all values of $Q$, and agrees with the RPY tensor to order $Q^{-3}$. The values of $A$ and $B$ for the regularized Oseen-Burgers (ROB) expression for the HI tensor are as follows, 
\begin{align}\label{eq:A_ROBT}
A & = \frac{Q^6+(7/2)p^2Q^4+(9/2)p^4Q^2}{\left(Q^2+p^2\right)^3} \\[10pt]
\label{eq:B_ROBT}
B & = \frac{Q^6+(3/2)p^2Q^4-(3/2)p^4Q^2}{\left(Q^2+p^2\right)^3}
\end{align}
where $p=2a/\sqrt{3}$. The quantity $\beta$ in Eq.~(\ref{eq:dim1}) is dimensionless and is defined as 
\begin{equation}
\beta=1-\frac{h}{Q}(A+B)
\end{equation}
with $h=\left(3/4\right)a$.

By substituting Eq.~(\ref{eq:dim1}) into the equation of continuity [Eq.~(\ref{eq:diff_eqn})], the Fokker-Planck equation for a FENE dumbbell with HI and IV is obtained as 
\begin{widetext}
\begin{equation}
\label{eq:fp_dim} 
\frac{\partial \psi}{\partial t}  = - \frac{\partial}{\partial \bm{Q}}\cdot\Biggl\{\left( \boldsymbol{\delta} - \frac{\epsilon\beta}{\epsilon\beta + 1}\frac{\bm{QQ}}{{Q}^2}\right)\cdot\left(\boldsymbol{\kappa}\cdot\bm{Q}   
- \frac{1}{\zeta}\bm{{M}}\cdot\frac{\partial \phi}{\partial \bm{Q}}\right)\psi\Biggr\} 
 + \frac{k_BT}{\zeta}\frac{\partial}{\partial \bm{Q}}\cdot\left\{\left[\left( \boldsymbol{\delta} - \frac{\epsilon\beta}{\epsilon\beta + 1}\frac{\bm{QQ}}{{Q}^2}\right)\cdot\bm{{M}}\right]\cdot\frac{\partial \psi}{\partial \bm{Q}}\right\}
\end{equation}
It is convenient to non-dimensionalize the equations obtained so far using the time-scale, $\lambda_{\text{H}}=\zeta/4H$, and the length-scale, $l_{\text{H}}=\sqrt{k_BT/H}$. Dimensionless quantities are denoted with an asterisk as a superscript, such that 
\begin{equation}\label{eq:scale}
t^* = \frac{t}{\lambda_{\text{H}}};\,\bm{Q}^* = \frac{\bm{Q}}{l_H};\,b = \frac{Q_0^2}{l_H^2};\,\boldsymbol{\kappa}^* = \lambda_{\text{H}}\boldsymbol{\kappa};\,\phi^* = \frac{\phi}{k_BT};\, \psi^* = \psi {l^3_H}
\end{equation}
Scaling the variables as shown above, the dimensionless Fokker-Planck equation for a FENE dumbbell with internal viscosity and hydrodynamic interactions is obtained as follows. 
\begin{align}\label{eq:fp_mid_rpy1}
 \frac{\partial \psi^*}{\partial t^*}  &= -\frac{\partial}{\partial \bm{Q}^*}\cdot\Biggl\{\left( \boldsymbol{\delta}  - \frac{\epsilon\beta^*}{\epsilon\beta^* + 1}\frac{\bm{Q^*Q^*}}{{Q}^{*2}}\right)\cdot 
\Biggl(\boldsymbol{\kappa^*}\cdot\bm{Q^*}  - \frac{\bm{M^*}}{2}\cdot\frac{\frac{1}{2}\bm{Q}^*}{1-{Q^{*2}}/{b}}\Biggr)\psi^*\Biggr\} \nonumber\\
& + \frac{1}{4}\frac{\partial}{\partial \bm{Q}^*}\cdot\left\{\left[\left( \boldsymbol{\delta} - \frac{\epsilon\beta^*}{\epsilon\beta^* + 1}\frac{\bm{Q^*Q^*}}{{Q}^{*2}}\right)\cdot\bm{M^*}\right]\cdot\frac{\partial \psi^*}{\partial \bm{Q}^*}\right\} 
\end{align}
\end{widetext}

Note that $\beta^*$ and $\beta$ are both dimensionless. However, 
\begin{equation}
\beta^*=1-\frac{\alpha}{Q^*}(A^*+B^*)
\end{equation}
where $A^*$ and $B^*$ can be obtained by recasting all the dimensional quantities in Eqs.~(\ref{eq:more_Q})\textemdash~(\ref{eq:B_ROBT}) into their non-dimensionalized form, and $\alpha$ is given by $\alpha=\left(3/4\right)\sqrt{\pi}h^*$, where $h^*=a/\left(\sqrt{\pi}l_H\right)$ is the hydrodynamic interaction parameter. Note that $\bm{M^*}$ is defined similarly to $\bm{M}$ in Eq.~(\ref{eq:Mtensor}), but with the components of the hydrodynamic interaction tensor written in terms of dimensionless quantities.

As shown in Appendix~\ref{sec:App_A}, the stochastic differential equation (SDE) equivalent to Eq.~(\ref{eq:fp_mid_rpy1}) can be derived using the It\^o interpretation as,  
\begin{widetext} 
\begin{equation}
\begin{split}
\label{eq:sde_final}
d\bm{Q}^* & = \biggl[\boldsymbol{\kappa}^*\cdot\bm{Q}^* +\frac{g_{2}}{2}\frac{\bm{Q^*}}{Q^*} - \left(\frac{Q^*-A^* \alpha}{Q^*}\right)\left(\boldsymbol{\delta}-g_1\frac{\bm{Q^*Q^*}}{{Q}^{*2}}\right) 
\cdot\left(\frac{\frac{1}{2}\bm{Q}^*}{1-{Q^{*2}}/{b}} + \epsilon\frac{\bm{Q^*Q^*}}{Q^{*2}}\cdot\boldsymbol{\kappa^*}\cdot\bm{Q^*}\right)\biggr]{d}t \\[10pt]
& + \sqrt{\frac{Q^*-A^* \alpha}{Q^*}} \left[\boldsymbol{\delta} - \left(1 - \sqrt{1-g_1} \right)\frac{\bm{Q^*Q^*}}{Q^{*2}}\right]\cdot {d}\bm{W}_t 
\end{split}
\end{equation}

In the second term of the above equation, the prefactor that multiplies ${d}\bm{W}_t$ is the diffusion tensor. It is worth noting that the functional form of the SDE remains the same, irrespective of the choice of the HI tensor. Definitions of the prefactors, $g_1$ and $g_2$, which are functions of $\left\{ \alpha, \epsilon, A^*, B^*, Q^*\right\}$, are given in Appendix~\ref{sec:App_A}. Eq.~(\ref{eq:sde_final}) is solved using a semi-implicit predictor-corrector algorithm, as outlined below. Note that, while Eqs.~(\ref{eq:pred}) to  (\ref{eq:cubic}) below are in their non-dimensionalized form, the asterisk superscript has been dropped from these equations for notational simplicity. 
\vskip5pt 
\noindent \textbf{Predictor Step}
\begin{align}\label{eq:pred}
\tilde{\bm{Q}}(t_{j+1})&=\bm{Q}(t_{j})+\Biggl[\boldsymbol{\kappa}(t_{j})\cdot\bm{Q}(t_{j})  -  \Biggl(\frac{\epsilon \beta(t_{j})}{\epsilon \beta(t_{j}) + 1}\Biggr) 
\frac{\bm{Q}(t_{j})\bm{Q}(t_{j})}{Q^2(t_{j})}\cdot\bigl(\boldsymbol{\kappa}(t_{j})\cdot\bm{Q}(t_{j})\bigr) \nonumber \\
 & - \frac{1}{2\left(1-{Q^2(t_{j})}/{b}\right)}\left(\frac{\beta(t_{j})}{\epsilon \beta(t_{j}) + 1}\right)\bm{Q}(t_{j}) 
   + \frac{g_2(t_{j})}{2}\frac{\bm{Q}(t_{j})}{Q(t_{j})}\Biggr]\Delta t_j + \Delta \bm{S}_{j}
\end{align}
where 
\begin{equation}
\Delta \bm{S}_{j} = \bm{b}_{j}\cdot\Delta \bm{W}_{j}
\end{equation}
and 
\begin{equation}
\bm{b}_{j} = \sqrt{\frac{Q(t_{j})-A(t_{j}) \alpha}{Q(t_{j})}}\Biggl[\boldsymbol{\delta}
- \left(1-\sqrt{1-g_1(t_{j})}\right)\frac{\bm{Q}(t_{j})\bm{Q}(t_{j})}{Q^2(t_{j})}\Biggr]
\end{equation}
$\Delta \bm{W}_j$ is a vector of three independent Wiener processes, each of mean zero and variance $\Delta t_j$.
\vskip5pt 
\noindent \textbf{Corrector Step} 
\begin{equation}
\begin{split}
\label{eq:corr}
& \Biggl[1+\frac{1}{4}\Biggl(\frac{\tilde{\beta}(t_{j+1})}{\epsilon \tilde{\beta}(t_{j+1}) + 1}\Biggr)\frac{\Delta t_j }{1-{Q^2(t_{j+1})}/{b}}\Biggr]\bm{Q}(t_{j+1})=
\bm{Q}(t_{j})\Biggl[1-\frac{1}{4}\Biggl(\frac{\beta(t_{j})}{\epsilon \beta(t_{j}) + 1}\Biggr)\Biggl(\frac{1}{1-{Q^2(t_{j})}/{b}}\Biggr)
+\frac{g_2(t_{j})}{4Q(t_{j})}\Biggr]\Delta t_j  \\[5pt] 
& + \tilde{\bm{Q}}(t_{j+1})\Biggl[\frac{\tilde{g}_2(t_{j+1})}{4\tilde{Q}(t_{j+1})}\Biggr]\Delta t_j  + \frac{1}{2}\Biggl[\boldsymbol{\kappa}(t_{j})\cdot\bm{Q}(t_{j})+\boldsymbol{\kappa}(t_{j+1})\cdot\tilde{\bm{Q}}(t_{j+1})-\left(\frac{\epsilon \beta(t_{j})}{\epsilon \beta(t_{j}) + 1}\right)
\frac{\bm{Q}(t_{j})\bm{Q}(t_{j})}{Q^2(t_{j})}\cdot\Bigl(\boldsymbol{\kappa}(t_{j})\cdot\bm{Q}(t_{j})\Bigr) \\[5pt]
& - \Biggl(\frac{\epsilon \tilde{\beta}(t_{j+1})}{\epsilon\tilde{\beta}(t_{j+1}) + 1}\Biggr)
\frac{\tilde{\bm{Q}}(t_{j+1})\tilde{\bm{Q}}(t_{j+1})}{\tilde{Q}^2(t_{j+1})}\cdot\Bigl(\boldsymbol{\kappa}(t_{j+1})\cdot\tilde{\bm{Q}}(t_{j+1})\Bigr)\Biggr]\Delta t_j  + \Delta \bm{S}_{j} 
\end{split}
\end{equation}
\end{widetext} 
Here, $\tilde{\bm{Q}}(t_{j+1})$ is the value of $\bm{Q}$ after the predictor step, evaluated in Eq.~(\ref{eq:pred}).

Note that only the FENE spring force term is treated implicitly in the corrector step. By setting the length of the vector on the RHS of Eqn.~(\ref{eq:corr}) to be $L$, and the length of $\bm{Q}(t_{j+1})$ to be $x$,  we get the following cubic equation in $x$
\begin{equation}\label{eq:cubic}
x^3-x^2L-xb\left[1+\frac{\Delta t_j }{4}\left(\frac{\tilde{\beta}(t_{j+1})}{\epsilon \tilde{\beta}(t_{j+1}) + 1}\right)\right]+Lb=0
\end{equation}
Eq.~(\ref{eq:cubic}) can be solved exactly using trigonometric functions ~\cite{Ottinger1996} and only one root lies in the interval $[0,\sqrt{b}]$, which is chosen as the physically relevant solution for $x$.

The Kramers expression for the stress-tensor is not thermodynamically consistent for dumbbells with internal viscosity~\cite{Schieber1994}, and the Giesekus expression cannot be used when hydrodynamic interactions are included~\cite{Bird1987b}. For the general case considered here, with both IV and HI, ~\citet{Hua19961473} suggest that the Kramers-Kirkwood expression can be used, since it is thermodynamically consistent~\cite{Schieber1994},  
\begin{equation}\label{eq:kram_kirk}
{\boldsymbol{\tau}_{\text{{p}}}}=-n_{\text{{p}}}\sum_{\nu}\left<\bm{R}_{\nu}\bm{F}_{\nu}^{(h)}\right>
\end{equation}
where $n_{\text{{p}}}$ is the number of polymer molecules per unit volume, and $\bm{R}_{\nu}$ is the position of the $\nu$th bead with respect to the centre-of-mass, $\bm{r}_{\text{{c}}}$. $\bm{F}_{\nu}^{(h)}$ is the hydrodynamic force acting on bead $\nu$. 

In terms of the connector vector $\bm{Q}$, the stress tensor is given by 
\begin{equation}
\begin{split}
\label{eq:stress_tensor}
{\boldsymbol{\tau^*}_p} =\frac{\boldsymbol{\tau}_p}{n_{\text{p}}k_BT} & = \boldsymbol{\delta} - \left<g_3\frac{\bm{Q^*Q^*}}{1-Q^{*2}/b}\right> -\epsilon\left<g_4\frac{\bm{Q^*Q^*}}{Q^{*2}}\right>  \\[5pt]
& - 2\epsilon\left<g_3\boldsymbol{\kappa^*}:\frac{\bm{Q^*Q^*Q^*Q^*}}{Q^{*2}}\right>
\end{split}
\end{equation}
The steps for arriving at Eq.~(\ref{eq:stress_tensor}) starting from Eq.~(\ref{eq:kram_kirk}) have been outlined in Appendix~\ref{sec:App_B}, along with the definitions of the prefactors $g_3$ and $g_4$. While the second term on the right hand side of Eq.~(\ref{eq:stress_tensor}), represents the elastic contribution to the stress tensor due to the presence of the FENE spring, the third and fourth terms arise due to the presence of IV. The last term on the right hand side is the viscous or dissipative component of the stress tensor, which disappears (appears) instantaneously when the flow is turned off (on). The jump in viscosity at the inception of shear flow is due to this last term, because none of the other terms in the equation contribute to the shear component of the stress tensor, $\tau_{p,yx}$, as the other averages are isotropic at equilibrium, i.e, $\left<Q_xQ_y\right>_{\text{eq}}=0$. Interestingly, bead-rod models and bead-rod chains also have a viscous contribution to stress~\cite{Bird1987b}. Flexible polymer models, with only an entropic spring connecting the beads, do not have this viscous contribution to the shear stress, and only the inclusion of a dashpot in parallel with the spring, through the incorporation of internal viscosity into such models, results in a stress jump.

For steady simple shear flow considered in this work,  
\begin{equation}\label{eq:kappa_shear}
\boldsymbol{\kappa^*}=\lambda_{\text{H}}\boldsymbol{\kappa}=\lambda_{\text{H}}\dot{\gamma}\begin{pmatrix}
0 & 1 & 0\\
0 & 0 & 0\\
0 & 0 & 0
\end{pmatrix}
\end{equation}
where $\lambda_{\text{H}}\dot{\gamma}$ is the dimensionless shear rate. The viscometric functions are given in terms of the components of the stress-tensor as follows, 
\begin{equation}
\eta^*_{\text{p}}=\frac{\eta_{\text{p}}}{n_{\text{p}}k_BT\lambda_{\text{H}}} = - \frac{\tau_{{\text{p}},xy}}{n_{\text{p}}k_BT\lambda_{\text{H}}\dot{\gamma}}
\end{equation}
\begin{equation}\label{eq:fnsd}
\Psi^*_1=\frac{\Psi_1}{n_{\text{p}}k_BT\lambda_{\text{H}}^2} = - \frac{\tau_{{\text{p}},xx}-\tau_{{\text{p}},yy}}{n_{\text{p}}k_BT\lambda_{\text{H}}^2\dot{\gamma}^2}
\end{equation}
\begin{equation}\label{eq:snsd}
\Psi^*_2=\frac{\Psi_2}{n_{\text{p}}k_BT\lambda_{\text{H}}^2} = - \frac{\tau_{{\text{p}},yy}-\tau_{{\text{p}},zz}}{n_{\text{p}}k_BT\lambda_{\text{H}}^2\dot{\gamma}^2}
\end{equation}
where $\tau^*_{p,xy}$ refers to the $xy$-component of the polymer contribution to the stress tensor, $\boldsymbol{\tau^*}_{\text{p}}$, $\eta^*_{\text{p}}$ is the polymer contribution to the viscosity, and $\Psi^*_1$ and $\Psi^*_2$ are the first and second normal stress difference coefficients, respectively. In the time dependent period before steady state is reached, in accord with conventional notation, the transient viscometric functions are denoted by $\eta_p^+$, $\Psi_1^+$, and $\Psi_2^+$.

{\noindent In all the  simulations, the initial ensemble of equilibrium FENE dumbbell configurations (typically of order $10^6$), was picked from a database of $10^7$ equilibrium configurations generated previously for the particular value of the FENE $b$ parameter of interest. The database was created by starting with an ensemble of $10^7$ Gaussian distributed dumbbell connector vectors $\bm{Q}$, and running simulations for thirty dimensionless times without flow ($\boldsymbol{\kappa^*}=0$), for various values of the FENE $b$ parameter and $\epsilon=0$ and $h^*=0$, and saving the final configurations of dumbbells at the end of the run. The probability distribution of multiple batches of $10^6$ samples picked from this database has been checked for agreement with the equilibrium distribution function for the lengths of the connector vectors, given by Eq.~(A20). Furthermore,  the value of $\left<Q^{*2}\right>$ for the initial ensemble has been verified by comparing with the the known analytical result for FENE dumbbells, $\left<Q^{*2}\right>=3b/(b+5)$. Since the equilibrium distribution is determined entirely by the value of $b$, and unaffected by the choice of $\epsilon$ and $h^*$, the database corresponding to the particular choice of $b$ in a simulation was used to pick the initial ensemble of dumbbells (regardless of values of $\epsilon$ and $h^*$), for all the simulations carried out in this work. Before subjecting the ensemble to flow, however, dumbbell configurations picked from such a database are equilibrated without flow (with the necessary values of $\epsilon$ and $h^*$) for an additional duration of one dimensionless time. For transient simulations,  $t^*=0$ denotes the time when the flow is switched on. }

Three different timesteps ($\Delta t$) were chosen for each simulation run, and the results extrapolated to zero time-step width. The choice of $\Delta t$ for a particular simulation depends on the shear-rate and the internal viscosity parameter. Larger values of $\lambda_{\text{H}}\dot{\gamma}$ or $\epsilon$ necessitate the use of smaller timesteps. Each set of $\Delta t$ was tested for convergence before carrying out the production runs. Unless otherwise mentioned, an ensemble of $\mathcal{O} (10^{6})$ dumbbells was used for generating the results.

\section{\label{sec:Results}Results}

\subsection{\label{sec:Validation} Code validation}

\begin{figure}[t]
\includegraphics[width=3.7in,height=!]{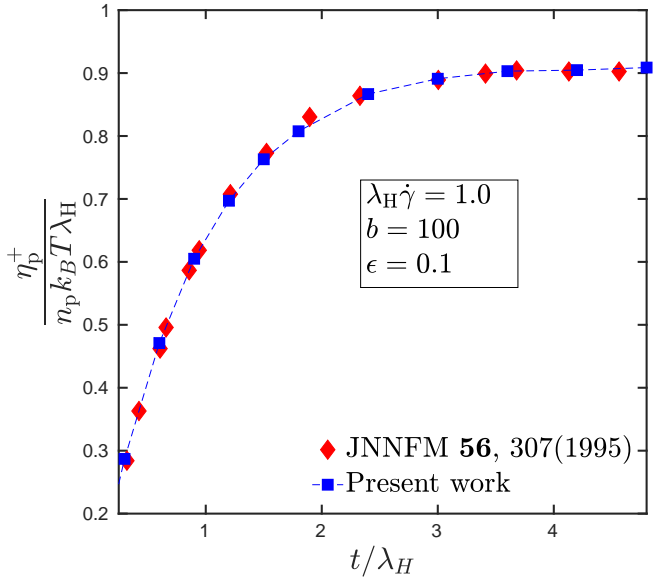}  
\caption{(Color online) Polymer contribution to shear viscosity, ${\eta^{+}_{\text{p}}}/{n_{\text{p}}k_BT\lambda_{\text{H}}}$, as a function of dimensionless time at a dimensionless shear rate of 1.0.  Error bars are smaller than symbol size.}
\label{fig:sr_valid}
\end{figure}

To test the validity of the code, an ensemble of $2\times10^5$ FENE dumbbells, with a dimensionless maximum allowable length of $b=100$ and internal viscosity parameter of $\epsilon=0.1$,  is subjected to shear flow at a dimensionless shear rate of $\lambda_{H}\dot{\gamma}=1.0$. In Fig.~\ref{fig:sr_valid}, the predicted shear viscosity of this ensemble is plotted against dimensionless time, and compared against data from the work by Hua and Schieber~\cite{Hua1995307}. The good agreement between the two results indicates the reliability of the current code. 

As mentioned in the introduction, Hua and Schieber~\cite{Hua19961473} have previously simulated a dumbbell model with internal viscosity and hydrodynamic interactions in startup of shear flow, using the CONFESSIT method. They conclude that hydrodynamic interactions have a negligible effect on the stress field. However, as shown in  Appendix~\ref{sec:App_A}, one of the functions in their governing stochastic differential equations is incorrect, and leads to erroneous predictions. As a consequence, there are no published results in the literature with which our simulations (which incorporate both internal viscosity and  hydrodynamic interactions) can be compared. Nevertheless, as will be seen in Section.~\ref{sec:jump}, our simulations agree with the analytical results of Manke and Williams~\cite{manke1992stress} (who used pre-averaged hydrodynamic interactions) for the stress jump at the onset of shear flow over a range of values of $\epsilon$ and $h^*$, providing some validation for our code in its most general form. 

Simulations which incorporate hydrodynamic interactions using the ROB expression take nearly twice as long to complete, for the same set of parameter values, as when the RPY expression is used. Though the two treatments yield results that are indistinguishable within error bars, the calculation of the shear viscosity of an ensemble of $2\times10^5$ FENE dumbbells with both internal viscosity and hydrodynamic interactions, on an Intel Core i7-6700 CPU, takes about 379 seconds for the RPY case, as opposed to nearly 768 seconds for the ROB case. As a consequence, all simulations which involve HI have been performed using the RPY tensor.
The RPY tensor has two branches and the terms $A$ and $B$ have to be evaluated separately for the two branches, as seen from Eqs.~(\ref{eq:more_Q}) and (\ref{eq:less_Q}). Nevertheless, its implementation is faster due to the higher number of function evaluations required for the ROB tensor, as can be seen from Eqs.~(\ref{eq:s_rho}) and (\ref{eq:magic_RPY}) in Appendix ~\ref{sec:App_A}. Essentially, the definitions of functions $g_1$ and $g_3$ are the same for both RPY and ROB, while the functions $g_2$ and $g_4$ entail more calculations for the ROB case.

\begin{figure}[h]
\begin{center}
\begin{tabular}{c}
{\includegraphics*[width=3.3in,height=!]{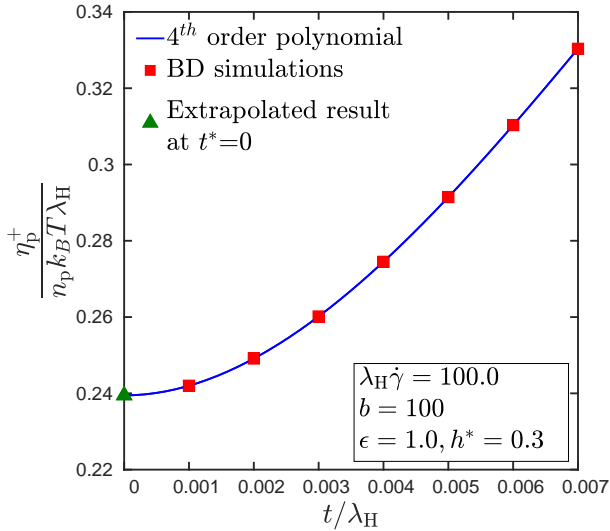}} \\
(a) \\
{\includegraphics*[width=3.3in,height=!]{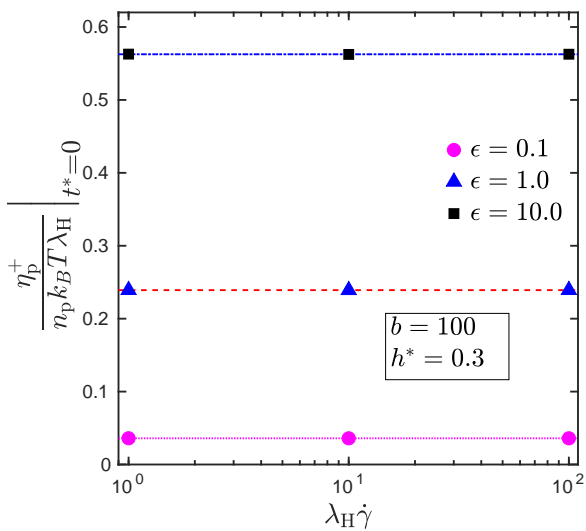}}\\
(b) \\
\end{tabular}
\end{center}
\caption{(Color online) (a) Methodology to find the stress jump by extrapolating to $t^*=0$ using a fourth order polynomial. (b) Shear rate independence of stress jump for various values of $\epsilon$ and a fixed value of $h^*=0.3$.  Dotted horizontal lines are error-weighted averages of the data points they traverse. Error bars are smaller than symbol size.}
\label{fig:jump_measure}
\end{figure}

\subsection{\label{sec:transient} Transient response to shear flow}

To calculate the viscometric functions, an initial equilibrium ensemble of $\mathcal{O} (10^{6})$ FENE dumbbells, picked from the equilibrated database as described in Section.~II, have been used in all the simulations. The transient response of the ensemble-averaged rheological properties recorded as a function of time, are presented in this section.

\subsubsection{\label{sec:jump} The stress jump}
The ``stress jump'' refers to the discontinuous jump in viscosity at the inception of flow, i.e, at  time, $t^*=0$. The methodology to find this quantity has been illustrated in  Fig.~\ref{fig:jump_measure}~(a) for a sample case where an ensemble of dumbbells with IV and HI is subjected to a dimensionless shear rate of 100. The viscosity is recorded as a function of time, and a fourth order polynomial is then fitted to this data (using the ``fit'' functionality of gnuplot, as described by ~\citet{2012arXiv1210.3781Y}), so as to find the stress jump by extrapolation to time $t^*=0$.

Analytical predictions~\cite{manke1992stress,Dasbach19924118,Gerhardt1994} indicate that the stress jump must be independent of shear rate. To test this prediction, FENE dumbbells with $b=100$ and various values of $\epsilon$ and $h^*$ are subjected to three different shear rates, and their corresponding stress jumps are calculated using the method outlined above. In Fig.~\ref{fig:jump_measure}~(b), stress jumps for such dumbbells with varying values of $\epsilon$ and a fixed value of $h^*=0.3$ are plotted against the respective shear rates, and it is seen that the jump is indeed independent of shear rate, thereby confirming the analytical predictions. A similar shear-rate independence is observed when the hydrodynamic interaction parameter is varied.

Manke and Williams~\cite{Manke1988} derived an analytical result for the stress jump exhibited by a Hookean dumbbell with IV. Schieber calculated the same quantity for a FENE dumbbell with IV and obtained the following formula~\cite{Hua1995307}, for the free-draining (FD) case, 
\begin{equation}\label{eq:fd_jump}
\left. \frac{\eta^+_{\text{p}}}{n_{\text{p}}k_BT\lambda_{\text{H}}}\right\vert_{t^*=0,\text{FD}}=\left(\frac{b}{b+5}\right)\frac{2\epsilon}{5(1+\epsilon)}
\end{equation} 
\citet{manke1992stress} have also derived a similar formula for the stress jump exhibited by a Hookean dumbbell with IV and pre-averaged HI. By following a procedure analogous to Schieber's, the analytical result of Manke and Williams can be extended to find the stress jump for a FENE dumbbell with IV and pre-averaged HI, and can be shown to be
 \begin{equation}
\left. \frac{\eta^+_{\text{p}}}{n_{\text{p}}k_BT\lambda_{\text{H}}}\right\vert_{t^*=0,\text{HI}}=\left(\frac{b}{b+5}\right)\frac{2\epsilon}{5(1+\epsilon(1-\sqrt{2}h^*))}
\end{equation} 
Recognizing that $\sigma(0)=-\eta^+_{\text{p}}(0)\dot{\gamma}$, the stress jump ratio is then given by taking a ratio of the above two equations.
\begin{equation}\label{eq:stress_jump}
\frac{\sigma_{\text{HI}}(0)}{\sigma_{\text{FD}}(0)}=\frac{1+\epsilon}{1+\epsilon(1-\sqrt{2}h^{*})}
\end{equation}

\begin{figure}[h]
\centering
\includegraphics[width=3.5in,height=!]{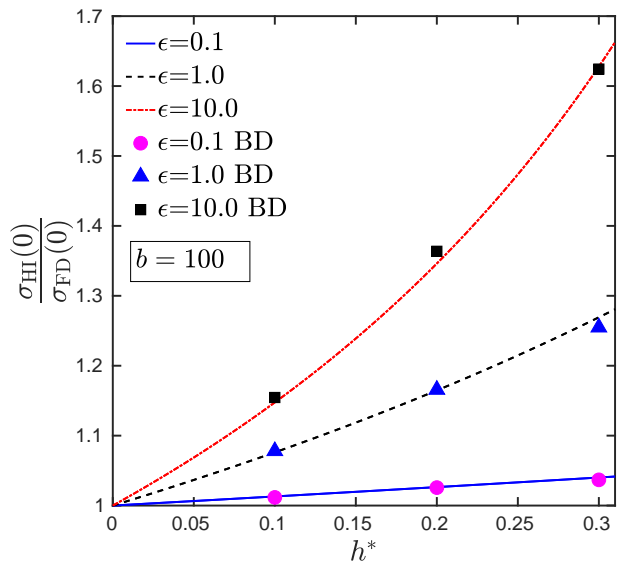}  
\caption{(Color online) Stress jump ratio calculations for FENE dumbbells with $b=100$ for various values of $\epsilon$, $h^*$. Each data point in the figure is obtained as an error-weighted average of the stress jump measured at the different shear rates as shown in Fig.~\ref{fig:jump_measure}~(b). Lines represent Eq.~(\ref{eq:stress_jump}). Error bars are smaller than symbol size.}
\label{fig:jump_ratio}
\end{figure}

This equation has been represented as lines in Fig.~\ref{fig:jump_ratio}, for various values of the internal viscosity parameter. The prediction from analytical theory is that the stress jump in the presence of hydrodynamic interactions and internal viscosity is higher than that due to IV alone. The stress jump ratio for the case with HI to the free draining case, obtained from BD simulations for various combinations of $\epsilon$ and $h^*$,  is plotted in~Fig.~\ref{fig:jump_ratio}. The simulations results agree well with the theoretical prediction, suggesting that fluctuations in HI do not play a significant role in determining the magnitude of the stress jump.

\begin{figure}[h]
\begin{center}
\begin{tabular}{c}
{\includegraphics*[width=3.3in,height=!]{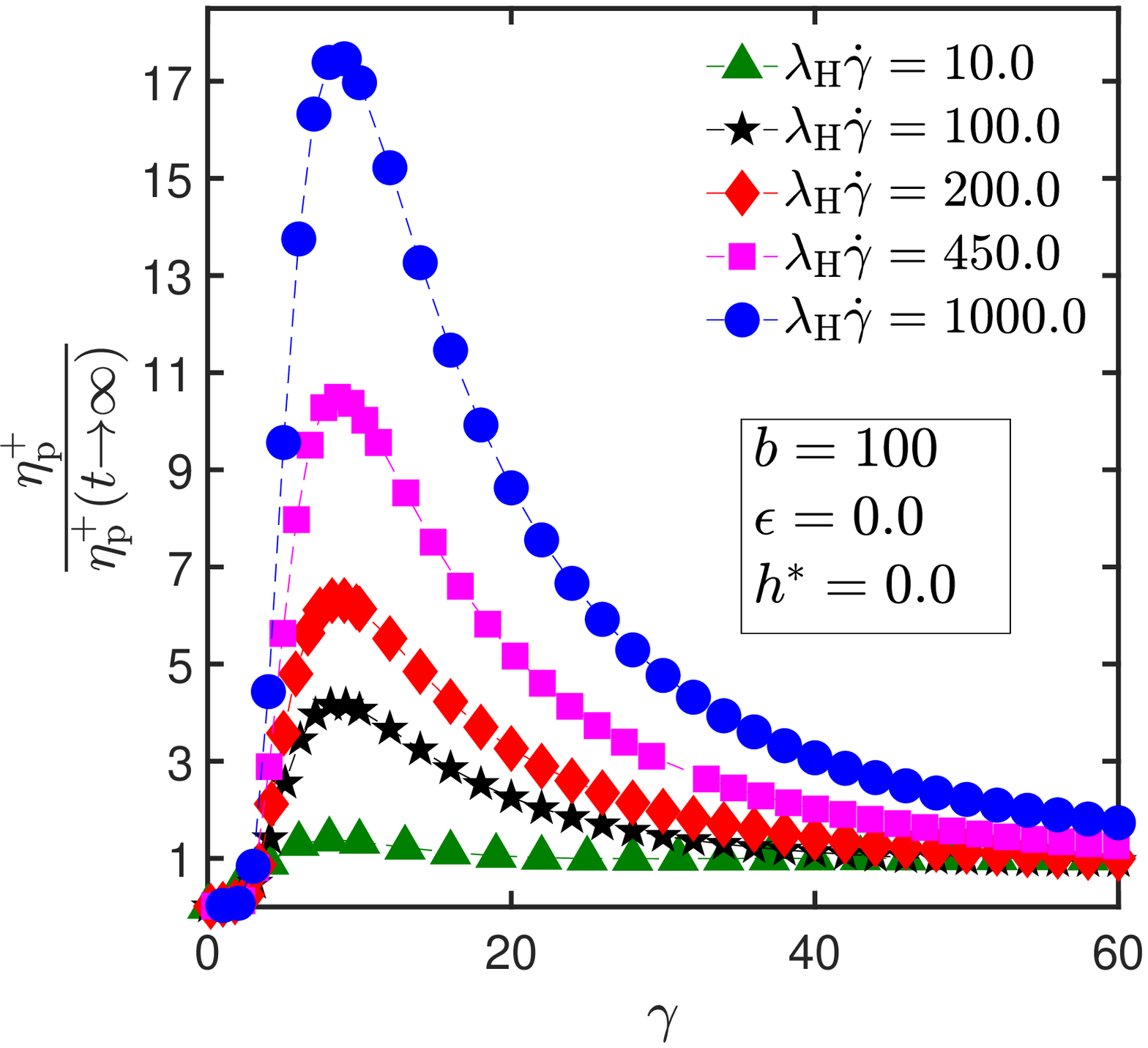}}\\
(a)\\
{\includegraphics*[width=3.3in,height=!]{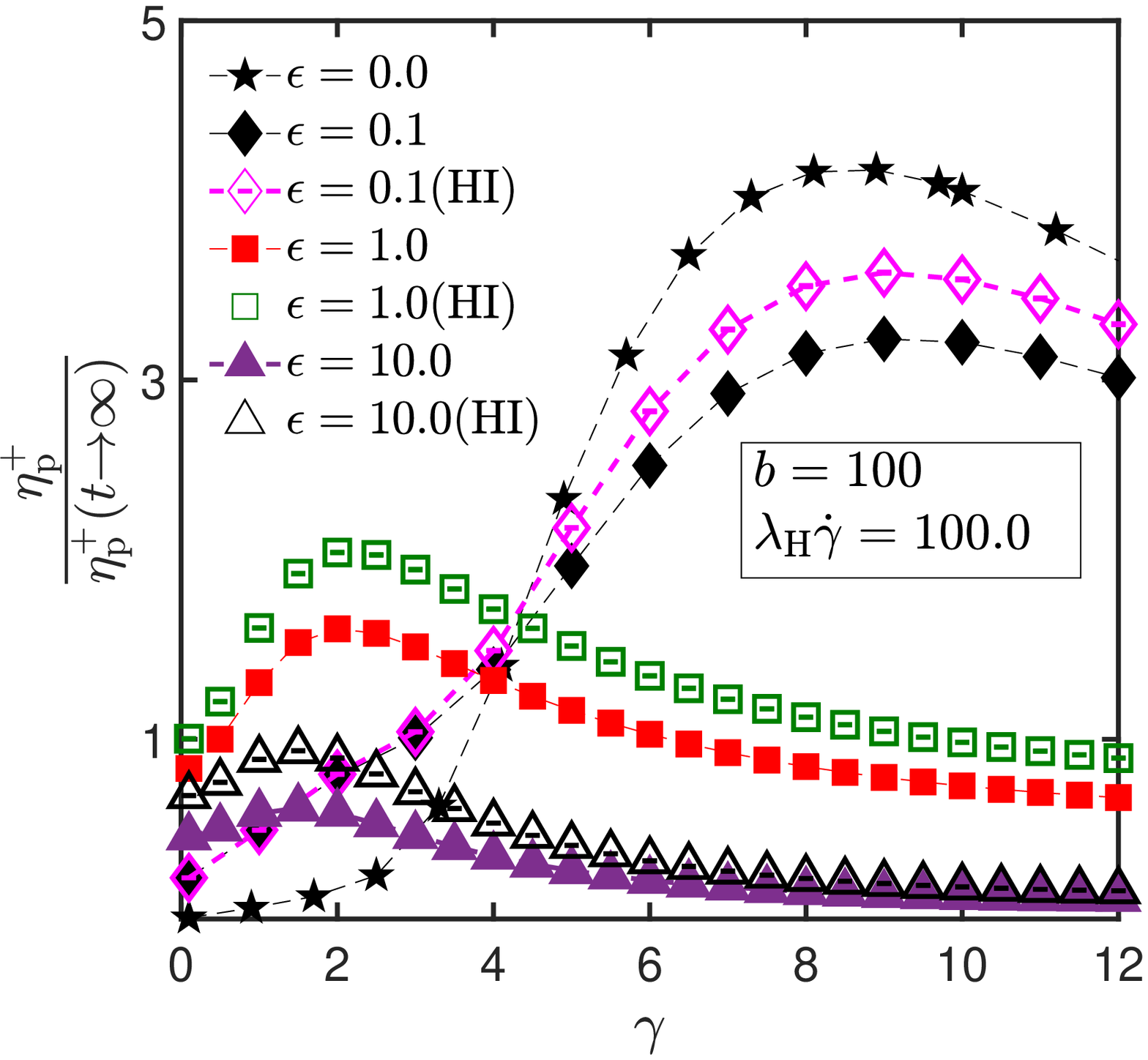}}\\
(b)\\
\end{tabular}
\end{center}
\caption{(Color online) Transient behavior of shear viscosity for FENE dumbbells with $b=100$ and various values of $\epsilon$ and $h^*$, as a function of strain units: (a) overshoot behavior for the pure FENE model, (b) effect of $\epsilon$ and $h^*$ on overshoot at a dimensionless shear rate of 100. {In each case, ${\eta}_{\mathrm{p}}^+$ is normalized by the steady-state value, ${\eta}_{\mathrm{p}}^+(t \to \infty)$, obtained at the same shear rate, and $\epsilon$ and $h^*$, as the transient data themselves.} The hydrodynamic interaction parameter in these simulations is $h^*=0.3$. Error bars are smaller than symbol size.}
\label{fig:eta_overshoot}
\end{figure}

Such a result also seems correct intuitively.  In the absence of IV or HI, the diffusion tensor in the stochastic differential equation is diagonal, i.e, there is no correlation between the motion of the two beads of the dumbbell. 
Introduction of IV adds off-diagonal terms to the diffusion tensor, introducing a coupling between the motion of the two beads. Hydrodynamic interactions also contribute to a coupling and this appears to enhance the effect brought on by IV. In a physical sense, the phenomenon of stress jump arises solely due to internal viscosity, as the ``dashpot" connecting the two beads responds instantaneously to a change in the displacement between the two beads. Hydrodynamic interactions add to this effect because it enhances the strength of the correlation between the motion of the two beads.

Interestingly, our BD simulations indicate that there is no jump in either $\Psi_1$ [defined in Eq.~(\ref{eq:fnsd})] or $\Psi_2$ [defined in Eq.~(\ref{eq:snsd})] at $t^*=0$ for models with internal viscosity. This is in line with prior observations on Hookean dumbbells that employed a Gaussian approximation~\cite{Schieber1993} for internal viscosity.

\subsubsection{\label{sec:trans_eta} Transient viscosity}
As observed in various experiments on polymer solutions and melts~\cite{Bird1987a}, simulations indicate that during the startup of shear flow, the viscosity shows a transient rise, above its steady-state value, commonly termed the ``overshoot", with the highest point of positive deviation from the steady-state value denoted as the magnitude of overshoot. To analyze this phenomenon, the viscosity of dumbbells (normalized by the steady-state value, {${\eta}_{\mathrm{p}}^+(t \to \infty)$}) is plotted as a function of strain units in Fig.~\ref{fig:eta_overshoot}. It is important to note that the phenomenon of overshoot is observed even for the case of FENE dumbbells, without the incorporation of internal viscosity or hydrodynamic interaction effects. 

Throughout the analysis in this section and the next, it must be noted that an ``overshoot" is said to occur only when the normalized quantity attains a value greater than one. Local peaks in the normalized quantity, such as the one observed for the $\epsilon=10$ case in Fig.~\ref{fig:eta_overshoot}~(b), are only considered as ``maxima" and not categorized as an ``overshoot".

In Fig.~\ref{fig:eta_overshoot}~(a), the overshoot behavior in FENE dumbbells with $b=100$ is recorded for various shear rates. At low shear rates, there is no overshoot, and viscosity reaches its steady-state value asymptotically. At a certain threshold shear rate, overshoot is first observed. From there on, the magnitude of the overshoot increases as the shear rate is increased. Interestingly, however, the location of the overshoot, i.e; the strain at which overshoot occurs ($\gamma_{\text{max}}$), is roughly constant over the entire range of shear rates examined in this work, in agreement with the experimentally observed trend for a range of different polymer solutions and melts~\cite{Bird1987a}. 

In Fig.~\ref{fig:eta_overshoot}~(b), the effect of internal viscosity and hydrodynamic interactions on the magnitude and location of overshoot is examined at a fixed dimensionless shear rate of 100. It is seen that a low value of $\epsilon$ (=0.1) dampens the magnitude of overshoot, but does not change $\gamma_{\text{max}}$ significantly.  At higher values of the internal viscosity parameter, the overshoot occurs at a lower strain, and its magnitude is significantly decreased. The inclusion of hydrodynamic interactions increases the magnitude of overshoot in comparison to the free-draining case ($h^*=0$) but does not affect $\gamma_{\text{max}}$ perceptibly. For a given value of the hydrodynamic interaction parameter, the coupling between IV and HI is enhanced at higher values of the IV parameter, at early times (or strains). At later times, the nature of the coupling is non-trivial. {While it appears that the $\epsilon=1.0$ and $\epsilon=10.0$ curves (with and without HI) settle to a value lower than 1 at the highest values of strain in the figure, as a matter of fact they do approach 1 at much larger strains, when the transient viscosity ${\eta}_{\mathrm{p}}^+$ attains its steady-state value, as can be seen from Fig.~\ref{fig:ltime}~(a)}.

It is observed that the dumbbell model with only FENE effects and that with a small value of  $\epsilon$ show nearly identical behaviour, qualitatively.  The onset of overshoot occurs at roughly the same shear rate, ${\lambda_{\text{H}} \dot{\gamma}}\sim10.0$, for both the cases. However, as the IV parameter is increased to higher values, say $\epsilon=1$ or $\epsilon=10$, there is a marked change in the transient response. This is clearly seen from Fig.~\ref{fig:eta_overshoot}~(b), where a shear rate of ${\lambda_{\text{H}} \dot{\gamma}}\sim100.0$ triggers an overshoot in a system with a lower value of $\epsilon$, but only causes a long wavelength oscillation in the shear viscosity for a system with $\epsilon=10$. {The occurrence of oscillations is discussed further in the context of Figs.~\ref{fig:ltime} below}.

\subsubsection{\label{sec:trans_psi1} Transient first normal stress difference coefficient}

\begin{figure}[h]
\begin{center}
\begin{tabular}{c}
 {\includegraphics*[width=3.3in,height=!]{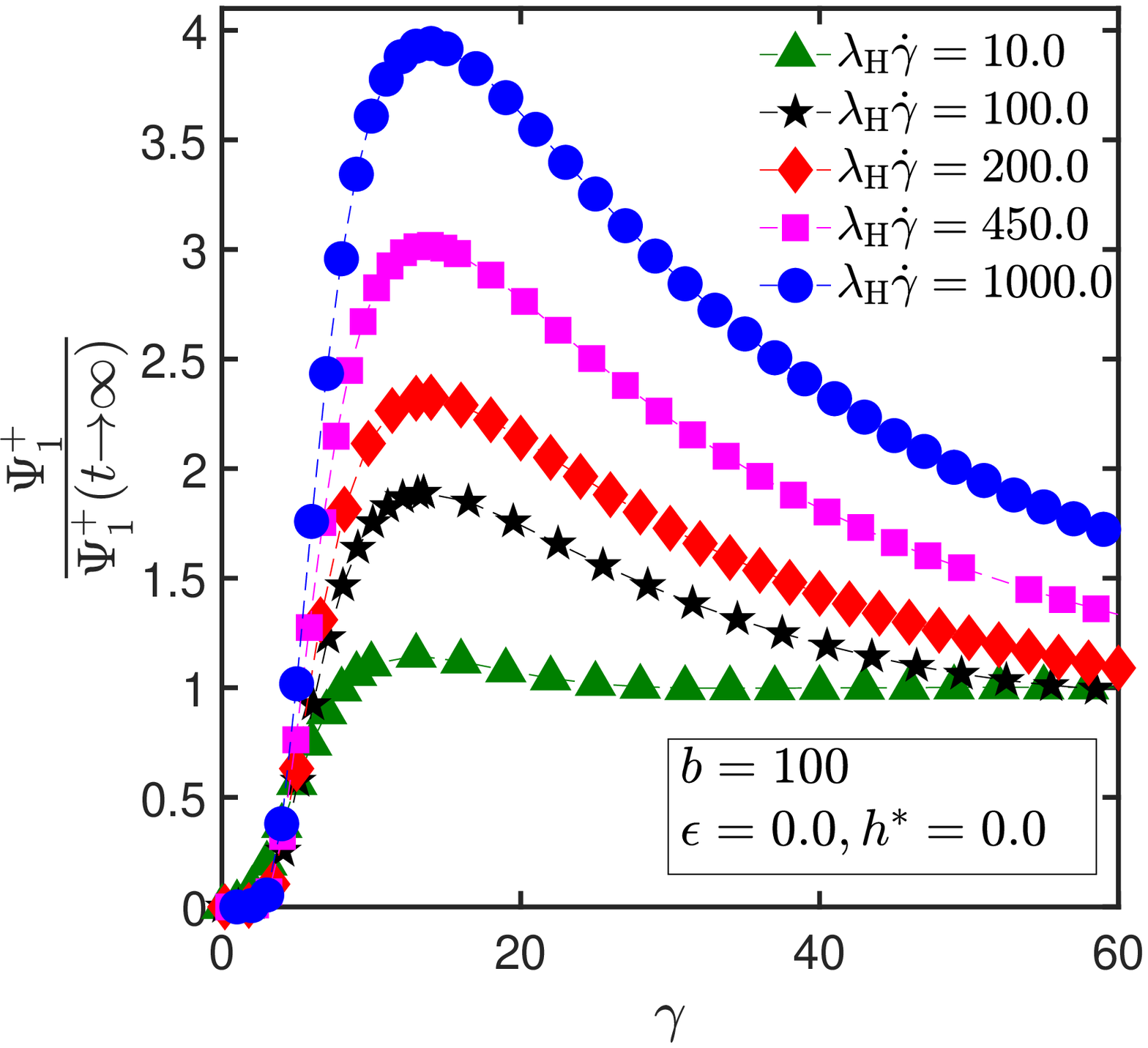}} \\
(a) \\
{\includegraphics*[width=3.3in,height=!]{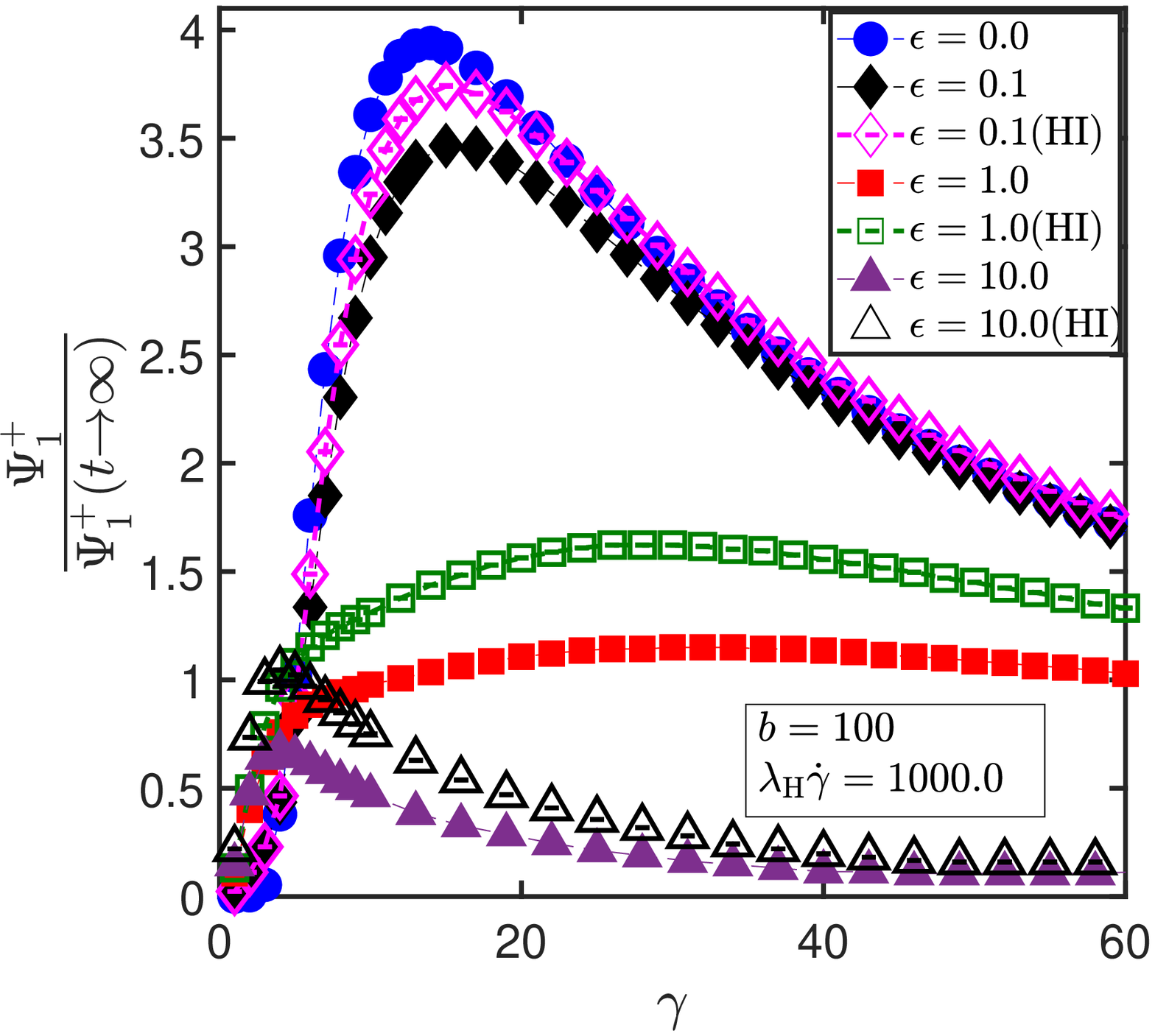}}\\
(b)  \\
\end{tabular}
\end{center}
\caption{(Color online) Transient behavior of the first normal stress difference coefficient for FENE dumbbells with $b=100$ and various values of $\epsilon$ and $h^*$, as a function of strain units: (a) overshoot behavior for the pure FENE model, (b) effect of $\epsilon$ and $h^*$ on overshoot at a dimensionless shear rate of 1000.  {In each case, $\Psi_1^{+}$ is normalized by the steady-state value, ${\Psi}_{\mathrm{1}}^+(t \to \infty)$, obtained at the same shear rate, and $\epsilon$ and $h^*$, as the transient data themselves.} The hydrodynamic interaction parameter in these simulations is $h^*=0.3$. Error bars are smaller than symbol size.}
\label{fig:psi1_overshoot}
\end{figure}

Overshoots in $\Psi_1$ [defined in Eq.~(\ref{eq:fnsd})] can also be analyzed using the framework developed in the previous section. The first normal stress difference coefficient (normalized by the steady-state value, {${\Psi}_{\mathrm{1}}^+(t \to \infty)$}) is plotted as a function of strain in Fig.~\ref{fig:psi1_overshoot}.    

In Fig.~\ref{fig:psi1_overshoot}~(a), the overshoot in $\Psi_1$ for FENE dumbbells with $b=100$ is plotted for various shear rates. The strain at which the overshoot occurs is slightly higher than that for the viscosity, in agreement with experiments~\cite{Bird1987a}. Nonetheless, $\gamma_{\text{max}}$ remains roughly constant over shear rates that span two orders of magnitude. However, the size of the overshoot in $\Psi_1$ is lower than that in $\eta_{\text{p}}$. This is in accordance with the predictions of the FENE-P dumbbell model~\cite{Mochimaru1981}, but in direct contrast with experimental observations~\cite{Huppler1967}. This deviation is perhaps an artefact of the dumbbell model, and simulating polymer chains using a larger number of beads may lead to predictions that are in better qualitative agreement with experimental results. 

\begin{figure}[h]
\begin{center}
\begin{tabular}{c}
{\includegraphics*[width=3.3in,height=!]{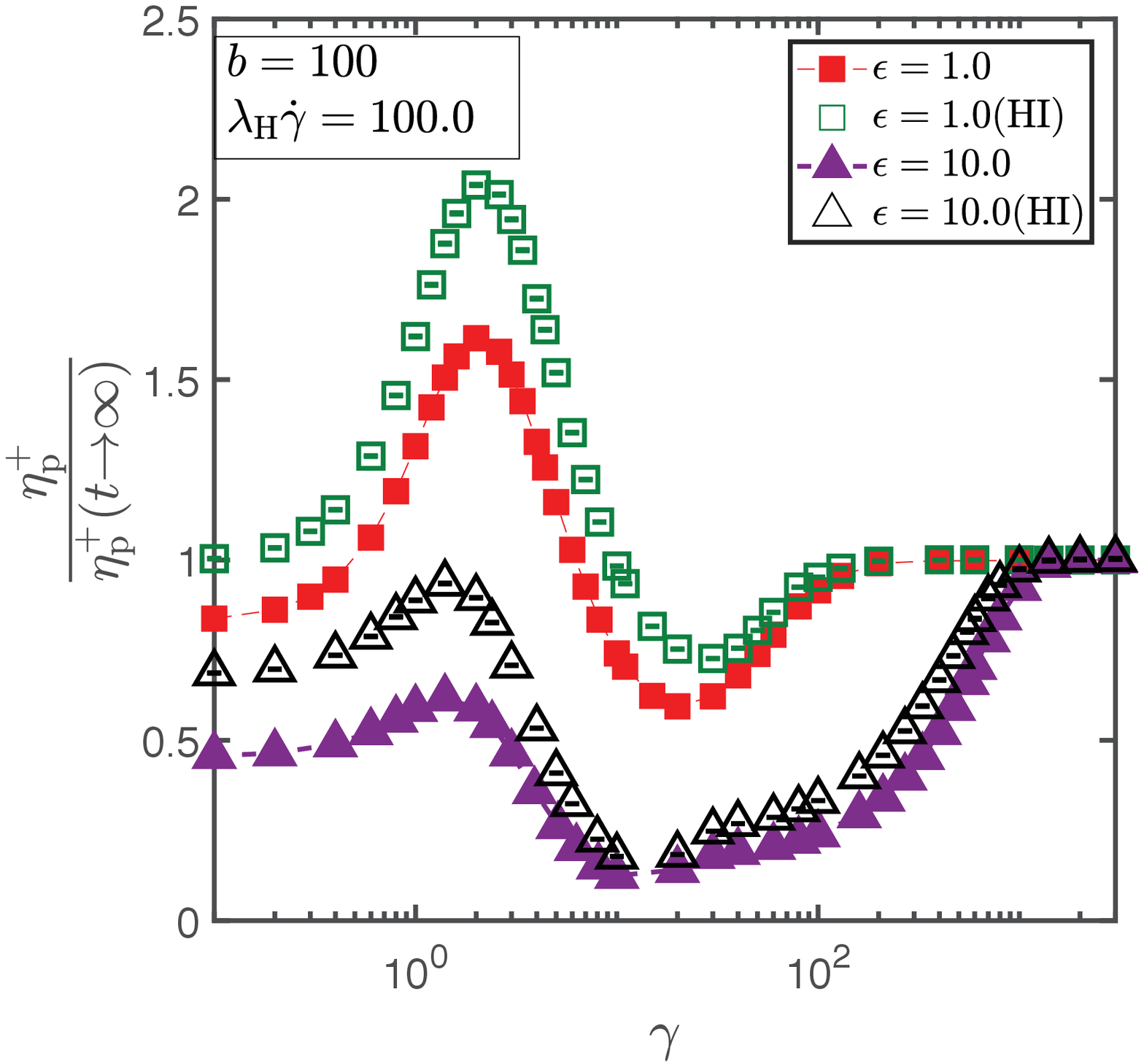}} \\
(a) \\
{\includegraphics*[width=3.3in,height=!]{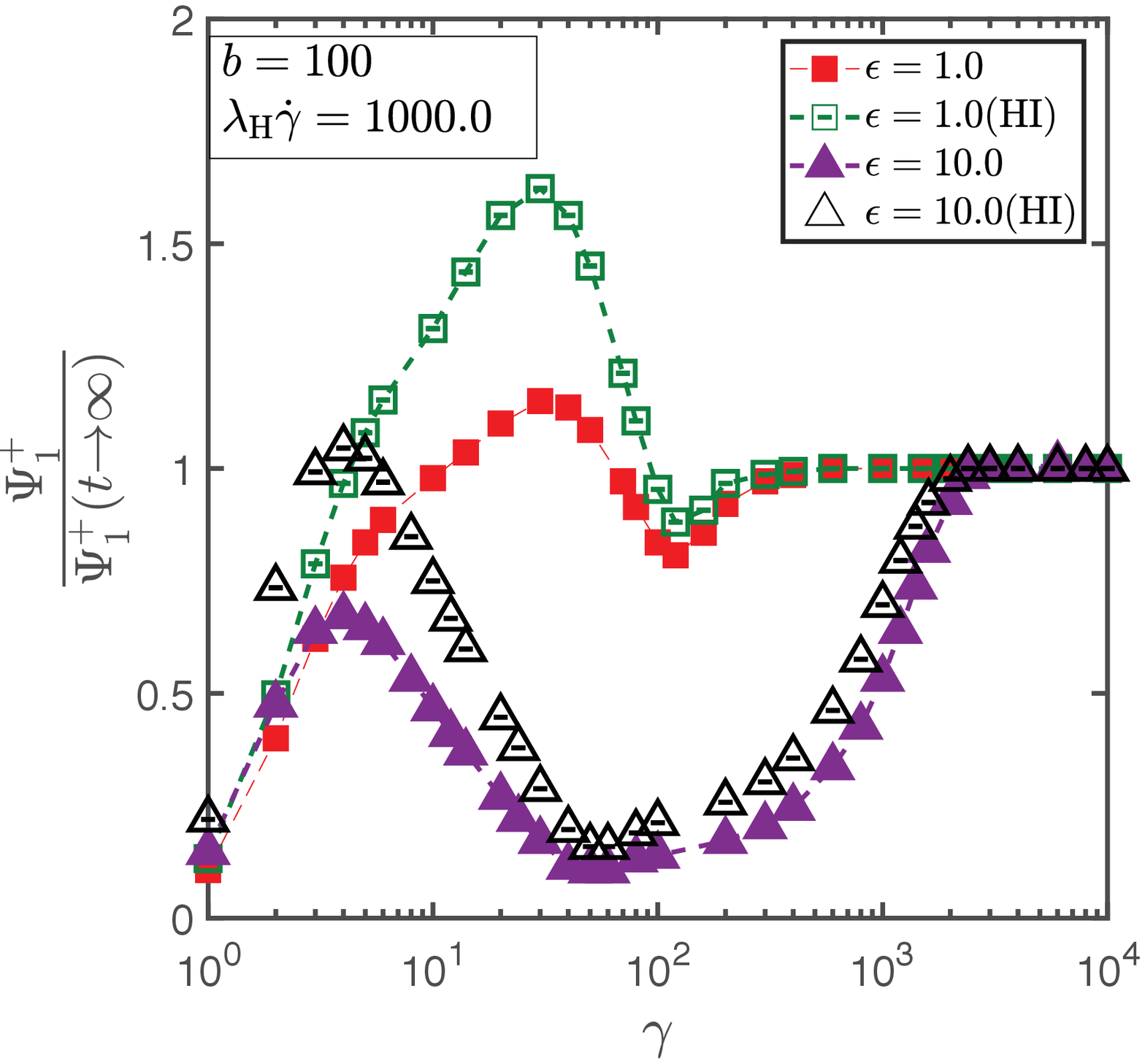}}\\
(b) \\
\end{tabular}
\end{center}
\caption{(Color online) Transient (a) viscosity, $\eta_{\text{p}}$  and (b) first normal stress difference coefficient, $\Psi_1$, normalized by the steady-value of the respective viscometric functions obtained at the same values of shear rate, and $\epsilon$ and $h^*$, as the transient data themselves, as a function of strain units.The hydrodynamic interaction parameter used in these simulations is $h^*=0.3$. Error bars are smaller than symbol size.}
\label{fig:ltime}
\end{figure}

In Fig.~\ref{fig:psi1_overshoot}~(b), the effect of internal viscosity and hydrodynamic interactions on the $\Psi_1$ overshoot is studied at a dimensionless shear rate of 1000. The addition of hydrodynamic interactions enhances the magnitude of overshoot but leaves $\gamma_{\text{max}}$ largely unperturbed. Similar to the trend observed in the transient behavior of viscosity, the time-variation of $\Psi_1$ for the pure FENE case and the case with low IV parameter are qualitatively comparable.  A low value of $\epsilon$ reduces the magnitude of the overshoot slightly, but high values of the IV parameter, say $\epsilon=10$, induce an overshoot only at the highest shear rate examined in this work $(\lambda_H\dot{\gamma}=1000)$, and that too, due to the enhancing effect of the hydrodynamic interactions mentioned above. In general, however, the inclusion of HI does not alter the shear rate at which overshoot is first observed. More simulations at higher shear rates need to be performed, in order to comment conclusively about the constancy of $\gamma_{\text{{max}}}$ for the high-$\epsilon$ cases.

\begin{figure}[t]
\centering
\includegraphics[width=3.5in,height=!]{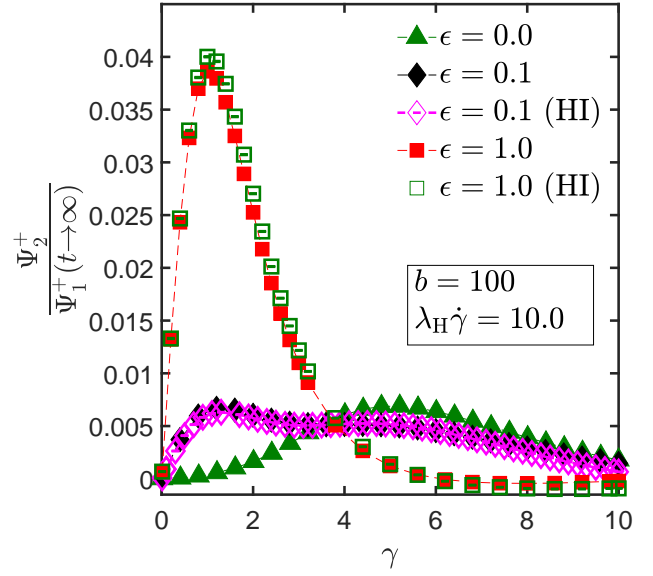}
\caption{(Color online) Influence of $h^*$ and $\epsilon$ on the second normal stress difference coefficient of FENE dumbbells with $b=100$. The hydrodynamic interaction parameter in these simulations is $h^*=0.3$.  {In each case, $\Psi_2^{+}$ is normalized by the steady-state value of the first normal stress difference coefficient, ${\Psi}_{\mathrm{1}}^+(t \to \infty)$, obtained at the same shear rate, and $\epsilon$ and $h^*$, as the transient data themselves.} Error bars are smaller than symbol size.}
\label{fig:psi2_overshoot}
\end{figure}

{The Gaussian approximation developed by ~\citet{Schieber1993} for the treatment of internal viscosity indicates that for $\epsilon=10$ and $\lambda_H\dot{\gamma}=100$, high-frequency oscillations are observed in $\eta_{\text{p}}$ and $\Psi_1$ (see Figs.~5 and~7 in Ref.~\citenum{Schieber1993}). For the same set of parameters studied by BD simulations, however, no such high-frequency oscillations are observed, as can be seen from Figs.~\ref{fig:ltime}, where the normalized transient viscosity and first normal stress difference coefficients of FENE dumbbells with $\epsilon=1$ and $\epsilon=10$ (with and without HI) have been plotted for a larger range of strain units than shown in Figs.~\ref{fig:eta_overshoot} and~\ref{fig:psi1_overshoot}. Both viscometric functions go through a local maximum and minimum before gradually attaining their steady state values. The inclusion of HI does not qualitatively alter this trend, and only slightly increases the maximum observed for the case with IV alone. Neither does the inclusion of HI affect the shear rate at which overshoot is first observed. While both $\eta_{\text{p}}$ and $\Psi_1$ exhibit an overshoot for the $\epsilon=1$ case, these observables vary differently for the higher value of the internal viscosity parameter. As seen from Fig.~\ref{fig:ltime}~(a), $\eta_{\text{p}}$ for the $\epsilon=10$ case (with and without HI) does not show any overshoot (the ordinate never crossing 1), but rather oscillates once, i.e, goes through a local maximum and a minimum, before a gradual approach to steady state. From Fig.~\ref{fig:ltime}~(b), it is seen that $\Psi_1$ for $\epsilon=10$ exhibits a slight overshoot only in the presence of HI, whereas for the free-draining case at the same value of the IV parameter, $\Psi_1$ goes through a single oscillation before attaining steady-state. While the Gaussian approximation includes fluctuations in the internal viscosity, it is an `uncontrolled' approximation~\cite{Prakash1999}, in the sense that though it is exact to first order in the perturbation parameter, it includes infinitely many unspecified higher order terms}. The results of BD simulations, on the other hand, are an exact solution of the governing equation.

\subsubsection{\label{sec:trans_psi2} Transient second normal stress difference coefficient}

In Fig.~\ref{fig:psi2_overshoot}, the second normal stress difference coefficient [defined in Eq.~(\ref{eq:snsd})] is plotted as a function of strain. It is observed that $\Psi_2$ evolves non-monotonically, and settles to zero as its steady-state value, within statistical error bars of the simulation. Therefore, $\Psi_2$ has been scaled using the steady-state value of the first normal stress difference coefficient,  {${\Psi}_{\mathrm{1}}^+(t \to \infty)$}. Increasing the value of the IV parameter increases the amplitude of oscillations in $\Psi_2$. Furthermore, it is seen that the effect of HI on $\Psi_2$ is minimal, becoming stronger as the value of the internal viscosity parameter is increased.

\subsection{\label{sec:steady_shear} Steady-shear results}

\subsubsection{\label{sec:zshearprop} Zero-shear rate properties}

The dimensionless zero-shear rate viscosity,  $\eta^*_{p,0}$, is obtained from BD simulations by taking an error-weighted mean of the viscosity at the four lowest shear-rates that were simulated, i.e, 0.01, 0.03, 0.05 and 0.07, after verifying that shear-thinning had not set in at these shear-rates.

\begin{figure}[h]
\begin{center}
\begin{tabular}{c}
{\includegraphics[width=3.5in,height=!]{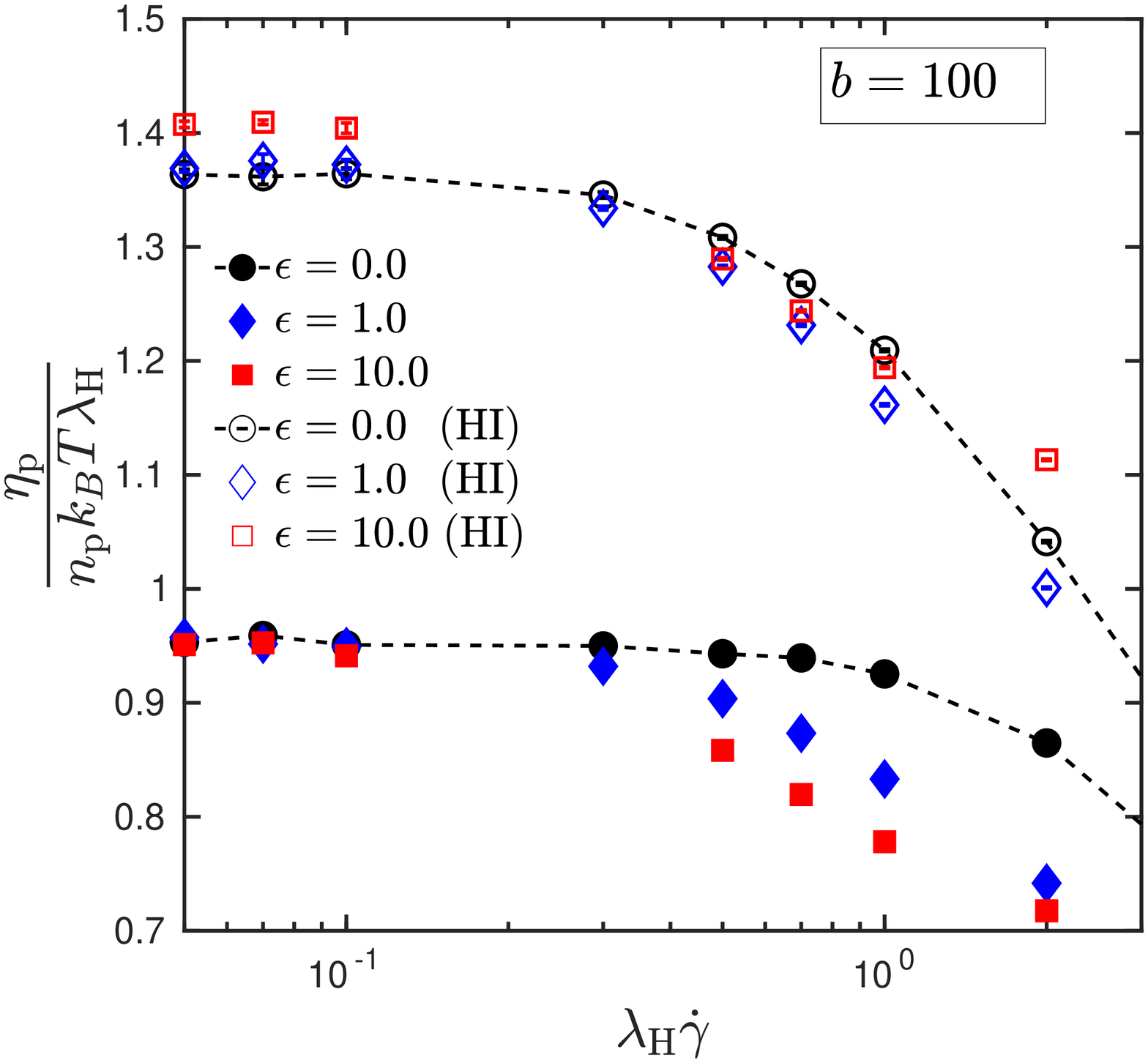}} \\
(a)\\
{\includegraphics[width=3.5in,height=!]{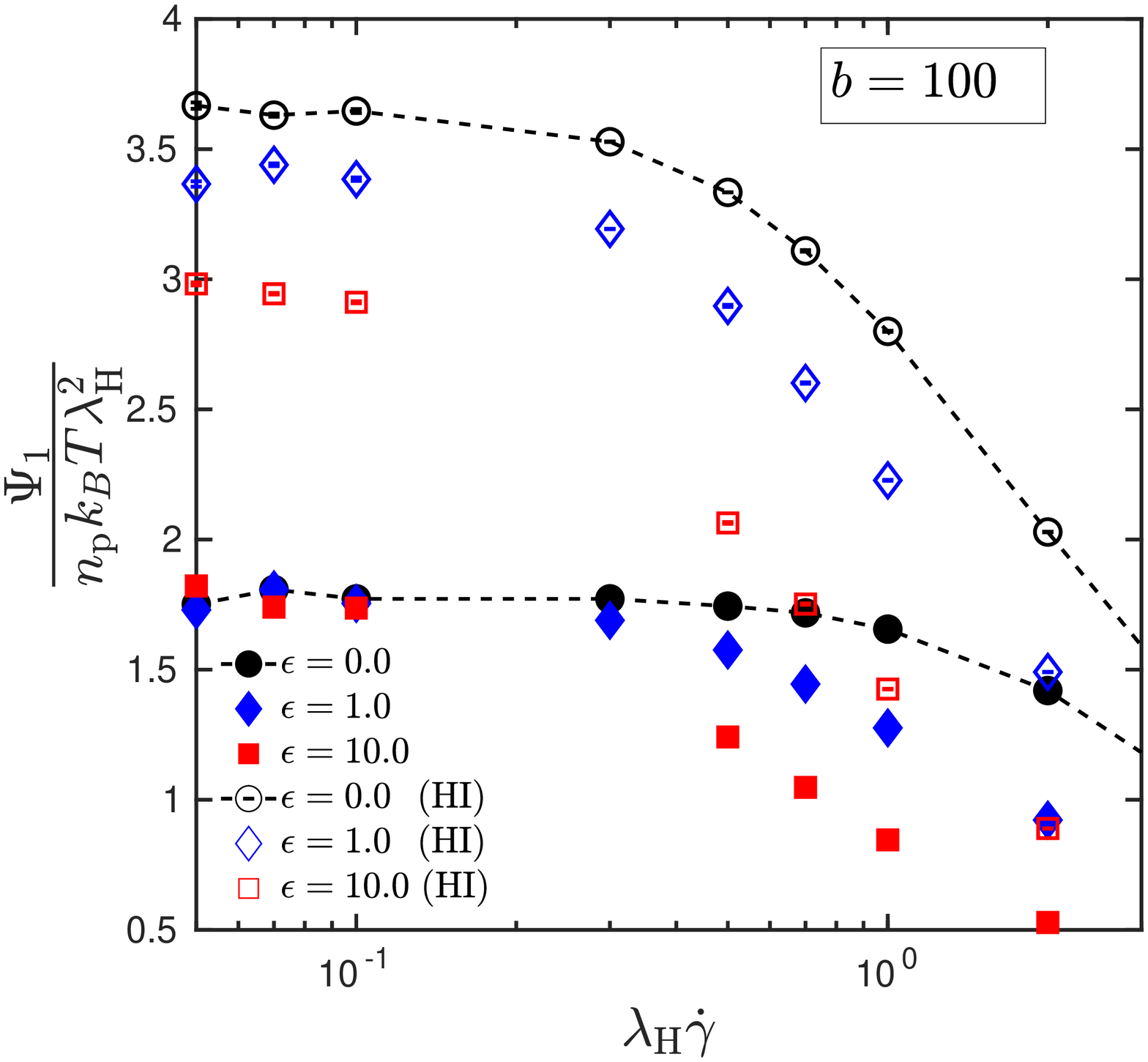}}\\
(b) \\
\end{tabular}
\end{center}
\caption{(Color online) (a) Polymer contribution to viscosity at low shear rates. (b) First normal stress difference coefficient at low shear rates. The hydrodynamic interaction parameter in these simulations is $h^*=0.3$. Error bars are smaller than symbol size.}
\label{fig:zshear}
\end{figure}

Using a Gaussian approximation (GA) analysis on Hookean dumbbells with internal viscosity, ~\citet{Schieber1993} has shown that zero-shear rate properties are unaffected by internal viscosity. 
To test this prediction for the case of FENE dumbbells, the viscosity and first normal stress difference coefficient are plotted as a function of dimensionless shear rate, in Fig.~\ref{fig:zshear}, for various values of $\epsilon$ and $h^*$. From Fig.~\ref{fig:zshear}~(a) it can be seen that for cases with internal viscosity alone, the zero-shear rate viscosity remains unaffected by $\epsilon$. However, in the presence of internal viscosity and hydrodynamic inteactions, the zero-shear rate viscosity no longer remains independent of $\epsilon$. Even though $h^*$ is constant, the coupling of HI and IV induces an internal viscosity dependence on $\eta^*_{p,0}$.

A plot for the first normal stress difference coefficient, as shown in Fig.~\ref{fig:zshear}~(b), seems to reveal a similar trend regarding the constancy of the zero-shear rate value in the presence of internal viscosity alone. With the inclusion of hydrodynamic interactions, however, $\Psi^*_{1,0}$ ceases to be independent of $\epsilon$. The coupling of HI and IV seems to lead to a more dramatic dependence of $\Psi^*_{1,0}$ on $\epsilon$, than that observed for $\eta^*_{p,0}$. It is worth noting that the fluctuations in $\Psi^*_1$ at $\lambda_{\text{H}}\dot{\gamma} < 0.1$ necessitate a time-averaging algorithm for calculating the ensemble average at low shear rates, as detailed below. 

\begin{figure}[ht]
\begin{center}
\begin{tabular}{c}
 {\includegraphics[width=3.3in,height=!]{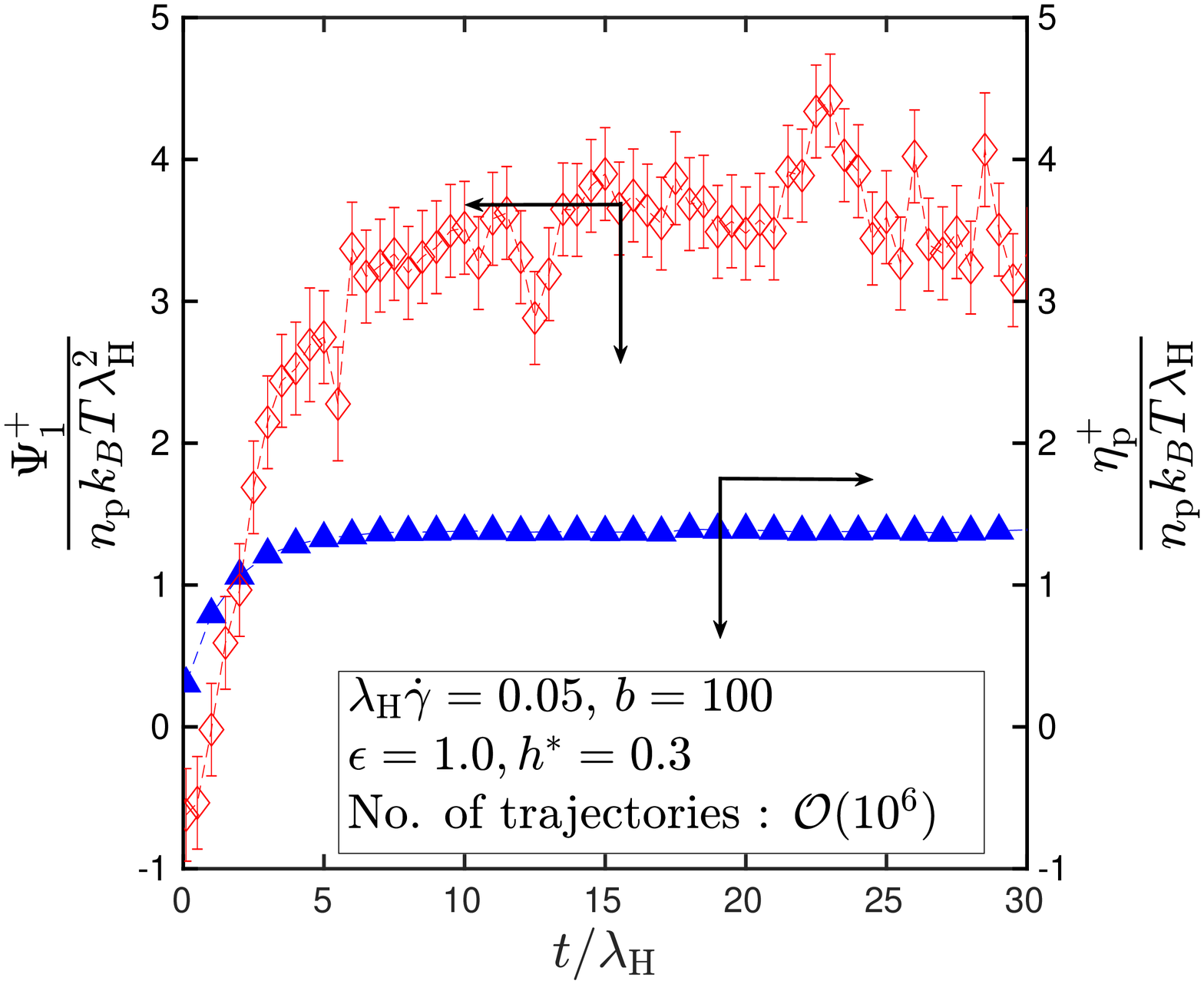}}\\
(a)\\
 {\includegraphics[width=3.3in,height=!]{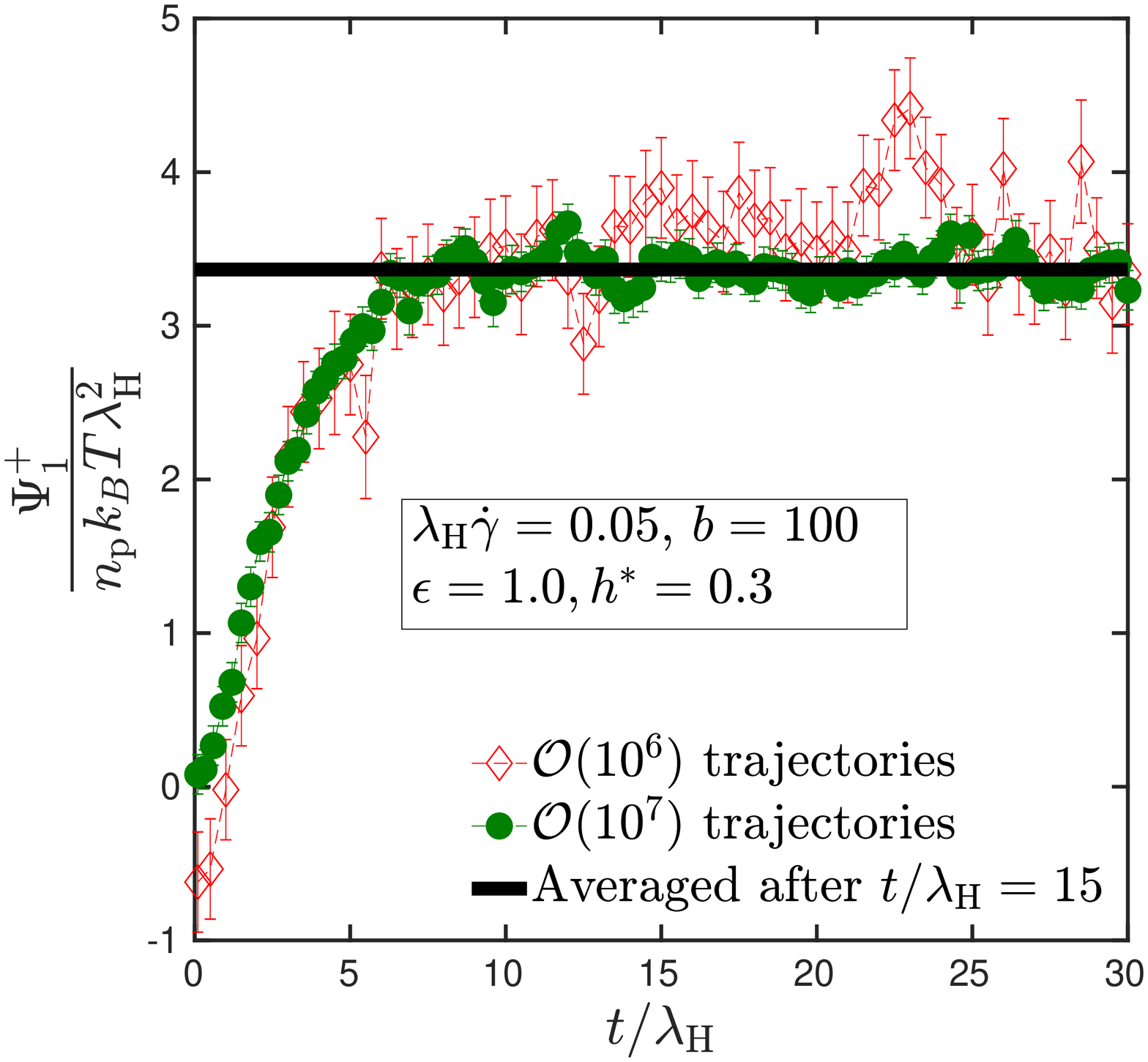}}\\
(b)\\
\end{tabular}
\end{center}
\caption{(Color online) (a) Transient viscosity and first normal stress difference coefficient estimated from an ensemble of approximately 1 million data points. (b) Time-averaging procedure to estimate the steady-state value of the first normal stress difference coefficient.}
\label{fig:fluc_psi1}
\end{figure}

For a dimensionless shear rate of 0.05 and representative values of $\epsilon$ and $h^*$, the transient behavior of the viscosity and the first normal stress difference coefficient is compared in Fig.~\ref{fig:fluc_psi1}(a). It can be seen that the standard deviation of the first normal stress difference coefficient is significantly higher than that for viscosity. Additionally, it was observed that the fluctuations in $\Psi^*_1$ are noticeable at both low and high values of the IV parameter. In Fig.~\ref{fig:fluc_psi1}~(b), the first normal stress difference coefficient calculated for the same conditions as in part (a) are compared for two different sizes of the ensemble. It is seen that even as the ensemble size is increased ten-fold, fluctuations in $\Psi^*_1$ persist. To obtain a reliable estimate of the steady-state value of $\Psi^*_1$, a time-averaging of the data points was carried out, after the stationary state is reached (the threshold value is 15 dimensionless time units for the case considered in Fig.~\ref{fig:fluc_psi1}~(b)). The horizontal line in Fig.~\ref{fig:fluc_psi1}~(b) represents the mean value obtained from such an averaging procedure, and the thickness of the line indicates the error in the mean. The steady-state value of the first normal stress difference coefficient for dimensionless shear rates lower than 0.1 was calculated in this manner, with an ensemble size of $\mathcal{O} (10^{7})$. For the viscosity, there is no statistically discernible difference between data sampled from an ensemble of the order of one million or ten million data points.

The steady-state value of the second normal stress difference coefficient is zero within statistical error bars of the simulation, for free-draining dumbbells with and without the inclusion of internal viscosity, as observed from the transient simulations at large times. With the inclusion of hydrodynamic interactions, however, small negative values for $\Psi_2$ are obtained at low shear rates, for cases with and without internal viscosity. At higher shear rates, $\Psi_2$ is zero for all the different parameters considered in this work. A similar trend in the variation of $\Psi_2$ with the shear rate, for models with fluctuating hydrodynamic interactions, has been noticed by ~\citet{Zylka1991} in his work on bead-spring chains.

\subsubsection{\label{sec:gt} Calculation of zero-shear rate viscosity from the relaxation modulus and the stress jump}

It is known from linear viscoelastic fluid theory that the zero-shear rate viscosity can be calculated by integrating the relaxation modulus, $G(t)$, with respect to time~\cite{Rubinstein2003}. That is, 
\begin{equation}\label{eq:eta0}
\eta_0=\int_{0}^{\infty}G(t)dt
\end{equation}
The general form of the relaxation modulus of  a polymer solution can be expressed as the sum of an elastic part, $G_{\text{el}}(t)$ (corresponding to a slow decay in stress),  and a singular part (accounting for the viscous portion) that captures the instantaneous response of the solution to a stress~\cite{Bird1987a,Hua1996},     
\begin{equation}\label{eq:gt_general}
G(t)=G_{\text{el}}(t)+2\eta_{\text{v}}\,\delta(t)
\end{equation}
For models without internal viscosity, $\eta_{\text{v}}=\eta_{\text{s}}$, but as will be discussed later in this section, $\eta_{\text{v}}>\eta_{\text{s}}$ for models that incorporate IV.

\begin{figure}[h]
\begin{center}
\begin{tabular}{c}
{\includegraphics[width=3.3in,height=!]{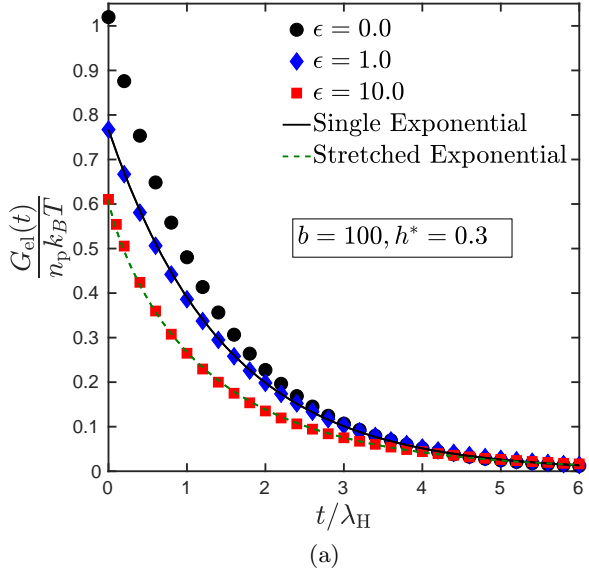}} \\
(a) \\
{\includegraphics[width=3.3in,height=!]{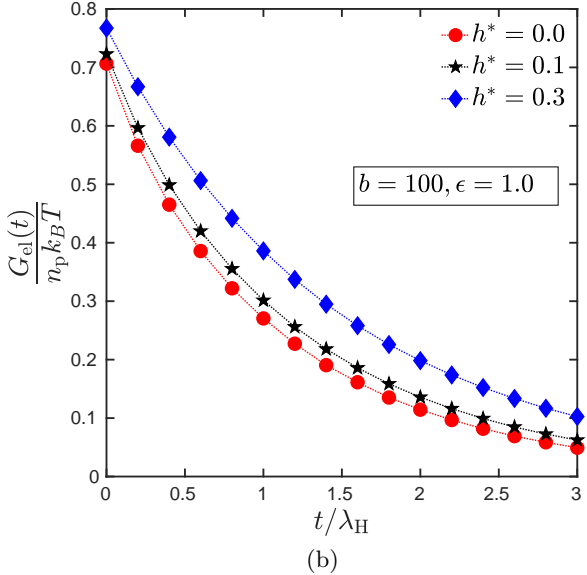}}\\
(b)  \\
\end{tabular}
\end{center}
\caption{(Color online) Elastic component of the relaxation modulus as a function of dimensionless time. (a) Effect of internal viscosity at a fixed value of the hydrodynamic interaction parameter, $h^*$. The solid line is a single exponential fit [see Eq.~(\ref{eq:simp_exp})] with $\tau=1.48$. The dashed line is a stretched exponential [see Eq.~(\ref{eq:stretch_exp})] with  $\tau_k=1.26$ and $m=0.85$. (b) Effect of hydrodynamic interactions at a fixed value of the internal viscosity parameter, $\epsilon$. Dotted lines are drawn to guide the eye. Error bars are smaller than symbol size.}
\label{fig:gtfit}
\end{figure}
Schieber~\cite{Hua1996} and coworkers have used BD simulations to estimate the relaxation modulus of Hookean dumbbells with IV, from the stress relaxation that follows from an instantaneous step-strain in shear~\cite{Hua1995307}. They find excellent agreement between the $G(t)$ obtained using such a procedure and that obtained from linear-response theory, and observe that the latter method is computationally more efficient and accurate. It is worth noting, however, that both the approaches are unable to capture the singular portion in $G(t)$.  In our simulations, ${G^*_{\text{el}}(t^*)}$ is obtained using linear response theory, i.e, the Green-Kubo relationship, which is based on the stress-autocorrelation of an ensemble of dumbbells at equilibrium, 
\begin{equation}\label{eq:gk}
G^*_{\text{el}}(t^*)=\frac{G_{\text{el}}(t^*)}{n_{\text{p}}k_BT}=\left<\tau^*_{{\text{p}},yx}(t^*)\tau^*_{{\text{p}},yx}(0)\right>_{\text{eq}}
\end{equation}
In Fig.~\ref{fig:gtfit}, the elastic component of the relaxation modulus calculated in this manner has been plotted against dimensionless time, for various values of $\epsilon$ and $h^*$. The area under the curve represents $\eta^{*,\text{el}}_{{\text{p}},0}$, the dimensionless elastic contribution to the zero shear viscosity.
In Fig.~\ref{fig:gtfit}~(a), the effect of varying $\epsilon$ on $G^*_{\text{el}}(t^*)$ at a fixed value of $h^*$ has been captured. As the internal viscosity is increased, it is seen that the area under the curve decreases, and the relaxation slows down. On the other hand, increasing the strength of hydrodynamic interactions while keeping $\epsilon$ constant also slows the decay of  $G^*_{\text{el}}(t^*)$, but increases $\eta^{*,\text{el}}_{{\text{p}},0}$, as seen from Fig.~\ref{fig:gtfit}~(b). 
The integration of the relaxation modulus required in Eq.~(\ref{eq:eta0}) is simplified if an analytical fit can be obtained to the $G^*_{\text{el}}(t^*)$ data. For small values of $\epsilon$, a single exponential of the form  
\begin{equation}\label{eq:simp_exp}
{G^*_{\text{el}}(t^*)}=\frac{G_{\text{el}}(0)}{n_{\text{p}}k_BT}\exp(-t^*/\tau)
\end{equation}
where $\tau$ is an adjustable parameter, produces a good fit, as seen in Fig.~\ref{fig:gtfit}~(a) for $\epsilon=1.0$.
For large values of $\epsilon$, it is found that a stretched exponential of the form, 
\begin{equation}\label{eq:stretch_exp}
{G^*_{\text{el}}(t^*)}=\frac{G_{\text{el}}(0)}{n_{\text{p}}k_BT}\exp\left[(-t^*/\tau_k)^{m}\right]
\end{equation}
provides a more accurate fit to the data than does the single exponential. Here, $\tau_k$ and $m$ are fitting parameters. This is seen from Fig.~\ref{fig:gtfit}~(b) for $\epsilon=10.0$.

Gerhardt and Manke~\cite{Gerhardt1994} have shown analytically that the instantaneous stress-jump observed as a polymer solution is subject to the start up of shear flow is identically equal to the singular portion in Eq.~(\ref{eq:gt_general}), for linear viscoelastic fluids. As  noted previously, since models with IV show a stress jump in $\eta^{+}_{{\text{p}}}$, it follows from Gerhardt and Manke's work that the numerical prefactor, $\eta_{\text{v}}$,  to the Dirac delta function in $G(t)$ must be greater than $\eta_s$, and in fact, the polymer contribution to the stress jump, $\eta_{\text{jump}}$, is given by
\begin{equation}\label{eq:eta_jump}
\eta_{\text{jump}}=\eta_{\text{v}}-\eta_{\text{s}}
\end{equation}
The polymer-contribution to zero-shear rate viscosity, $\eta_{{\text{p}},0}$, can then be written as 
\begin{align}
\eta_{{\text{p}},0}\equiv\eta_{0}-\eta_{\text{s}}&=\left[\int_{0}^{\infty}G(t)dt\right]-\eta_{\text{s}}\nonumber\\[5pt]
&=\int_{0^+}^{\infty}G_{\text{el}}(t)dt + 2\int_{0}^{\infty}\eta_{\text{v}}\,\delta(t)dt-\eta_{\text{s}}\nonumber\\[5pt]
&=\eta^{\text{el}}_{{\text{p}},0} + \eta_{\text{jump}}\label{gjv_rel}
\end{align}
where $\eta^{\text{el}}_{{\text{p}},0}=\int_{0^+}^{\infty}G_{\text{el}}(t)dt$, and $2\int_{0}^{\infty}\eta_{\text{v}}\,\delta(t)dt = \eta_{\text{v}}$.  

The zero-shear rate viscosity calculated in this manner is compared against the value obtained from direct BD simulations in Table~\ref{zshear}, for various values of the internal viscosity and the hydrodynamic interaction parameter. We see that there is a good agreement between the values obtained by these two different approaches, and in some sense validates both the estimation of the zero-shear rate viscosity from BD simulations, and the estimation of the relaxation modulus.  

\begin{figure}[t]
\begin{center}
\begin{tabular}{c}
{\includegraphics[width=3.3in,height=!]{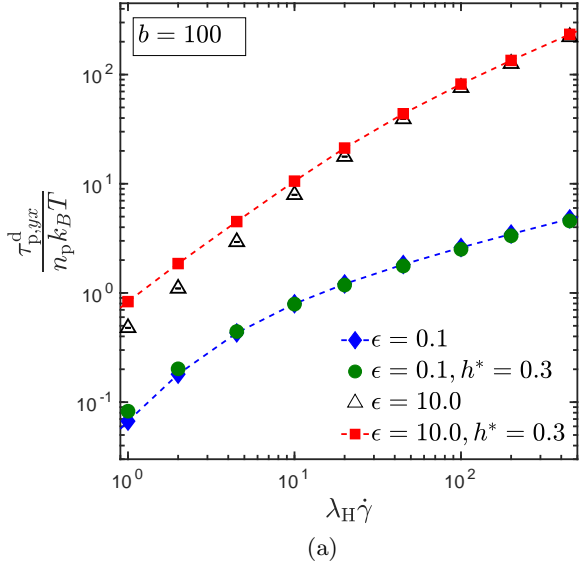}}\\
(a)\\
{\includegraphics[width=3.3in,height=!]{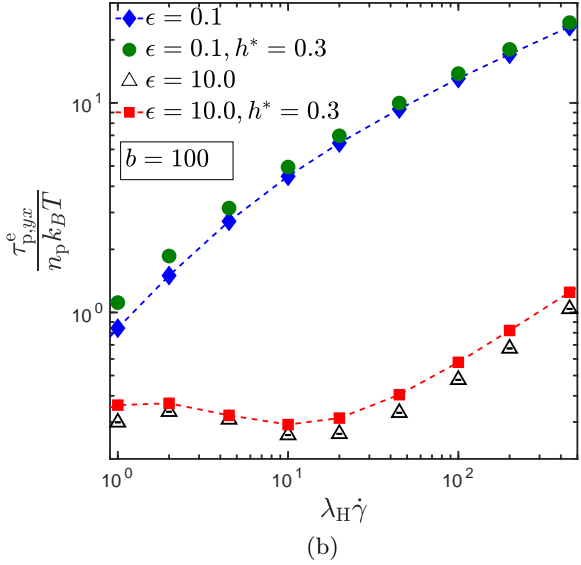}}\\
(b)\\
\end{tabular}
\end{center}
\caption{(Color online) Effect of internal viscosity and hydrodynamic interactions on (a) dissipative and (b) elastic portions of the $yx$-component of the polymer contribution to the stress tensor, as a function of shear-rate. Error bars are smaller than symbol size.}
\label{fig:tau_components}
\end{figure}

\subsubsection{\label{sec:tau_comp} Stress tensor components}

Even away from the linear viscoelastic regime, there exist separate contributions to the stress tensor for models with internal viscosity. As briefly discussed in section \ref{sec:SM}, the second and third terms on the right hand side of Eq.~(\ref{eq:stress_tensor}) together represent the elastic contribution to the stress tensor, while the last term represents the viscous (dissipative) contribution.     
Liang et al.~\cite{Mackay1992} have devised experimental techniques involving the cessation of flow to separately identify elastic and dissipative contributions to the total polymer shear stress, and present results for semidilute xanthan gum solutions~\cite{liang1993stress}. The predictions of our model, however, are valid only for dilute polymer solutions, and hence cannot be compared directly with their experimental results. Nonetheless, it is worthwhile examining the two contributions, as they provide interesting insights into the roles played by internal viscosity and hydrodynamic interactions in affecting the steady-shear stress field of dilute polymer solutions.

In Fig.~\ref{fig:tau_components}~(a),  the steady-state dissipative contribution to the shear-component of the stress tensor $\left(\tau^{\text{d}}_{\text{p},yx}\right)$ is plotted as a function of shear rate. For a fixed value of $h^*$, it is seen that an increase in the internal viscosity parameter from $\epsilon=0.1$ to $\epsilon=10.0$ results in a significant increase in the dissipative contribution to shear stress. 

\begin{figure}[t]
\centering
\includegraphics[width=3.5in,height=!]{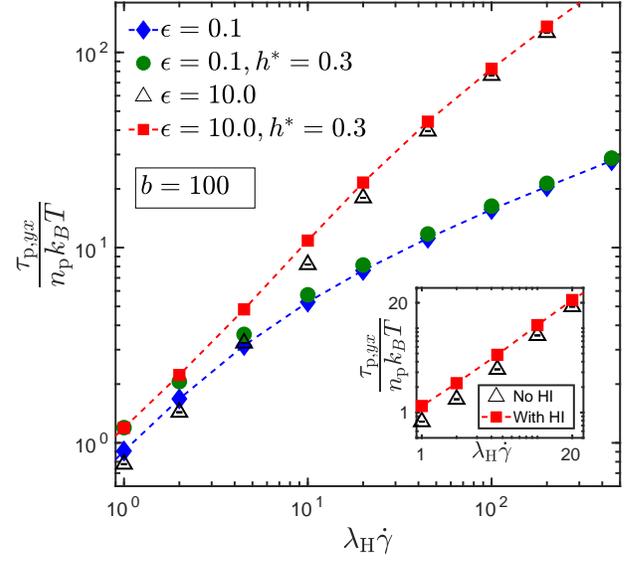}  
\caption{(Color online) Total shear stress, $\tau_{\text{p},yx}$, as a function of shear rate for various values of $\epsilon$ and $h^*$. Error bars are smaller than symbol size.}
\label{fig:tau_tot}
\end{figure}

\begin{savenotes}
\begin{table*}[t]
\setlength{\tabcolsep}{12pt}
\centering
\caption{\label{zshear} Zero-shear rate viscosity of FENE dumbbells with $b=100$ for various values of the internal viscosity parameter, $\epsilon$ and the hydrodynamic interaction parameter, $h^{*}$, calculated from Eq.~\eqref{gjv_rel} and  BD simulations. Dimensionless quantities are denoted with an asterisk as a superscript.}
\begin{center}
\begin{tabular}{c c c c c }
\hline
\hspace{12pt} &  & $h^*=0$ \\
\cline{2-5}
 \hline \hline $\epsilon$ & $\eta^{*,\text{el}}_{{\text{p}},0}$ & $\eta^*_{\text{jump}}$ & $\eta^{*}_{{\text{p}},0}=\eta^{*,\text{el}}_{{\text{p}},0}+\eta^*_{\text{jump}}$ & $\eta^{*}_{{\text{p}},0}$ from BD  \\
\hline
0.0 & $0.951 \pm 0.003$ & 0& $0.951 \pm 0.003$& $0.954 \pm 0.003$ \\
0.1 & $0.918 \pm 0.001$ & $0.0346 \pm 0.0006$ & $0.952 \pm 0.001$& $0.956 \pm 0.003$ \\
0.5 & $0.8275 \pm 0.0009$ & $0.1270 \pm 0.0003$ & $0.954 \pm 0.001$& $0.953 \pm 0.003$\\
1.0  & $0.7580 \pm 0.0009$ & $0.1905 \pm 0.0002$ & $0.9485 \pm 0.0009$& $0.956 \pm 0.002$\\
10.0 & $0.5982 \pm 0.0005$ & $0.3464 \pm 0.0003$ & $0.9445 \pm 0.0006$& $0.952 \pm 0.001$\\
\hline
{} &{}& { $h^*=0.3$} & {} & {}\\
\cline{2-5}
 \hline \hline $\epsilon$ & $\eta^{*,\text{el}}_{{\text{p}},0}$ & $\eta^*_{\text{jump}}$ & $\eta^{*}_{{\text{p}},0}=\eta^{*,\text{el}}_{{\text{p}},0}+\eta^*_{\text{jump}}$ & $\eta^{*}_{{\text{p}},0}$ from BD  \\
\hline
0.0&  $1.358 \pm 0.002$ & $0$ & $1.358 \pm 0.002$&  $1.363 \pm 0.003$\\
0.1& $1.326 \pm 0.002$ & $0.0359 \pm 0.0001$ & $1.362 \pm 0.002$& $1.363 \pm 0.003$\\
0.5& $1.223 \pm 0.002$ & $0.1466 \pm 0.0002$ & $1.369 \pm 0.002$& $1.371 \pm 0.002$\\
1.0& $1.137 \pm 0.001$ & $0.2391 \pm 0.0002$ & $1.376 \pm 0.001$& $1.373 \pm 0.002$\\
10.0& $0.8353 \pm 0.0006$ & $0.5626 \pm 0.0004$ & $1.3979 \pm 0.0007$& $1.409 \pm 0.001$\\
\hline
\end{tabular}
\end{center}
\vspace{-15pt}
\end{table*} 
\end{savenotes}

The effect of hydrodynamic interactions on $\tau^{\text{d}}_{\text{p},yx}$, however, is less pronounced. At a constant value of $\epsilon$, an increase in $h^*$ (from 0 to 0.3) increases the stress marginally at lower shear rates, and ceases to affect the stress field significantly at higher shear rates. The reason for this behavior is well established~\cite{Zylka1989,Zylka1991} and can be understood by examining the role of hydrodynamic interactions at low and high shear rates. At low shear rates, the beads of the dumbbell are closer to each other, and there is a significant contribution to the hydrodynamic interaction force, as is clear from the form of the HI tensor in Eq.~(\ref{eq:hitensor}). At higher shear rates, the connector between the beads is expanded, resulting in lower contributions to the HI tensor, and the viscometric functions tend to approach their free-draining values. This trend for the effect of HI is also true for the other steady-shear observables measured in our work, as will be seen in the following sections.

In Fig.~\ref{fig:tau_components}~(b), the variation of  the steady-state elastic contribution to the shear-component of the stress tensor $\left(\tau^{\text{e}}_{\text{p},yx}\right)$ is plotted as a function of shear rate. When the value of $h^*$ is fixed, and $\epsilon$ is increased from 0.1 to 10, the elastic contribution to shear stress decreases markedly. The inclusion of hydrodynamic interactions, at a constant value of the IV parameter, has a less perceptible effect on $\tau^{\text{e}}_{\text{p},yx}$.

Fig.~\ref{fig:tau_tot} shows the variation of the total shear stress, which is a sum of the elastic and dissipative components examined in Fig.~\ref{fig:tau_components}, with the shear rate. At low shear rates,  the total stress for $\epsilon=0.1$ slightly exceeds that for $\epsilon=10.0$. There is a crossover region at approximately $\lambda_{\text{H}}\dot{\gamma}\approx 4$, after which the total stress for a higher value of $\epsilon$ exceeds that for lower $\epsilon$. Such a crossover appears to indicate that the predominant contribution to the total stress at lower shear rates arises from the elastic component (which is higher for low values of $\epsilon$), whereas at higher shear rates, the dissipative component (which is higher for higher values of $\epsilon$) contributes more significantly to the total stress. The inset in Fig.~\ref{fig:tau_tot} shows the effect of hydrodynamic interactions (with $h^*=0.3$) on the total stress, at $\epsilon=10.0$. The trend is similar to that observed for the effect of HI on $\tau^{\text{d}}_{\text{p},yx}$ and $\tau^{\text{e}}_{\text{p},yx}$, with HI contributing to a small increase in stress compared to the free-draining case at lower shear rates, and weakening in effect at higher shear rates.

\subsubsection{\label{sec:srthin} Shear-thinning of $\eta_{\textrm{p}}$ and $\Psi_{1}$}

\begin{figure}[t]
\begin{center}
\begin{tabular}{c}
{\includegraphics[width=3.5in,height=!]{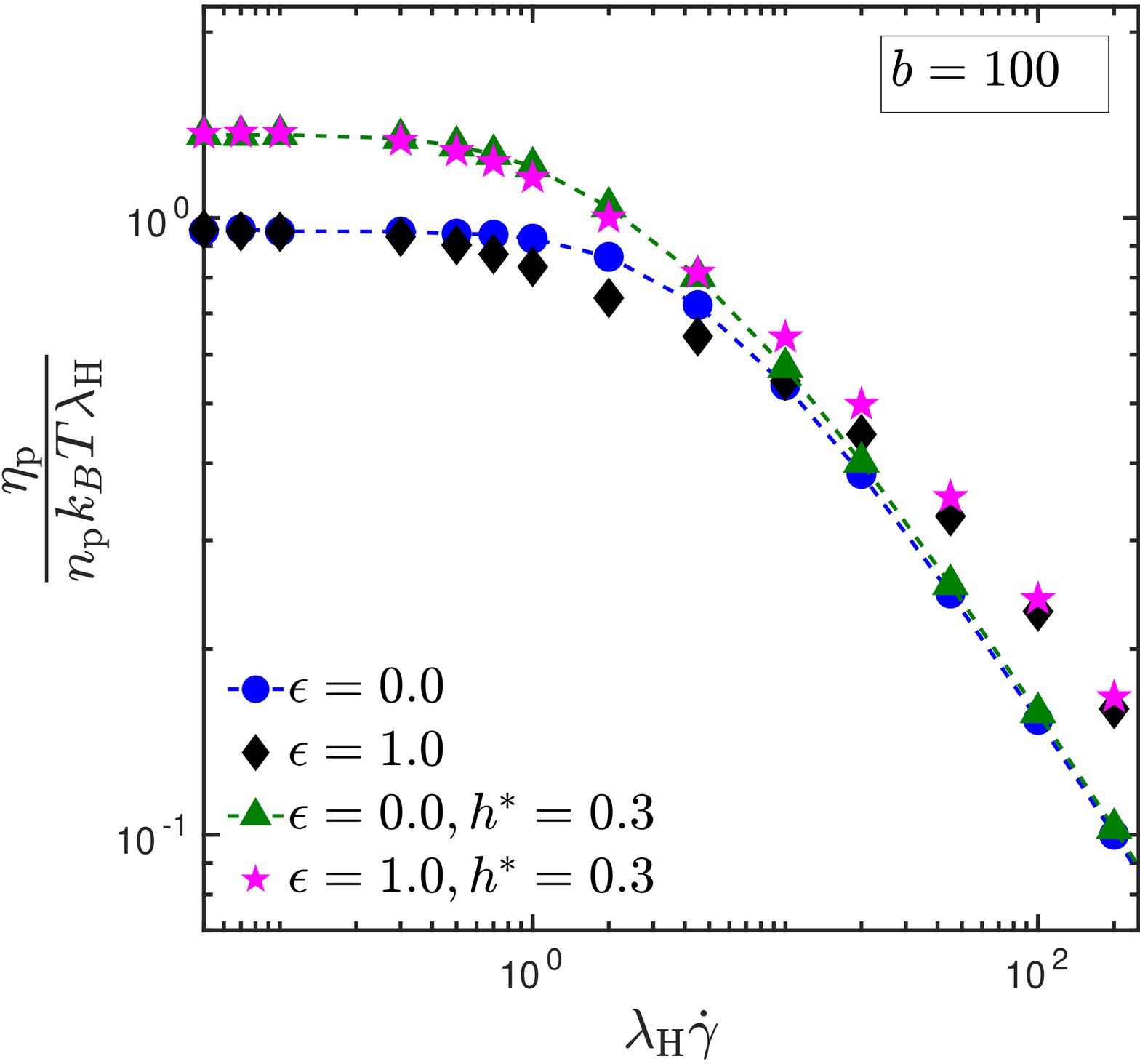}}\\
(a) \\
{\includegraphics[width=3.1in,height=!]{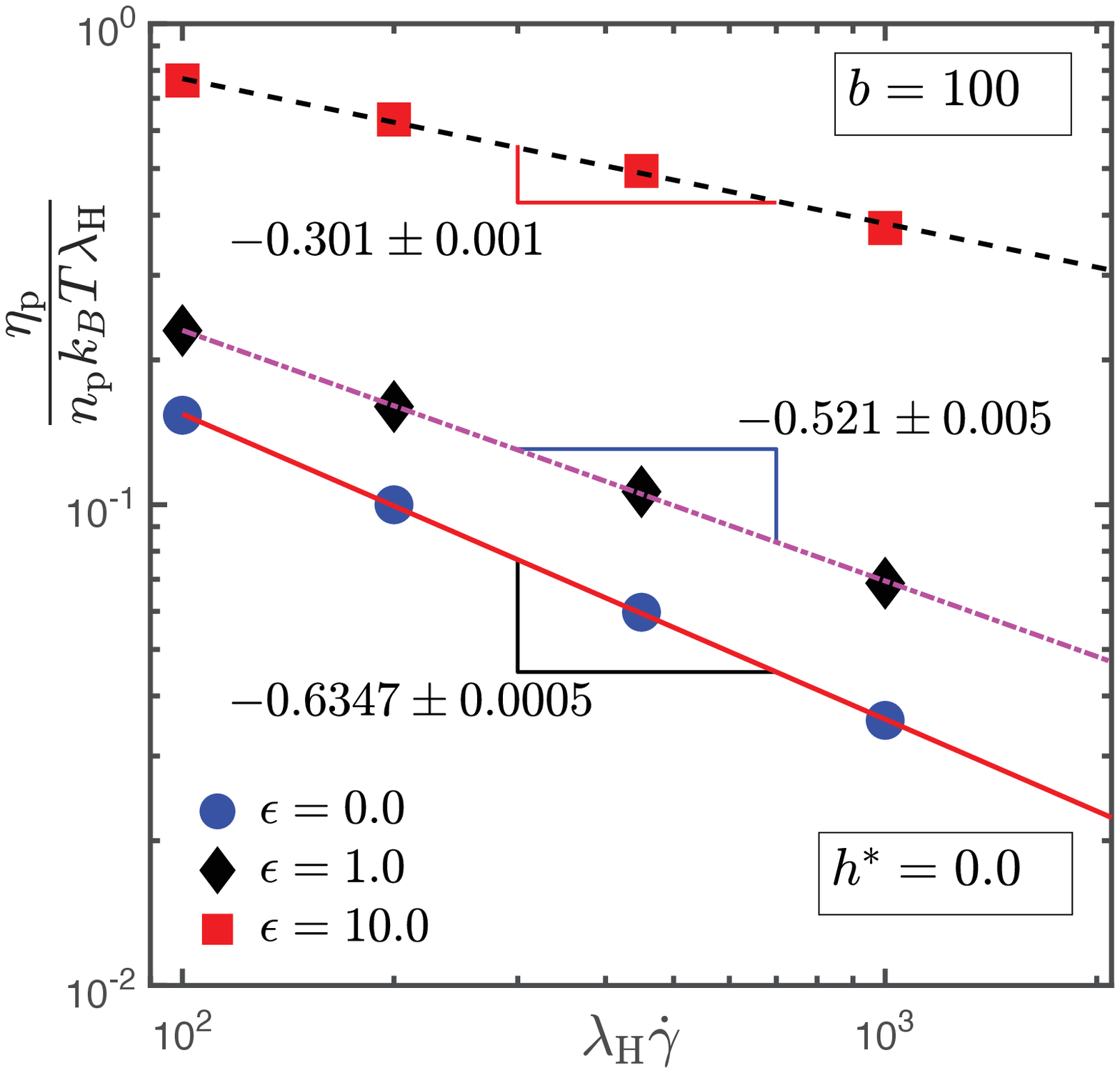}}\\
(b)  \\
\end{tabular}
\end{center}
\caption{(Color online) Plots of steady-shear viscosity for FENE dumbbells with $b=100$. (a) Effect of IV and HI on shear-thinning and zero-shear rate viscosity at low and moderate shear rates. (b) Effect of IV on the shear-thinning exponent, at high shear rate. Errorbars are smaller than symbol size.}
\label{fig:eta_steady}
\end{figure}

\begin{figure}[t]
\begin{center}
\begin{tabular}{c}
{\includegraphics[width=3.5in,height=!]{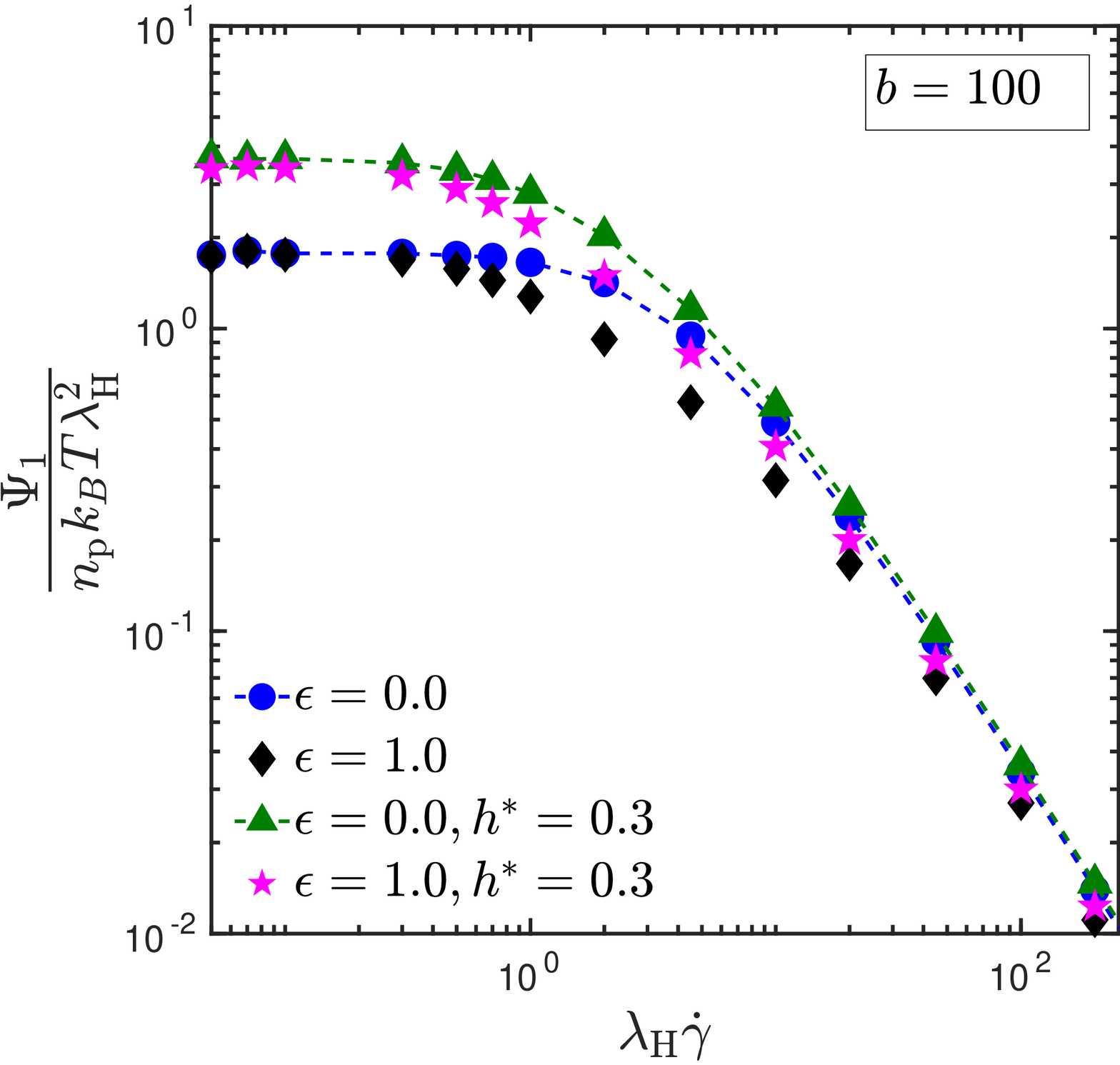}}\\
(a)\\
{\includegraphics[width=3.1in,height=!]{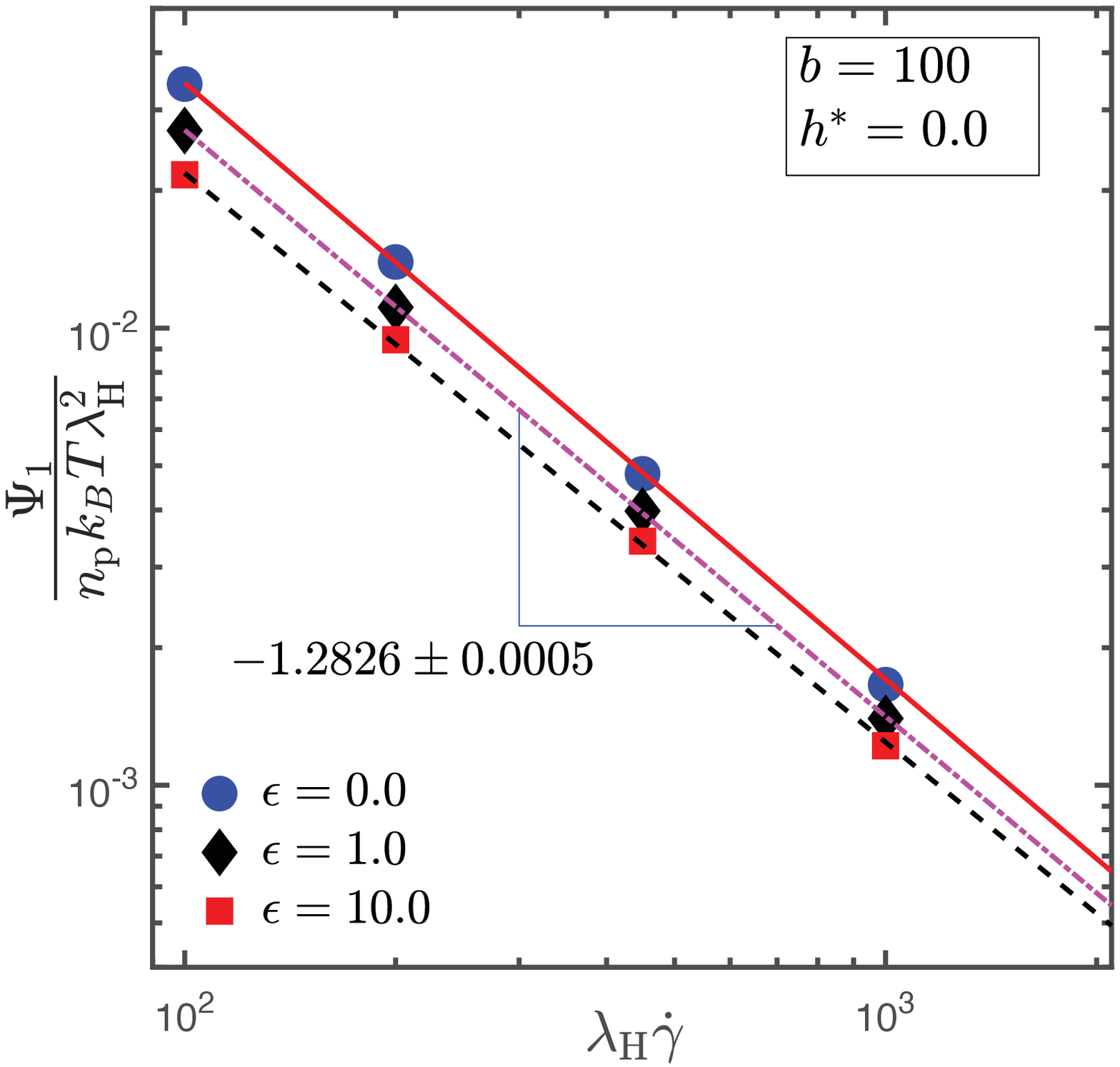}}\\
(b) \\
\end{tabular}
\end{center}
\caption{(Color online) Plots for steady-shear first normal stress difference coefficient. (a) Effect of IV and HI on the zero-shear rate value, and shear-thinning at low and moderate shear rates. (b) Effect of IV on the shear-thinning exponent, at high shear rates. Error bars are smaller than symbol size.}
\label{fig:psi1_steady}
\end{figure}

To examine the steady-state behavior of the viscosity and first normal stress difference coefficient, both these quantities are plotted as a function of dimensionless shear rate in Figs.~\ref{fig:eta_steady} and \ref{fig:psi1_steady}, respectively, for various values of $\epsilon$ and $h^*$. 

Figure~\ref{fig:eta_steady}~(a) shows that hydrodynamic interactions enhance the zero-shear rate viscosity and push the onset of shear-thinning to lower shear rates. At higher shear rates, however, hydrodynamic interactions do not have a significant effect on the viscosity. Internal viscosity, on the other hand, leaves the zero-shear rate viscosity unperturbed, but quickens the onset of shear-thinning to low shear rates. At higher shear rates, however, internal viscosity is seen to slow down shear-thinning, as evidenced by the crossover between the $\epsilon=1$ and the pure FENE case.

In Fig.~\ref{fig:eta_steady}~(b), the viscosity at high shear rates is plotted for three values of the internal viscosity parameter, to quantify the rates of shear thinning for the various cases. The FENE-P model predicts that the shear-thinning of viscosity follows a power law behavior~\cite{Bird1987b}, with an exponent of $-\left(2/3\right)$. Exact BD simulations of a FENE dumbbell lead to an exponent of $-0.634(7)$, as seen from  Fig.~\ref{fig:eta_steady}(b). As the IV parameter is increased, viscosity decreases less steeply with the shear rate, and the magnitude of the shear-thinning exponent decreases. A coupling between internal viscosity and finite extensibility appears to determine the rate at which the viscosity decreases with shear rate in the asymptotic limit. The inclusion of hydrodynamic interactions does not significantly change the shear-thinning exponent (not shown in figure). { The shear-thinning exponent for the $\epsilon=10$ case is roughly $-1/3$, which corresponds to the exponent for the rigid dumbbell model. It is worth noting that the $-1/3$ exponent is observed at sufficiently high shear rates, even for larger values of the FENE parameter. Section~\ref{sec:rigid_compare} contains a more detailed comparison between the FENE dumbbells with high IV and the rigid dumbbell model.}

Similar to the trend observed for the polymer contribution to the viscosity, we see from Fig.~\ref{fig:psi1_steady}~(a) that hydrodynamic interactions increase the zero-shear rate first normal stress difference coefficient and quicken the onset of shear-thinning. Also, the effect of hydrodynamic interactions becomes less perceptible at higher shear rates. In a deviation from the trend observed for viscosity, the first normal stress difference coefficient for the cases with and without internal viscosity scale identically with shear rate at higher shear rates, as observed from the lack of a crossover between the $\epsilon=1$ and the pure FENE case.

\begin{figure}[t]
\begin{center}
\begin{tabular}{c}
{\includegraphics[width=3.3in,height=!]{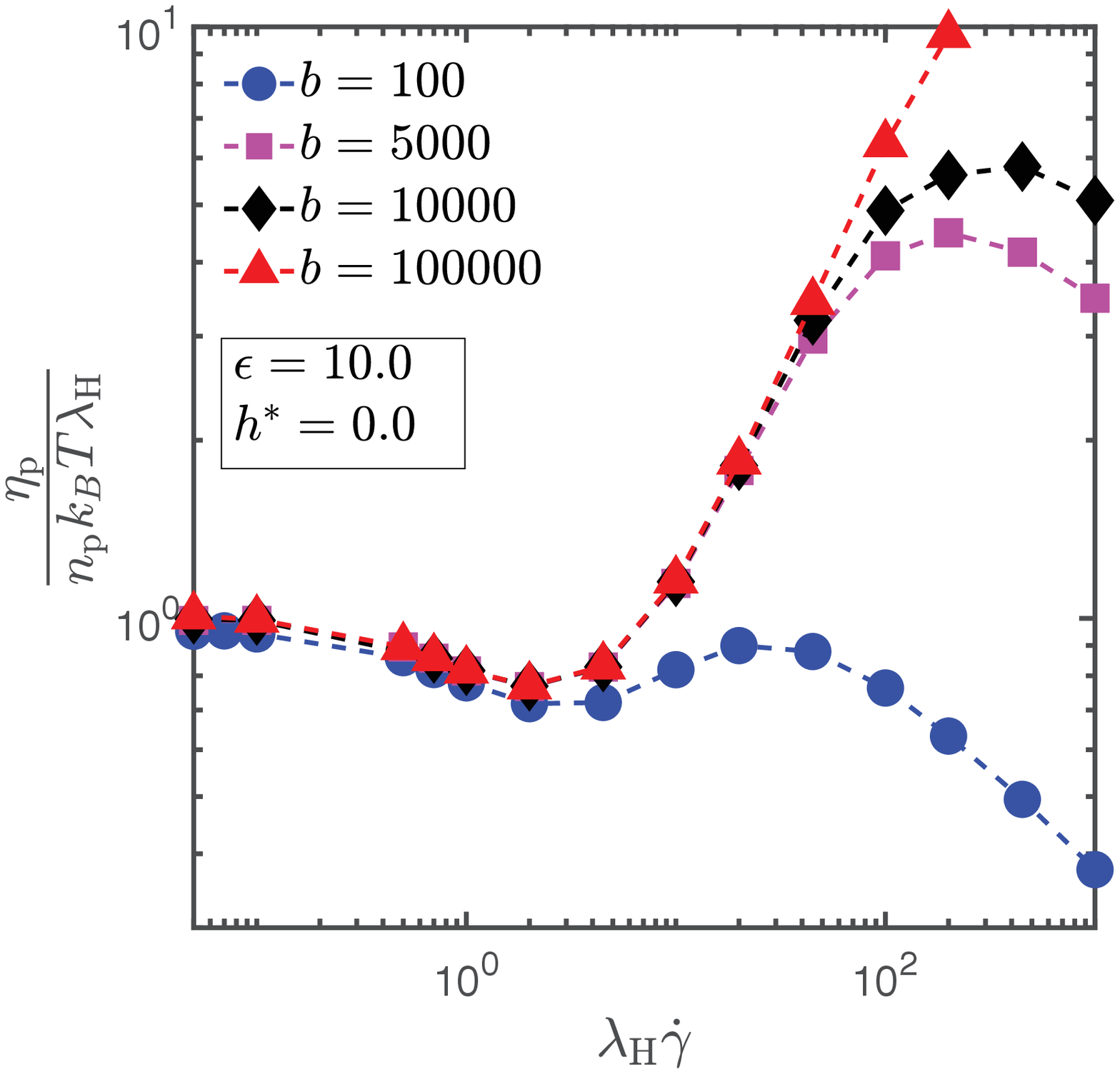}}\\
(a)\\
{\includegraphics[width=3.3in,height=!]{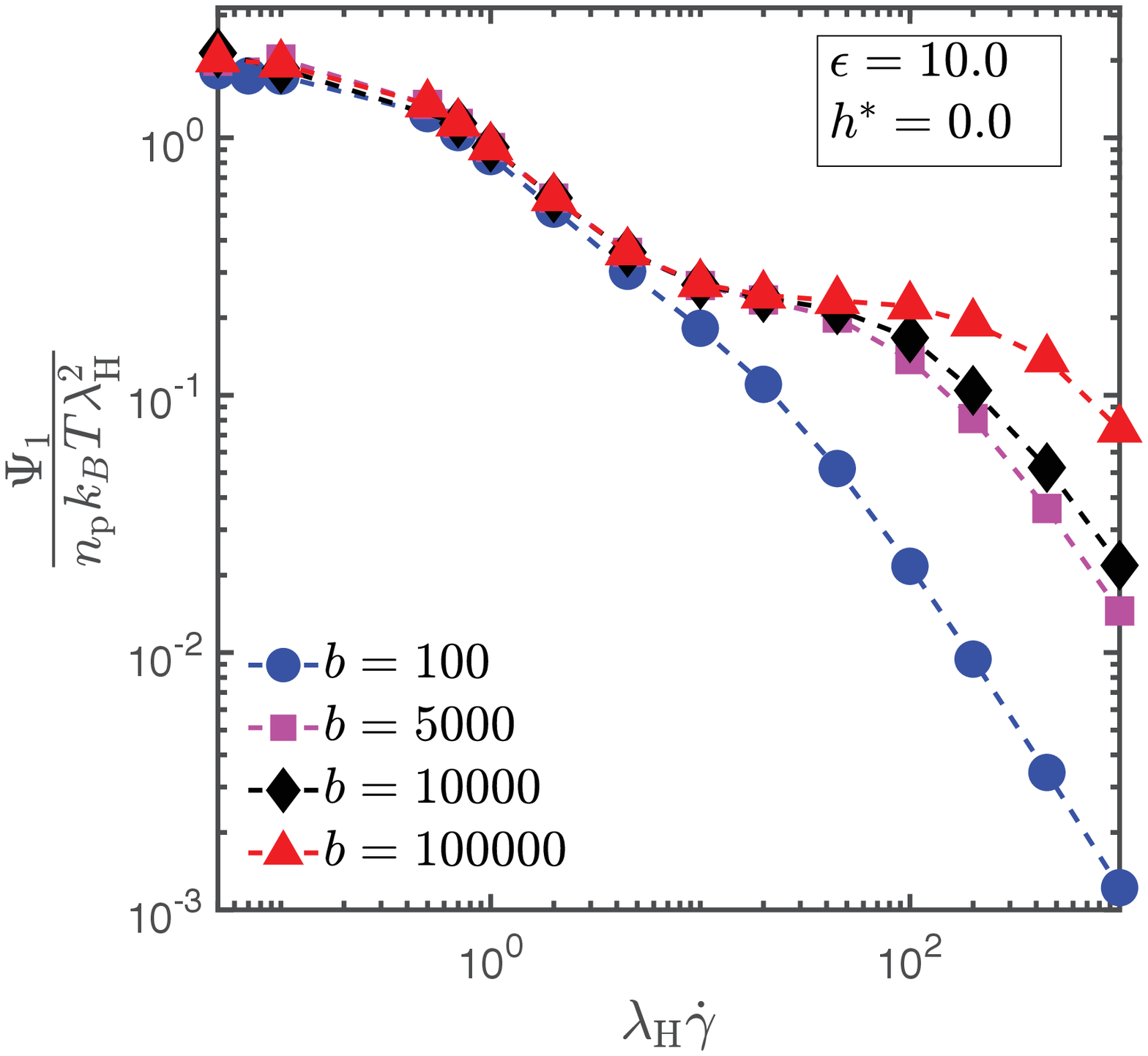}}\\
(b)  \\
\end{tabular}
\end{center}
\caption{(Color online) Effect of FENE parameter on the steady-shear viscometric functions {for free-draining dumbbells} with $\epsilon=10$: (a) dimensionless viscosity as a function of dimensionless shear rate, (b) dimensionless first normal stress difference coefficient as a function of dimensionless shear rate. Error bars are smaller than symbol size.}
\label{fig:bvar_eta_psi1}
\end{figure}

The asymptotic scaling of $\Psi_1$ with shear rate is captured more clearly in Fig.~\ref{fig:psi1_steady}~(b), where the shear-thinning at high shear rates has been quantified using the slope of the curve. It is seen that there is no  significant effect of the IV parameter on the shear thinning exponent for $\Psi_1$, in stark contrast to the dependence of the shear-thinning exponent on IV for viscosity. The FENE-P model predicts a shear-thinning exponent~\cite{Bird1987b} of $-\left(4/3\right)$ for $\Psi_1$, which is twice the shear-thinning exponent for the viscosity. As seen from the figure, a slope of $-1.282(6)$ is obtained for the $\epsilon=1.0$ case, which is also roughly the slope for the FENE dumbbell without internal viscosity, and twice the value of the shear-thinning exponent for the viscosity of the FENE dumbbell. Furthermore, the inclusion of hydrodynamic interactions is not seen to distinctly affect the shear-thinning exponent.
 
Figures~\ref{fig:bvar_eta_psi1}, which examine the shear rate-dependence of both the viscosity and the first normal stress difference coefficient on the FENE parameter {(for free-draining dumbbells)}, display significant differences in the behavior of both viscometric functions. The steady-shear values of these two viscometric functions are plotted for various values of the FENE parameter, at a fixed value of the internal viscosity parameter $(\epsilon=10)$. {Two limiting conditions arise in this discussion, that of the high shear rate limit ($\dot{\gamma}\to\infty$), and the Hookean limit ($b\to\infty$). Since the $b\to\infty$ limit is a singular one, the order in which these limits are taken matters.} Note that the averaging technique described in Fig.~\ref{fig:fluc_psi1}~(b) has been used to obtain the steady-state value for the first normal stress difference coefficient at $\lambda_{\text{H}}\dot{\gamma} < 0.1$ for all the values of the FENE parameter plotted in Fig.~\ref{fig:bvar_eta_psi1}~(b).

For the lowest value of the FENE parameter $(b=100)$ examined in Fig.~\ref{fig:bvar_eta_psi1}~(a), two shear-thinning regimes are observed for the viscosity, separated by a shear-thickening regime. {Since any finite value of the FENE parameter, no matter how large, is indicative of a nonlinear spring, its finite extensibility will cause both $\eta_{\text{p}}$ and $\Psi_1$ to eventually shear thin at large enough shear rates. This can be seen in Fig.~\ref{fig:bvar_eta_psi1}~(a), where it is clear that increasing the value of $b$ causes the power law region to be pushed to larger $\dot{\gamma}$. It appears that the shear-thinning exponent at sufficiently large $\dot{\gamma}$ is likely to be independent of $b$. If the FENE dumbbell is allowed to approach the Hookean limit, i.e; $b\to\infty$ (which, as noted is singular), we anticipate that there would be no shear-thinning at high shear rates, but only shear-thickening. }

Internal viscosity-induced shear thickening has also been observed by Hua and Schieber~\cite{Hua1995307}, who noticed that the inclusion of internal viscosity for Hookean dumbbells results in shear-thinning at lower shear rates followed by shear-thickening at higher shear rates.  {They do not observe a second shear thinning regime, as expected for Hookean dumbbells}. They also observe that an increase in the IV parameter shifts the onset of shear-thickening to lower shear rates. 

The pattern of shear-thinning-thickening-thinning is strikingly similar to that previously reported by ~\citet{Kishbaugh1990} and ~\citet{Prabhakar2006} in their work on multi-bead chains with finitely extensible springs in the presence of hydrodynamic interactions. In the Hookean limit, and in the presence of hydrodynamic interactions, they also observe an indefinite shear-thickening of viscosity which follows the initial shear-thinning. The thinning of viscosity at high shear rates has been attributed by these authors to the finite extensibility of the spring. It is known that the inclusion of hydrodynamic interactions results in a shear-thinning for Hookean dumbbells~\cite{Zylka1989}, and only induces a shear-thickening in bead-spring chains which have six or greater beads~\cite{Zylka1991}. As discussed in section \ref{sec:tau_comp}, for models with HI, the viscometric functions at high shear rates tend towards their values in the free-draining limit. For dumbbells $(N=2)$, the Rouse viscosity (free-draining) is lower than the Zimm viscosity (pre-averaged hydrodynamic interactions), and hence there is a shear-thinning when hydrodynamic interactions are included. For $N\geq6$, the Rouse viscosity is higher than the Zimm viscosity, and consequently, the viscosity tends towards the higher Rouse value at higher shear rates, resulting in shear-thickening. 

In contrast, internal viscosity is seen to cause shear-thickening even for the dumbbell case. The physical mechanism behind internal viscosity-induced shear thickening remains unclear. Manke and Williams have analyzed the transient stress response of multi-bead models with internal viscosity~\cite{Manke1989} using the LRV approximation, in the low $\epsilon$ regime. However, they do not present the steady-shear viscosity of these dumbbells as a function of shear rate, and we are currently unable to comment on the existence of shear-thickening of viscosity in multi-bead chains with internal viscosity. Furthermore, the failings of the LRV approximation are well-documented~\cite{Manke1988,Dasbach19924118}, and a rigorous treatment of the multi-bead model with internal viscosity is needed to draw meaningful conclusions about its steady-shear response.

In Fig.~\ref{fig:bvar_eta_psi1}~(b), for the lowest value of the FENE parameter $(b=100)$, it is seen that the first normal stress difference coefficient exhibits a continuous shear-thinning. An increase in the extensibility of the spring results in the appearance of a plateau in $\Psi_1$, followed by a second shear-thinning regime. Higher values of the FENE parameter pushes the onset of the second shear-thinning regime to higher shear rates, while widening the range of shear-rates over which the plateau is observed. In the Hookean limit, i.e;  $b\to\infty$, we expect the second shear-thinning regime to vanish completely. This is in accord with the results of Hua and Schieber~\cite{Hua1995307}, who observe a plateauing in $\Psi_1$ for Hookean dumbbells with internal viscosity. They also notice that an increase in the IV parameter causes the plateau to appear at lower shear rates.

\subsection{\label{sec:rigid_compare} Comparison of a model with large IV parameter with a rigid dumbbell model}
There has been significant interest in the literature, ~\cite{Manke1986,Manke1989,Manke1991,Manke1993} in approximating rheological properties\textemdash such as shear and complex viscosity \textemdash of rigid dumbbells with flexible polymer models using an infinitely high value of the IV parameter. ~\citet{Manke1986} argue that with an increase in the value of the IV parameter, the timescale for the stretching of the dumbbell also increases. As $\epsilon\to\infty$, the dumbbells rotate and orient themselves much quicker than the time needed for their stretching. Essentially, such a high value of $\epsilon$ ``freezes'' the stretching of the dumbbell's connector vector, so their lengths retain the original distribution they were sampled from. 

In this section, predictions by a FENE dumbbell model with a high value of the IV parameter ($\epsilon=10$) is compared against that by a rigid dumbbell for three observables, namely, the relaxation modulus, the stress jump, and the steady-shear viscosity. 

For a monodisperse ensemble of rigid dumbbells of length $L$, the relaxation modulus, $G^{\text{{uni}}}(L,t)$, is given by~\cite{Bird1987a}
\begin{equation}\label{eq:g_uniform}
\frac{G^{\text{{uni}}}(L,t)}{n_{\text{p}}k_BT}=2\left(\frac{\eta_{\text{s}}}{n_{\text{p}}k_BT}+\frac{2}{5}\lambda_{\text{R}}\right)\delta(t)+\frac{3}{5}\exp\left(-t/\lambda_{\text{R}}\right)
\end{equation}
where the rod relaxation time, $\lambda_{\text{R}} = \zeta L^2/12k_BT$.
~\citet{Hua1996} showed that the relaxation modulus of a Hookean dumbbell model with a large value of the IV parameter $(\epsilon=10)$ is well approximated by  $G^{\text{{H,mix}}}(t)$, which is the relaxation modulus of a rigid dumbbell model with a mixture of lengths. $G^{\text{H,mix}}(t)$ is calculated by convolving $G^{\text{{uni}}}(L,t)$ with the equilibrium distribution function for Hookean dumbbells. 

\begin{figure}[ht]
\centering
\includegraphics[width=3.5in,height=!]{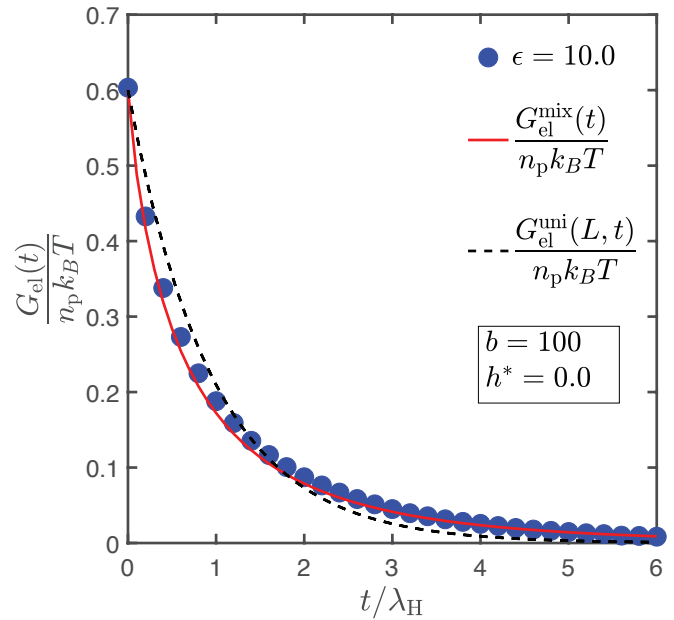}  
\caption{(Color online) Elastic component of the relaxation modulus {for free-draining} FENE dumbbells with $b=100$ and $\epsilon=10$. The dotted line corresponds to Eq.~(\ref{eq:g_uniform}) with $L=\sqrt{<Q^2>_{\text{eq}}}$, without the singularity. The solid line is a plot of Eq.~\eqref{gmix}, without the singularity. Error bars are smaller than symbol size. }
\label{fig:gtrigid}
\end{figure}

Following the approach of Hua, Schieber, and Manke~\cite{Hua1996}, the relaxation modulus for a system of rigid dumbbells whose lengths are sampled from a FENE distribution, $G^{\text{{mix}}}(t)$, is given by the following expression
\begin{equation}\label{eq:gmix_1}
\frac{G^{\text{{mix}}}(t)}{n_{\text{p}}k_BT} =\int \frac{G^{\text{{uni}}}(L,t)}{n_{\text{p}}k_BT}\psi^{*}_{\text{eq}}(\bm{Q^*})d\bm{Q^*} 
\end{equation}
where $\psi^{*}_{\text{eq}}(\bm{Q^*})$ is the equilibrium configurational distribution function for an ensemble of FENE dumbbells. The steps for evaluating the integral in Eq.~(\ref{eq:gmix_1}) and the resultant analytical expression for $G^{\text{{mix}}}(t)$ have been outlined in Appendix~\ref{sec:App_C}. 

In Fig.~\ref{fig:gtrigid}, the elastic component of the relaxation modulus {for free-draining} FENE dumbbells with an internal viscosity parameter of $\epsilon=10$ is plotted as a function of dimensionless time. There appears to be good agreement with the elastic component of $G^{\text{{mix}}}(t)$. Furthermore, it is observed that an attempt to fit the elastic portion of $G^{\text{{uni}}}(L,t)$, with $L$ equal to the equilibrium length of the FENE dumbbells with $\epsilon=10$, does not produce a good fit. Thus, at equilibrium, an internal viscosity parameter $\epsilon=10$ appears to be sufficient to capture rigid dumbbell behavior.  

In the presence of flow, however, the dumbbell lengths are not completely frozen when an IV parameter of $\epsilon=10$ is used, and a comparison with rigid dumbbell models is harder to draw. From Eq.~\eqref{gjv_rel} and \eqref{gmix}, the stress jump for a rigid dumbbell system, with a FENE distribution of lengths, is given by
\begin{equation}
\frac{\eta_{\text{jump,rigid}}}{n_{\text{p}}k_BT\lambda_{\text{H}}}=\frac{2}{5}\left(\frac{b}{b+5}\right)
\end{equation}
The analytical prediction of the same quantity for an ensemble of FENE dumbbells with IV is given by Eq.~(\ref{eq:fd_jump}). Comparing the two equations shows that the stress jump predictions will be identical for the two models only in the limit of $\epsilon \gg 1$. Using 
$\epsilon=10$, the stress jump prediction for the model with IV $[\eta^{*}_{\text{jump,IV}} = 0.346(3)]$ is within $10\%$ of that predicted by the rigid dumbbell model $[\eta^{*}_{\text{jump,rigid}} = 0.381]$, for $b=100$.

Using an approximate analytical model, Manke and Williams establish a rheological equivalence~\cite{Manke1986} between a  rigid dumbbell system and an ensemble of monodisperse dumbbells with an infinite value of the IV parameter, and argue that an ensemble of Hookean dumbbells with an infinite value of the internal viscosity parameter should resemble the viscometric functions of an ensemble of rigid dumbbells, at least qualitatively. Exact BD simulations on dumbbells with a high value of the IV parameter undergoing shear flow, however, paint a different picture~\cite{Hua1995307}. 

\begin{figure}[th]
\centering
\includegraphics[width=3.4in,height=!]{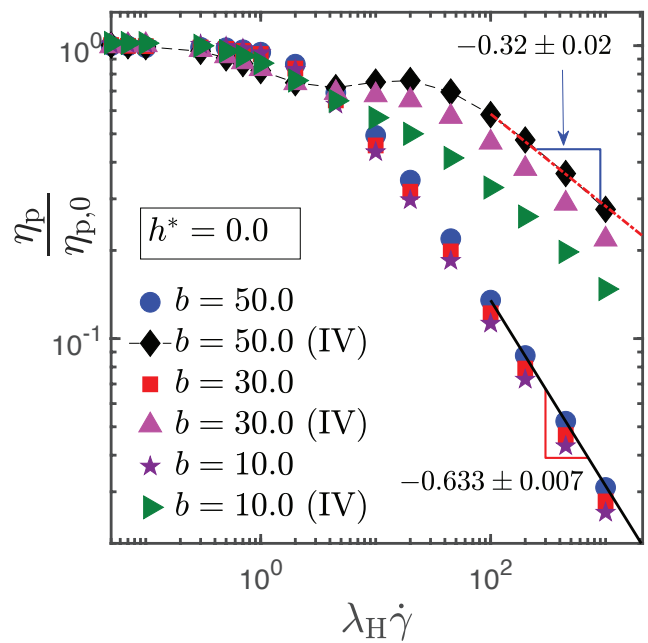}  
\caption{(Color online) Normalized shear viscosity as a function of dimensionless shear rate, for different values of the FENE parameter, {for free-draining dumbbells} with and without internal viscosity. The internal viscosity parameter used in these simulations is $\epsilon=10$. Error bars are smaller than symbol size.}
\label{fig:bvar_epsvar}
\end{figure}

We know from the work of ~\citet{Stewart1972} that the steady-shear viscosity profile of rigid dumbbells has an initial Newtonian plateau followed by shear-thinning with an exponent of $-\left(1/3\right)$. Hookean dumbbells with a high value of the IV parameter, on the other hand, display shear-thickening at high shear rates, as shown in Fig.~\ref{fig:bvar_eta_psi1}~(a). Therefore, we do not have sufficient reason to believe that increasing the value of the IV parameter to higher values for a Hookean dumbbell model would bring a qualitative similarity with rigid dumbbell behavior.

With reference to the discussion surrounding Fig.~\ref{fig:bvar_eta_psi1}~(a), ~\citet{Kishbaugh1990} and ~\citet{Prabhakar2006} observe for multi-bead chains (with finitely extensible springs and hydrodynamic interactions) that as the FENE parameter is decreased below a threshold value, the inflection point in the curve is seen to vanish, and only a continuous shear-thinning regime is observed, after an initial Newtonian plateau at low shear rates. The similarity between their system and that of a FENE dumbbell model with IV in the limit of a high value of $b$ suggests that a similar trend would be observed in the low $b$ limit as well. 

In Fig.~\ref{fig:bvar_epsvar}, the polymer contribution to shear viscosity (normalized by the zero-shear rate value) is plotted against the dimensionless shear rate, for various values of the FENE parameter, {for free-draining dumbbells} with and without internal viscosity. It is observed that at high shear rates, the viscosity of FENE dumbbells without internal viscosity scales roughly as $-0.63(3)$ with respect to the shear rate, for the three values of the FENE parameter considered here. As noted previously in the discussion surrounding Fig.~\ref{fig:eta_steady}~(b), this is in agreement with the FENE-P model prediction~\cite{Bird1987b}, which assigns a shear-thinning exponent of $-\left(2/3\right)$ at high shear rates, irrespective of the value of the FENE parameter. For models with internal viscosity $\left(\epsilon=10\right)$, the shear-thinning exponent for viscosity in all the three cases is roughly $-0.3(2)$. Interestingly, decreasing the value of the FENE parameter from $b=50$ to $b=10$ leads to the disappearance of the inflection point, and a smooth shear-thinning of viscosity is observed in the asymptotic limit of high shear rates. 

\begin{figure}[th]
\begin{center}
\begin{tabular}{c}
{\includegraphics*[width=2.8in,height=!]{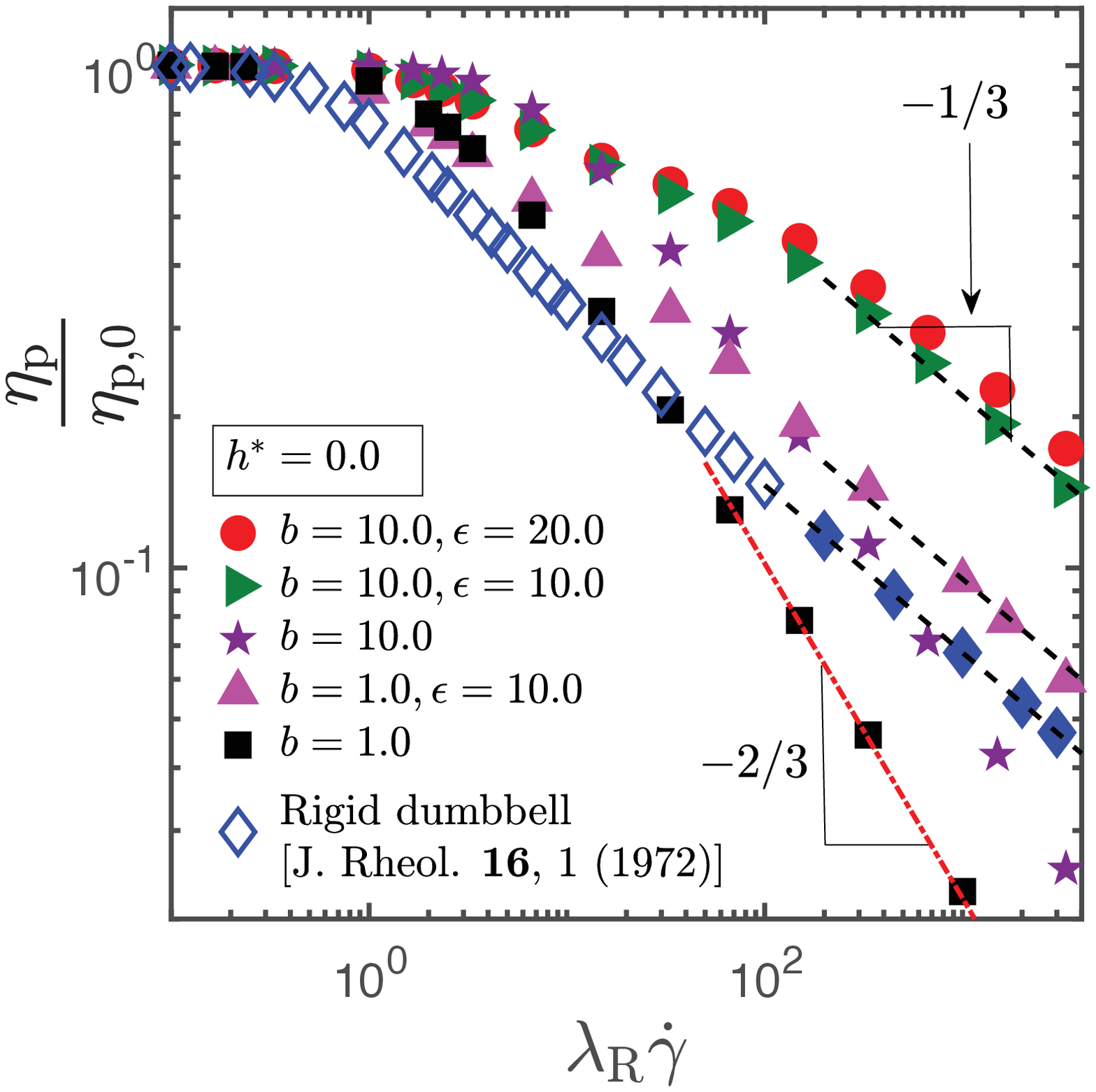}} \\
(a) \\
{\includegraphics*[width=3.1in,height=!]{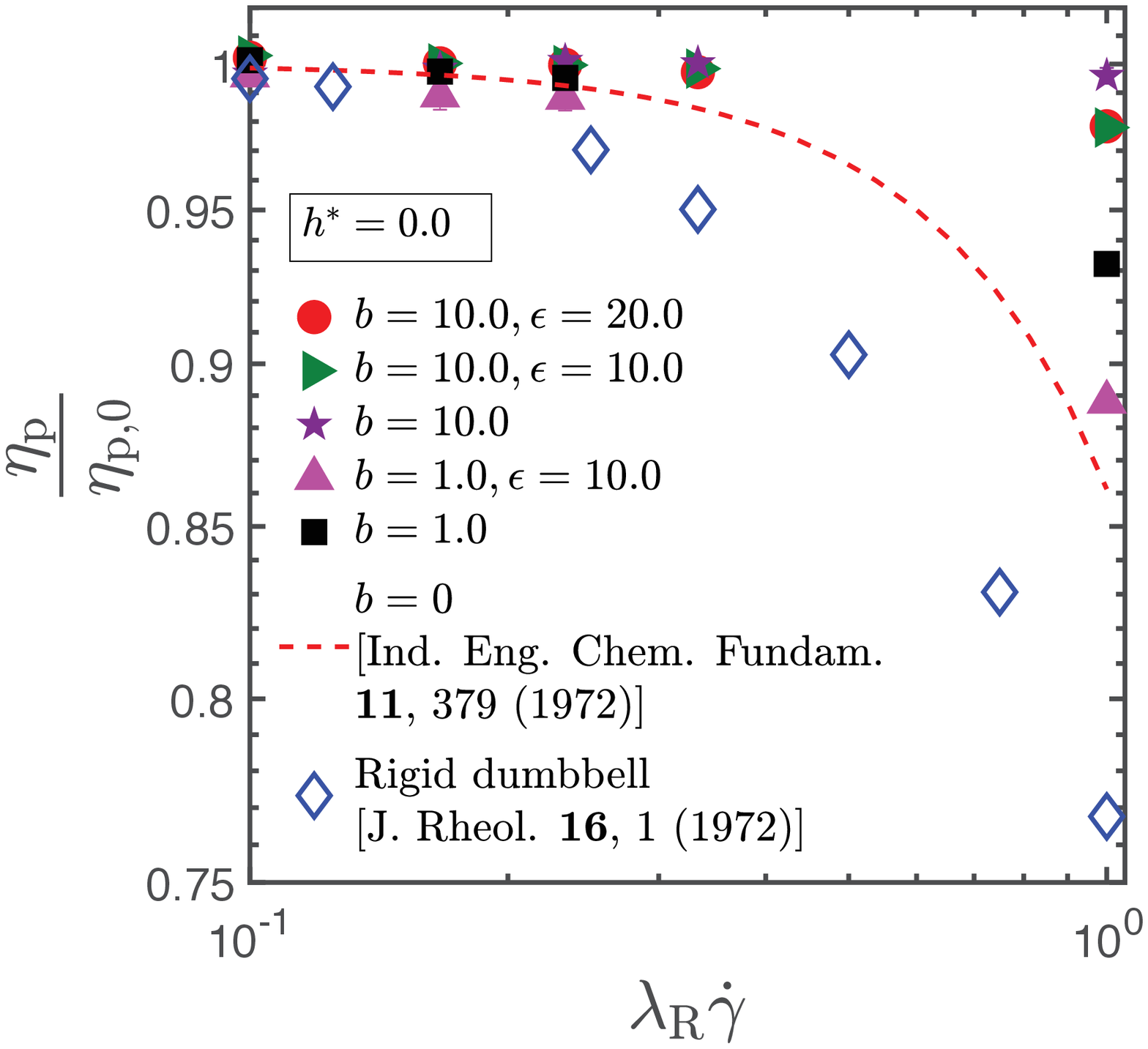}}\\
(b) \\
\end{tabular}
\end{center}
\vspace{-9pt}
\caption{(Color online)  Normalised shear viscosity as a function of dimensionless shear rate (non-dimensionlised with the bead-rod relaxation time), for free-draining FENE dumbbells with $b=1$ and $b=10$, with and without internal viscosity. (a) Comparison with rigid dumbbell results reported in Ref.~\citenum{Stewart1972}. Data corresponding to the empty blue-diamonds (\protect\markerfour) have been taken from Table II, while the filled blue-diamonds (\protect\markerfive) have been obtained using the expression for the asymptotic values of the viscosity ratio at high shear rates (Eq. (17)). (b) Comparison with the results of the bead-string model ($b=0$) reported in Ref.~\citenum{Warner1972}, at small values of shear rate. Error bars are smaller than symbol size.}
\label{fig:rig_comp}
\end{figure}

{While sufficiently decreasing the value of the FENE $b$ parameter, and increasing the IV parameter $\epsilon$, leads to the disappearance of the shear thickening regime, and to a power law shear thinning exponent that is the same as that for a bead-rod model, it is not clear however, if the entire curve for the FENE dumbbell model coincides with that for the bead-rod model over all values of shear rate. Bead-rod results cannot be reported  in Fig.~\ref{fig:bvar_epsvar} since the relaxation time $\lambda_{\text{H}}$ is not appropriate for non-dimensionalizing the shear rate in that case. The relevant relaxation time is the bead-rod relaxation time $\lambda_{\text{R}}\,(= \zeta Q_0^2/12k_BT)$, introduced previously in Eq.~(\ref{eq:g_uniform}), but defined here for a rod of length $Q_0$ in place of $L$. Using $\lambda_{\text{R}}$ to non-dimensionalize FENE dumbbell results does not pose a problem, and as a result, the viscosity ratio for the FENE dumbbell model for various values of $b$ and $\epsilon$ can be compared with bead-rod results over a wide range of shear rates, as shown in Figs.~\ref{fig:rig_comp}. The values of the viscosity ratio for the bead-rod model are taken from ~\citet{Stewart1972} who report data in the form of a Table for $\lambda_{\text{R}} \dot \gamma \lesssim 100$ (see Table II in Ref.~\citenum{Stewart1972}) and as an analytical expression for asymptotic values of the ratio at high shear rates (see Eq. (17) in Ref.~\citenum{Stewart1972}).}

{There are several points that are worth noting in Fig.~\ref{fig:rig_comp}~(a), and they are discussed in turn. Both the $b=10$ and $b=1$ curves have the same asymptotic shear thinning slope of $-(1/3)$ as the bead-rod model for $\epsilon \ge 10$ (as noted previously), but they lie above the bead-rod curve at all values of the shear rate. When $\epsilon$ is set equal to zero, the asymptotic shear thinning exponent for the FENE dumbbell model becomes $-(2/3)$, and as a result, curves for both $b=10$ and $b=1$ cross the bead-rod curve at large enough shear rates due to the enhanced shear thinning. An important point to note here is that the notion of the FENE dumbbell model as a coarse-grained representation of a bead-rod chain is no longer tenable for small values of $b$~\cite{Underhill:2004p1150,Pham}, and their use here should be viewed as merely an examination of a  phenomenological model that could serve as a substitute for the bead-rod model. A problem with decreasing values of $b$ is, however, that accurate results require smaller and smaller values of the time step.} 

{In his early investigation of FENE dumbbell models,\ ~\citet{Warner1972} introduced the notion of a \emph{bead-string} model that corresponds to the limit $b=0$, in which the force is negligible until a finite length. This is clearly as small a value as $b$ can take, and in this limit ~\citet{Warner1972} derived a power series expansion for the polymer contribution to viscosity up to second order in the non-dimensional shear rate (see Eq. (23) in Ref.~\citenum{Warner1972}). Fig.~\ref{fig:rig_comp}~(b) compares the viscosity ratio predicted by the bead-string, bead-rod and FENE dumbbell models for small values of shear rate. It is apparent that the extent of shear thinning is nearly independent of the value of $b$ at very small shear rates, and that the bead-rod model shear thins more rapidly than the FENE dumbbell model at any value of $b$. One can consequently conclude that while the asymptotic shear thinning exponent is identical for bead-rod and FENE dumbbell models for small values of $b$ and large values of $\epsilon$, the onset of shear thinning occurs at smaller shear rates for the bead-rod model, causing the viscosity curves for the two models to diverge.}

The rigid dumbbell model~\cite{Stewart1972} predicts a shear-thinning exponent of $-\left(4/3\right)$ for $\Psi_1$, which is four times the shear-thinning exponent for the viscosity. From the variation in the first normal stress difference coefficient as a function of shear rate (not plotted here), a shear-thinning exponent of approximately $-1.2(6)$ is obtained for the entire parameter set used in Fig.~\ref{fig:bvar_epsvar}, which is roughly four times the shear-thinning exponent for viscosity.

\section{\label{sec:conc}Conclusions}
Using a dumbbell model of a polymer that accounts for the finite extensibility of the spring, internal viscosity and hydrodynamic interaction effects, we have examined the effect of treating wet and dry friction exactly on a range of rheological properties. This model may be viewed as a preliminary pedagogical tool for capturing the influence of these phenomena on key rheological observables. 
The important results of our study are summarised below.  
\begin{enumerate}
\item
The most significant effect of hydrodynamic interactions is the magnification of the stress jump in dilute poymer solutions in comparison to the free-draining case, a finding that concurs with analytical predictions. 
\item
The zero-shear rate viscosity can be expressed as the sum of an elastic component (the integral of the relaxation modulus) and a viscous component (the stress jump). Values calculated in this manner have been compared against BD simulations and are in good agreement.
\item
The zero-shear rate viscometric functions are practically independent of the internal viscosity parameter, for free-draining dumbbells. The inclusion of hydrodynamic interactions, however, induces a non-trivial coupling with internal viscosity, and consequently, the zero-shear rate properties display a dependence on both the HI and the IV parameter. 
\item
Hydrodynamic interactions alter the transient viscometric functions perceptibly. However, their effect on the steady-shear properties are less marked, with the effect of HI weakening at higher shear rates. 
\item
Overshoots in viscosity and the first normal stress difference coefficient occur at progressively earlier times as the shear rate is increased, yet the strain at which overshoot occurs remains roughly constant over a wide range of shear rates, in agreement with experimental observations on polymer solutions and melts.
\item
In the asymptotic limit of high shear rates, internal viscosity significantly affects the shear-thinning exponent in viscosity, with the magnitude of the slope decreasing with an increase in the magnitude of the internal viscosity parameter. The shear-thinning exponent for the first normal stress difference coefficient, however, remains practically unaffected by the inclusion of internal viscosity.
\item
There is a remarkable, but unexplained, similarity in the steady-shear viscosity profiles of FENE-IV dumbbells and multi-bead chains with finitely extensible springs and hydrodynamic interactions.
\item
The relaxation modulus of an ensemble of FENE dumbbells with $\epsilon=10$ is identical to that of a mixture of rigid dumbbells.  For the stress jump, increasing values of the IV parameter give results that are in closer agreement with the rigid dumbbell case. In steady shear flow, the asymptotic shear thinning exponent is identical for bead-rod and FENE dumbbell models with small values of $b$ and large values of $\epsilon$. However, the onset of shear thinning occurs at smaller shear rates for the bead-rod model than for FENE dumbbells with any value of $b$, no matter how small.\end{enumerate}

A comparison with biophysical experiments that determine the reconfiguration time of proteins, or the energy landscape of polysaccharides, would necessitate the use of a multi-bead spring chain that incorporates IV and HI. ~\citet{Dasbach19924118} have obtained an approximate analytical solution for such a chain model. The use of BD simulations to solve the bead-spring-dashpot chain model exactly is rendered difficult by the fact that formulating the correct Fokker-Planck equation for such a system, and finding the equivalent set of stochastic differential equations, is non-trivial, and is the subject of our future study.

\section{\label{sec:Ak}Acknowledgements}

The authors would like to thank Burkhard D\"{u}nweg for valuable input regarding the validation of the code. The authors gratefully acknowledge CPU time grants from the National Computational Infrastructure (NCI) facility hosted by the Australian National University, the MonARCH facility maintained by Monash University, {and the computational resources granted by IIT Bombay}.\ We appreciate the funding and support from the IITB-Monash Research Academy. {The anonymous referee whose comments helped to improve the paper is also gratefully acknowledged.}

\appendix 
\section{\label{sec:App_A} Fokker-Planck and stochastic differential equations for a FENE dumbbell with IV and HI}

A force balance on the $\nu^{{\textrm{th}}}$ bead of a multi-bead chain with internal viscosity, hydrodynamic interactions, and a FENE spring can be written as follows, 
\begin{equation}\label{eq:fbal}
\bm{F}=m\textbf{a}=\bm{F}_{\nu}^{(h)} + \bm{F}_{\nu}^{(b)} + \bm{F}_{\nu}^{(c)}
\end{equation}
 where $\bm{F}_{\nu}^{(h)}$ is the hydrodynamic force,  $\bm{F}_{\nu}^{(b)}$ is the Brownian force, and $\bm{F}_{\nu}^{(c)}$ is the force due to the spring-dashpot system for any arbitrary spring force. Expressions for each of these forces are given in Bird et al.~\cite{Bird1987b} On substituting the expressions, and neglecting the masses of the beads, the force balance can be recast as
\begin{widetext}
\begin{equation}
\begin{split}
\label{eq:fbal_mod}
 - \zeta\Biggl[\llbracket{\dot {\bm{r}}}_{\nu}\rrbracket-\bm{v}_{0}-\boldsymbol{\kappa}\cdot\bm{r}_{\nu}+\sum_{\mu}\boldsymbol{\Omega}_{\nu \mu}\cdot\bm{F}_{\mu}^{(h)}\Biggr]  & -k_BT\frac{\partial \ln \psi}{\partial \bm{r}_{\nu}}  - \frac{\partial \phi}{\partial \bm{r}_{\nu}}
 + K\Biggl(\frac{\bigl(\bm{r}_{\nu+1} - \bm{r}_{\nu}\bigr)\bigl(\bm{r}_{\nu+1} - \bm{r}_{\nu}\bigr)}{|\bm{r}_{\nu+1} - \bm{r}_{\nu}|^2}\Biggr)\cdot\llbracket{\dot {\bm{r}}}_{\nu+1}-\dot{\bm{r}}_{\nu}\rrbracket  \\[5pt]
& - K\Biggl(\frac{\bigl(\bm{r}_{\nu} - \bm{r}_{\nu - 1}\bigr)\bigl(\bm{r}_{\nu} - \bm{r}_{\nu - 1}\bigr)}{|\bm{r}_{\nu} - \bm{r}_{\nu - 1}|^2}\Biggr)\cdot\llbracket{\dot {\bm{r}}}_{\nu}-\dot{\bm{r}}_{\nu - 1}\rrbracket = 0
\end{split} 
\end{equation}

For a dumbbell, the time-rate of change of the position vectors of the two beads can then be written as 

\begin{equation}\label{eq:r_1}
\llbracket{\dot {\bm{r}}}_{1}\rrbracket = \bm{v}_{0}+\boldsymbol{\kappa}\cdot\bm{r}_{1}+\boldsymbol{\Omega}\cdot\Biggl(-k_BT\frac{\partial \ln \psi}{\partial \bm{r}_{2}}-\frac{\partial \phi}{\partial \bm{r}_{2}} 
- K\frac{\bm{QQ}}{Q^2}\cdot\llbracket{\dot {\bm{Q}}}\rrbracket\Biggr) -\frac{k_BT}{\zeta}\frac{\partial \ln \psi}{\partial \bm{r}_{1}} 
- \frac{1}{\zeta}\frac{\partial \phi}{\partial \bm{r}_{1}}+\frac{\epsilon}{2}\frac{\bm{QQ}}{Q^2}\cdot\llbracket{\dot {\bm{Q}}}\rrbracket 
\end{equation}
and
\begin{equation}\label{eq:r_2}
\llbracket{\dot {\bm{r}}}_{2}\rrbracket = \bm{v}_{0}+\boldsymbol{\kappa}\cdot\bm{r}_{2}+\boldsymbol{\Omega}\cdot\Biggl(-k_BT\frac{\partial \ln \psi}{\partial \bm{r}_{1}}-\frac{\partial \phi}{\partial \bm{r}_{1}} 
+ K\frac{\bm{QQ}}{Q^2}\cdot\llbracket{\dot {\bm{Q}}}\rrbracket\Biggr) -\frac{k_BT}{\zeta}\frac{\partial \ln \psi}{\partial \bm{r}_{2}} 
 - \frac{1}{\zeta}\frac{\partial \phi}{\partial \bm{r}_{2}}-\frac{\epsilon}{2}\frac{\bm{QQ}}{Q^2}\cdot\llbracket{\dot {\bm{Q}}}\rrbracket 
\end{equation}
Subtracting Eq.~(\ref{eq:r_2}) from Eq.~(\ref{eq:r_1}) yields the following equation for the time-rate of change of the connector vector, $\llbracket{\dot {\bm{Q}}}\rrbracket$, 
\begin{equation}
\llbracket{\dot {\bm{Q}}}\rrbracket = \left[\boldsymbol{\kappa}\cdot\bm{Q}\right] - \frac{k_BT}{\zeta}\bm{M}\cdot\frac{\partial}{\partial \bm{Q}}\ln \psi 
- \frac{1}{\zeta}\bm{M}\cdot\frac{\partial \phi}{\partial \bm{Q}}-\frac{\epsilon}{2}\bm{M}\cdot\frac{\bm{QQ}}{Q^2}\cdot\llbracket{\dot {\bm{Q}}}\rrbracket
\end{equation}
Grouping together the terms containing $\llbracket{\dot {\bm{Q}}}\rrbracket$, the equation can be rewritten as 
\begin{equation}\label{eq:qdot_part1}
\llbracket{\dot {\bm{Q}}}\rrbracket = \left[\boldsymbol{\delta} + \epsilon\beta\frac{\bm{QQ}}{Q^2}\right]^{-1}\cdot\Biggl(\left[\boldsymbol{\kappa}\cdot\bm{Q}\right] - \frac{k_BT}{\zeta}\bm{M}\cdot\frac{\partial}{\partial \bm{Q}}\ln \psi 
- \frac{1}{\zeta}\bm{M}\cdot\frac{\partial \phi}{\partial \bm{Q}}\Biggr)
\end{equation}
\end{widetext}
One can find the inverse of the first bracketed term on the RHS of Eq.~(\ref{eq:qdot_part1}) analytically with the Sherman-Morrison formula~\cite{press2007numerical}, which states that for a matrix $\bm{Z}$ whose inverse $\bm{Z}^{-1}$ is known, 
\begin{equation}
\left[\bm{Z}+\bm{u}\bm{v}\right]^{-1}=\bm{Z}^{-1} - \frac{(\bm{Z}^{-1}\cdot\bm{u})(\bm{Z}^{-1}\cdot\bm{v})}{1 + \bm{v}\cdot\bm{Z}^{-1}\cdot\bm{u}}
\end{equation}
where $\bm{u}$ and $\bm{v}$ are vectors. Identifying $\boldsymbol{\delta}$ as $\bm{Z}$,  $\sqrt{\epsilon\beta}\left({\bm{Q}}/{Q}\right)$ as $\bm{u}$,  and $\sqrt{\epsilon\beta}\left({\bm{Q}}/{Q}\right)$ as $\bm{v}$, we get
\begin{equation}\label{eq:inverse}
\left[\boldsymbol{\delta} + \epsilon\beta\frac{\bm{QQ}}{Q^2}\right]^{-1} = \boldsymbol{\delta} - \frac{\epsilon\beta}{\epsilon\beta + 1}\frac{\bm{QQ}}{Q^2}
\end{equation}
Once the inverse has been found in this manner, the equation for $\llbracket{\dot {\bm{Q}}}\rrbracket$ can be written as shown in Eq.~(\ref{eq:dim1}). Substituting the expression for the time-rate of change of the connector vector into the equation of continuity yields the appropriate Fokker-Planck equation for the system, as given by Eq.~(\ref{eq:fp_dim}). The dimensionless form of the same has been given in Eq.~(\ref{eq:fp_mid_rpy1}).

Using It\^o's interpretation, any Fokker-Planck equation of the following form
\begin{equation}\label{eq:fp_ref}
\frac{\partial{\psi}}{\partial t} = -\frac{\partial}{\partial{\bm{Q}}}\cdot\left(\bm{a}\psi\right) + \frac{1}{2}\frac{\partial}{\partial{\bm{Q}}}\frac{\partial}{\partial{\bm{Q}}}:\left[\bm{D}\psi\right]
\end{equation}
has its equivalent SDE~\cite{Ottinger1996} given by
\begin{equation}
d\bm{Q} = \bm{a}dt + \bm{b}\cdot d\bm{W}_t
\end{equation}
where $\bm{W}_t$ is a Wiener process and $\bm{b}\cdot\bm{b^T} = \bm{D}$. Using the identity, 
\begin{equation*}\label{eq:stod}
\frac{\partial}{\partial \bm{Q}}\cdot\left[\bm{L}\cdot\frac{\partial f}{\partial \bm{Q}}\right] =  \frac{\partial}{\partial{\bm{Q}}}\frac{\partial}{\partial{\bm{Q}}}:\left[\bm{L}^{\text{T}}f\right] - \frac{\partial}{\partial \bm{Q}}\cdot\left[f\frac{\partial}{\partial{\bm{Q}}}\cdot\bm{L}^{\text{T}}\right] 
\end{equation*}
where $\bm{L}$ is a tensor and $f$ is a scalar, the second term on the RHS of Eq.~(\ref{eq:fp_mid_rpy1}) can be written as 
\begin{widetext}
\begin{equation}
\begin{split}
\label{eq:converted_dot}
\frac{1}{4}\frac{\partial}{\partial \bm{Q}^*}\cdot\left\{\left[\left( \boldsymbol{\delta} - \frac{\epsilon\beta^*}{\epsilon\beta^* + 1}\frac{\bm{Q^*Q^*}}{{Q}^{*2}}\right)\cdot\bm{M^*}\right]\cdot\frac{\partial \psi^*}{\partial \bm{Q}^*}\right\} & = 
\frac{1}{4}\frac{\partial}{\partial{\bm{Q^*}}}\frac{\partial}{\partial{\bm{Q^*}}}\bm{:}\left[\Bigl(\boldsymbol{\delta} - \frac{\epsilon\beta^*}{\epsilon\beta^* + 1}\frac{\bm{Q^*Q^*}}{{Q}^{*2}}\Bigr)\cdot\left(\bm{M^*}\right)\psi^*\right]  \\[5pt]
& - {\frac{\partial}{\partial{\bm{Q^*}}}\cdot\left[\frac{g_2}{2}\frac{\bm{Q^*}}{Q^*}\psi^*\right]}
\end{split}
\end{equation}
With the above conversion, the Fokker-Planck can be rewritten in a form that is amenable for applying the It\^o interpretation, as shown below, 
\begin{align}\label{eq:fp_fin_rpy}
\frac{\partial \psi^*}{\partial t^*}  &= -\frac{\partial}{\partial \bm{Q}^*}\cdot\Biggl\{\frac{g_2}{2}\frac{\bm{Q^*}}{Q^*}  + \Biggl( \boldsymbol{\delta} - \frac{\epsilon\beta^*}{\epsilon\beta^* + 1}\frac{\bm{Q^*Q^*}}{{Q}^{*2}}\Biggr)\cdot 
\left(\boldsymbol{\kappa^*}\cdot\bm{Q^*} - \frac{\bm{M^*}}{2}\cdot\frac{\frac{1}{2}\bm{Q^*}}{1-{Q^{*2}}/{b}}\right)\psi^*\Biggr\} \nonumber \\
& + \frac{1}{4}\frac{\partial}{\partial{\bm{Q^*}}}\frac{\partial}{\partial{\bm{Q^*}}}\bm{:}\Biggl[\Biggl(\boldsymbol{\delta} - \frac{\epsilon\beta^*}{\epsilon\beta^* + 1}\frac{\bm{Q^*Q^*}}{{Q}^{*2}}\Biggr)\cdot\left(\bm{M^*}\right)\psi^*\Biggr] 
\end{align}
\end{widetext}

The SDE corresponding to this Fokker-Planck equation has been given in Eq.~(\ref{eq:sde_final}), and has the same functional form, irrespective of whether the hydrodynamic interaction tensor is described using the RPY expression or the Regularized Oseen Burgers expression. The definitions of $g_1$ and $g_2$ are as follows.
\begin{equation}\label{eq:g_1}
g_1  = \frac{\alpha B^*Q^*+\epsilon(Q^*-A^*\alpha)[Q^*-\alpha(A^*+B^*)]}{(Q^*-A^*\alpha)\left\{Q^*+\epsilon[Q^*-\alpha(A^*+B^*)]\right\}} 
\end{equation}
and
\begin{eqnarray}\label{eq:g_2}
g_2 & =& \frac{2\alpha B^*}{\left\{Q^*+\epsilon[Q^*-\alpha(A^*+B^*)]\right\}^2}
\nonumber \\[5pt]
& - & \rho\,\frac{2\epsilon \alpha [Q^*-\alpha(A^*+B^*)]\left[2Q^*+\epsilon[Q^*-\alpha(A^*+B^*)]\right]}{Q^{*2}\left\{Q^*+\epsilon[Q^*-\alpha(A^*+B^*)]\right\}^2}\nonumber \\[5pt]
&-&2g_1\left(\frac{Q^*-A^*\alpha}{Q^{*2}}\right) 
\end{eqnarray}
where
\begin{equation}\label{eq:s_rho}
\rho = (s-B^*),  \, s=\frac{1}{2}\left[(A^*+B^*) - Q^*\frac{\partial}{\partial Q^*}(A^*+B^*)\right]
\end{equation}
Interestingly, for the RPY tensor, the following property holds for both its branches.
\begin{equation}\label{eq:magic_RPY}
(A^*+B^*) - Q^*\frac{\partial}{\partial Q^*}(A^*+B^*) = 2B^*
\end{equation}
This results in a simplification and we have
\begin{equation}\label{eq:g_2_rpy}
g^{\text{{RPY}}}_2 = \frac{2\alpha B^*}{\left\{Q^*+\epsilon[Q^*-\alpha(A^*+B^*)]\right\}^2} - 2g_1\left(\frac{Q^*-A^*\alpha}{Q^{*2}}\right)
\end{equation} 

\begin{figure}[b]
\centering
\includegraphics[width=3.5in,height=!]{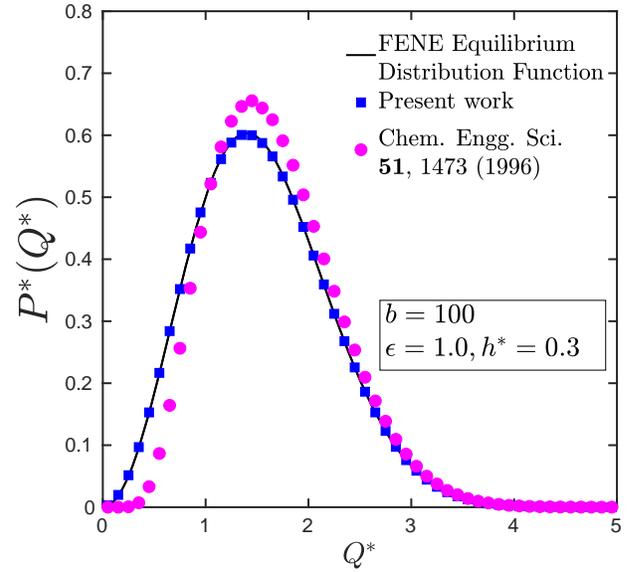}  
\caption{(Color online) Probability distribution of $Q^*$, the dimensionless length of the connector vector. Solid line corresponds to the analytical function given by Eq.~(\ref{eq:pofq}) for $b=100$. Error bars are smaller than symbol size.}
\label{fig:prob_dist}
\end{figure}

Though the SDE obtained in this work is identical to the one derived by ~\citet{Hua19961473} , the definition of $g_2$ in our work is different from that obtained in theirs. Since it is known that IV and HI do not affect the equilibrium probability distribution of the dumbbell configurations, we can test the correctness of the SDE by comparing the probability distribution of the lengths of the connector vector obtained from simulations against its analytically known expression for FENE dumbbells. The equilibrium configurational distribution function for an ensemble of FENE dumbbells has the following form~\cite{Hua1995307}, 
\begin{dmath}\label{eq:fen_eqb}
\psi^{*}_{\text{eq}}(\bm{Q}^*)=\frac{1}{J^*_{\text{eq}}}\left(1-\frac{Q^{*2}}{b}\right)^{b/2}
\end{dmath}
where $J^*_{\text{eq}}=2\pi b^{3/2}B(3/2,((b+2)/2))$. By averaging over the orientations of the dumbbells in spherical coordinates, the probability distribution of the \textit{lengths} of the connector vector can be obtained as 
\begin{dmath}\label{eq:pofq}
P^*\left(Q^*\right)=\frac{4\pi}{J^*_{\text{eq}}}Q^{*2}\psi^{*}_{\text{eq}}(\bm{Q}^*)
\end{dmath}
In Fig.~\ref{fig:prob_dist}, the probability distribution generated by our code and that which results when the SDE from the work by Hua and Schieber~\cite{Hua19961473} is used, are plotted alongside the function given by Eq.~(\ref{eq:pofq}) for $b=100$. The good agreement between the simulation results obtained by our code with the analytical result establishes the validity of our SDE.

\section{\label{sec:App_B} Stress Tensor Expression}
The Kramers-Kirkwood expression for the stress tensor, written in terms of the bead vectors, is given by Eq.~(\ref{eq:kram_kirk}), where 
\begin{equation}
\bm{R}_{\nu} = \bm{r}_{\nu} - \bm{r}_{c}\,; \quad \bm{r}_{c} = \frac{\bm{r}_{1}+\bm{r}_{2}}{2}
\end{equation}
and 
\begin{dmath}
\bm{F}_{\nu}^{(h)}=\zeta\bm{C}\cdot\left[\llbracket{\dot {\bm{r}}_{\mu}}\rrbracket-\bm{v}_{o} - \boldsymbol{\kappa}\cdot\bm{r}_{\mu}\right]
\end{dmath}
with
\begin{dmath}
\bm{C}=\left(\frac{Q}{Q-Ah}\right)\left[\boldsymbol{\delta}+\frac{hB}{Q\beta}\frac{\bm{QQ}}{Q^2}\right]
\end{dmath}
In terms of the connector vector $\bm{Q}$, the stress tensor expression is as follows
\begin{equation}\label{eq:tau_p}
\boldsymbol{\tau}_{\text{{p}}}=-\frac{n_{\text{p}}}{2}\left<\bm{Q}\left[\zeta\bm{C}\cdot\left[\boldsymbol{\kappa}\cdot\bm{Q}-\llbracket{\dot {\bm{Q}}}\rrbracket\right]\right]\right>
\end{equation}
On substituting the expression for $\llbracket{\dot{\bm{Q}}}\rrbracket$ from Eq.(\ref{eq:dim1}) into Eq.(\ref{eq:tau_p}) and simplifying, 
\begin{widetext}
\begin{equation}
\label{eq:inter_tau}
\boldsymbol{\tau}_{\text{{p}}} = -\frac{n_{\text{p}}\zeta \epsilon}{2}\Bigg \langle{g_3\boldsymbol{\kappa}:\frac{\bm{QQQQ}}{Q^{2}}}\Bigg \rangle-{n_{\text{p}}H}\Bigg \langle g_3\frac{\bm{QQ}}{1-(Q/Q_0)^2}\Bigg \rangle 
 + \frac{n_{\text{p}}\zeta}{2}\Bigg \langle\biggl(\frac{Q}{Q-Ah}\biggr)\Biggl[\biggl(\delta -\theta \frac{\bm{QQ}}{Q^2}\biggr)\cdot\biggl(-\frac{k_BT}{\zeta}\bm{A}
\cdot \frac{\partial}{\partial \bm{Q}}\ln \psi\biggr)\Biggr]\Bigg \rangle
\end{equation}
where
\begin{equation}
\theta=\frac{\epsilon \beta}{\epsilon \beta + 1} - \frac{hB}{Q\beta (\epsilon \beta + 1)}
\end{equation}
Using the following identity 
\begin{equation}\label{eq:iv_prefac}
\frac{\partial }{\partial Q}\left(\frac{\epsilon\beta}{\epsilon\beta + 1}\right) = \frac{\epsilon h}{\left\{Q+\epsilon[Q-h(A+B)]\right\}^2} \Biggl[(A+B) 
- Q\frac{\partial}{\partial Q}(A+B)\Biggr]
\end{equation}
to simplify Eq.~(\ref{eq:inter_tau}), the equation for the stress tensor  given in dimensionless form in Eq.~(\ref{eq:stress_tensor}) can be obtained. The prefactors $g_3$ and $g_4$ in Eq.~(\ref{eq:stress_tensor}) are defined as follows, 
\end{widetext}
\begin{align}
g_3 & = \frac{1}{\epsilon\beta^* + 1} \nonumber \\
& =\frac{Q^*}{Q^*+\epsilon[Q^*-\alpha(A^*+B^*)]}
\end{align} 
and
\begin{align}
g_4 & =\frac{2s\alpha Q^*}{\left\{Q^*+\epsilon[Q^*-\alpha(A^*+B^*)]\right\}^2} 
\nonumber \\[10pt]
 & + \frac{3[Q^*-\alpha(A^*+B^*)]}{\left\{Q^*+\epsilon[Q^*-\alpha(A^*+B^*)]\right\}}
\end{align} 
where $s$ has already been defined in Eq.~(\ref{eq:s_rho}).

\section{\label{sec:App_C} Relaxation modulus for rigid dumbbells with a distribution of lengths}
The integral in Eq.~(\ref{eq:gmix_1}) can be evaluated by converting to spherical co-ordinates, and using the definition for the FENE equilibrium distribution function given in Eq.~(\ref{eq:fen_eqb}), as shown below, 
\begin{align} \label{gmix}
\frac{G^{\text{{mix}}}(t)}{n_{\text{p}}k_BT} &= 4\pi\int_{0}^{\infty} \frac{G^{\text{{uni}}}(L,t)}{n_{\text{p}}k_BT}\psi^{*}_{\text{eq}}(\bm{Q^*}) Q^{*2}d{Q}^*  \nonumber \\[10pt]
&= 2\left[\left(\frac{\eta_{\text{s}}}{n_{\text{p}}k_BT}\right)+\frac{2}{5}\left(\frac{b}{b+5}\right)\lambda_{\text{H}}\right]\delta(t) \nonumber \\[10pt]
&+\frac{3}{5}\left[y_1(t)+c_2q^{3/2}(t)y_2(t)\right] \nonumber \\[10pt]
&= 2\left(\frac{\eta_{\text{s}}+\eta_{\text{jump,rigid}}}{n_{\text{p}}k_BT}\right)\delta(t)+\frac{G^{\mathrm{mix}}_{\mathrm{el}}(t)}{n_{\mathrm{p}}k_BT}
\end{align}
where $\lambda_{\text{H}}=\zeta/4H$, $q(t)={3t}/{\lambda_{\text{H}} b}$ and
\begin{equation}
\begin{aligned}\label{eq:gmix_form}
y_1(t) &= {}_1F_1\left(-\left(\frac{b+3}{2}\right);-0.5;-q(t)\right)\\
y_2(t) &= {}_1F_1\left(-\frac{b}{2};2.5;-q(t)\right)\\
c_2      &= \frac{4\sqrt{\pi}}{3B(3/2,((b+2)/2))}
\end{aligned}
\end{equation}
${}_1F_1$ is the confluent hypergeometric function of the first kind, defined by
\begin{equation}\label{eq:h1gf1}
{}_1F_1(c;d;x)=\sum_{k=0}^{\infty}\frac{(c)_k}{(d)_k}\frac{x^k}{k!}
\end{equation}
$(c)_k$ and $(d)_k$ are Pochhammer symbols defined by
\begin{equation}\label{eq:pochammer}
(y)_m=\frac{\Gamma(y+m)}{\Gamma(y)}
\end{equation}
where $\Gamma(.)$ is the gamma function~\cite{handbook-mathematical}.

 \newcommand{\noop}[1]{}

\end{document}